\documentclass[prd,nofootinbib,preprint,superscriptaddress]{revtex4}

\usepackage{amsmath, amssymb, amsthm, graphicx, epsfig, fancyhdr,epsfig, slashed}

\usepackage{tikzsymbols}
\usepackage[compat=1.0.0]{tikz-feynman}
\usepackage{natbib}
\usepackage{float}
\usepackage{xcolor}

\usepackage{enumitem}
\usepackage{amsmath}

\usepackage{tikz,xcolor,hyperref}
 \usepackage{%ulem,
 soul,
 cancel,%footmisc,
 enumitem,
 multirow,
 slashed,
 upgreek, 
 physics, cases}
 \usepackage{easyReview}
\usepackage{bbold}
\usepackage[normalem]{ulem}

\usepackage{slashed}

\usepackage{subcaption}
\usepackage{braket}

\usepackage{hhline}
\usepackage{multirow}

\definecolor{lime}{HTML}{A6CE39}
\DeclareRobustCommand{\orcidicon}{
	\begin{tikzpicture}
	\draw[lime, fill=lime] (0,0) 
	circle [radius=0.2] 
	node[white] {{\fontfamily{qag}\selectfont \tiny ID}};
	\draw[white, fill=white] (-0.0625,0.095) 
	circle [radius=0.007];
	\end{tikzpicture}
	\hspace{-2mm}
}

\foreach \x in {A, ..., Z}{\expandafter\xdef\csname orcid\x\endcsname{\noexpand\href{https://orcid.org/\csname orcidauthor\x\endcsname}
			{\noexpand\orcidicon}}
}

\newcommand{\be}{\begin{equation}}
\newcommand{\ee}{\end{equation}}
\newcommand{\bea}{\begin{eqnarray}}
\newcommand{\eea}{\end{eqnarray}}

\newcommand{\ba}{\begin{eqnarray}}
\newcommand{\ea}{\end{eqnarray}}
\newcommand{\bi}{\begin{itemize}}
\newcommand{\ei}{\end{itemize}}
\newcommand{\x}{\star}
\begin{document}
%%%%%%%%%%%%%%%%%%%%%%%%%%%%%%%%%%%%%%%%%%%%%%%%%%%%%%%%%%%%%%%%%%%%%%%%%%%%%%
%%%%%%%%%%%%%%%%%%%%%%%%%%%%%%%%%%%%%%%%%%%%%%%%%%%%%%%%%%%%%%%%%%%%%%%%%%%%%%
%%%%%%%%%%%%%%%%%%%%%%%%%%%%%%%%%%%%%%%%%%%%%%%%%%%%%%%%%%%%%%%%%%%%%%%%%%%%%%

\title{Induced Multi-phase Inflation with Reheating: \\ \it{Leptogenesis and Dark Matter Production in Metric versus Palatini}}
%%%%%%%%%%%%%%%%%%%%%%%%%%%%%%%%%%%%%%%%%%%%%%%%%%%%%%%%%%%%%%%%%%%%%%%%%%%%%%
%%%%%%%%%%%%%%%%%%%%%%%%%%%%%%%%%%%%%%%%%%%%%%%%%%%%%%%%%%%%%%%%%%%%%%%%%%%%%%
%%%%%%%%%%%%%%%%%%%%%%%%%%%%%%%%%%%%%%%%%%%%%%%%%%%%%%%%%%%%%%%%%%%%%%%%%%%%%%
\author{Nilay Bostan\orcidA}
\affiliation{Department of Physics, Faculty of Science, Marmara University, 34722 Istanbul, Türkiye}
%%%%%%%%%%%%%%%%%%%%%%%%%%%%%%%%%%%%%%%%%%%%%%%%%%%%%%%%%%%%%%%%%%%%%%%%%%%%%%
\author{Rafid H. Dejrah\orcidB}
\affiliation{Department of Physics, Faculty of Science, Ankara University,  06100 Ankara, Türkiye}
%%%%%%%%%%%%%%%%%%%%%%%%%%%%%%%%%%%%%%%%%%%%%%%%%%%%%%%%%%%%%%%%%%%%%%%%%%%%%%
\author{Anish Ghoshal\orcidC}
\affiliation{Department of Physics and Astronomy, University of Sussex, \\
Brighton, BN1 9RH, United Kingdom}
%%%%%%%%%%%%%%%%%%%%%%%%%%%%%%%%%%%%%%%%%%%%%%%%%%%%%%%%%%%%%%%%%%%%%%%%%%%%%%
\author{Zygmunt Lalak\orcidD}
\affiliation{Institute of Theoretical Physics, Faculty of Physics,\\ University of Warsaw, ul. Pasteura 5, 02-093 Warsaw, Poland}
%%%%%%%%%%%%%%%%%%%%%%%%%%%%%%%%%%%%%%%%%%%%%%%%%%%%%%%%%%%%%%%%%%%%%%%%%%%%%%
%%%%%%%%%%%%%%%%%%%%%%%%%%%%%%%%%%%%%%%%%%%%%%%%%%%%%%%%%%%%%%%%%%%%%%%%%%%%%%
%%%%%%%%%%%%%%%%%%%%%%%%%%%%%%%%%%%%%%%%%%%%%%%%%%%%%%%%%%%%%%%%%%%%%%%%%%%%%%
\begin{abstract}
{
\renewcommand{\thefootnote}{\fnsymbol{footnote}}%
\footnotemark[0]%
}
We study non-minimally coupled scalar-induced multi-phase inflation in metric and Palatini gravity, considering linear, Brans-Dicke-like, and Higgs-like sectors. The scalar spectral index lies in the range 
\( n_s \simeq 0.93 \ \text{--} \ 0.98 \), consistent with \textit{Planck} and combined \textit{Planck}+ACT data. The tensor-to-scalar ratio can reach 
\( r \sim 0.03  \) in metric, whereas Palatini models generically predict \( r \lesssim 10^{-5} \). In the Palatini case, field excursions remain sub-Planckian, and the perturbative unitarity cutoff is raised. Reheating proceeds via perturbative inflaton decays into Higgs bosons and fermionic dark matter (DM) through the portal coupling \( \lambda_{12} \) and Yukawa coupling \( y_\chi \). 
Radiative stability of the inflationary plateau constrains the couplings to \( y_\chi, \lambda_{12} \sim 10^{-7} \ \text{--} \ 10^{-3} \), implying 
\( 4\,\mathrm{MeV} \lesssim T_{\rm rh} \lesssim 10^{15}\,\mathrm{GeV} \). Palatini realizations require smaller couplings and thus a narrower reheating window. Non-thermal DM production  $\chi$ from inflaton decays is viable for DM mass \( m_\chi \sim \mathrm{keV} \ \text{--} \ \mathrm{PeV} \) with \( y_\chi \lesssim 10^{-6} \) over large parameter regions. 
We estimate the inflaton-right-handed neutrino (RHN) Yukawa coupling \( y_N \) required for successful baryogenesis via non-thermal leptogenesis within a Type-I seesaw framework, for the lightest RHN mass 
\( M_{N_1} \sim 10^{9} \ \text{--} \ 10^{14}\,\mathrm{GeV} \), provided 
\( M_{N_1} > T_{\rm max} \), where \( T_{\rm max} \) follows from radiatively consistent reheating. In Palatini scenarios, the lower maximal temperature and tighter stability bounds further restrict the leptogenesis parameter space. 
\end{abstract}
{
\renewcommand{\thefootnote}{\fnsymbol{footnote}}%
\footnotetext[0]{Email addresses: nilay.bostan@marmara.edu.tr, rafid.dejrah@gmail.com (he/him), anish.ghoshal@fuw.edu.pl, zygmunt.lalak@fuw.edu.pl}%
\renewcommand{\thefootnote}{\arabic{footnote}}%
}
\maketitle
\tableofcontents
\flushbottom
%%%%%%%%%%%%%%%%%%%%%%%%%%%%%%%%%%%%%%%%%%%%%%%%%%%%%%%%%%%%%%%%%%%%%%%%%%%%%%
%%%%%%%%%%%%%%%%%%%%%%%%%%%%%%%%%%%%%%%%%%%%%%%%%%%%%%%%%%%%%%%%%%%%%%%%%%%%%%
%%%%%%%%%%%%%%%%%%%%%%%%%%%%%%%%%%%%%%%%%%%%%%%%%%%%%%%%%%%%%%%%%%%%%%%%%%%%%%
\section{Introduction}\label{sec:intro}
The inflationary paradigm constitutes a cornerstone of modern cosmology, resolving the horizon and flatness problems of the hot Big Bang framework while providing a dynamical origin for the primordial curvature perturbations that seed large-scale structure~\cite{Starobinsky:1980te,Guth:1980zm,Linde:1981mu,Albrecht:1982wi}. Precision measurements of the cosmic microwave background (CMB), primarily from \emph{Planck} in combination with BICEP/\emph{Keck}, indicate a nearly scale-invariant scalar spectrum with spectral index $n_s \simeq 0.965$ and impose a stringent upper bound on the tensor-to-scalar ratio, $r_{0.05} < 0.036$ at $95\%$ CL~\cite{BICEP:2021xfz}.

More recent analyses combining data from the Atacama Cosmology Telescope (ACT)~\cite{ACT:2020gnv}, and large-scale structure (LSS) surveys, including the Dark Energy Spectroscopic Instrument (DESI)~\cite{DESI:2024mwx}, have further sharpened constraints on the scalar spectrum, indicating $n_s \sim 0.97$ with percent-level precision. Current limits on primordial tensor modes continue to be driven primarily by CMB $B$-mode polarization measurements, which constrain the tensor-to-scalar ratio to $r_{0.05} < 0.036$ at $95\%$ CL~\cite{BICEP:2021xfz}. The steadily increasing observational accuracy substantially enhances the discriminative power among competing inflationary models, particularly those that predict suppressed tensor amplitudes and small but controlled departures from exact scale invariance. Despite this empirical progress, the microscopic origin of the inflaton and its consistent embedding within a complete particle physics framework remain open theoretical questions.

A pivotal distinction in our analysis arises from the choice of gravitational formulation. While the standard metric formalism assumes the Levi-Civita connection, the Palatini formulation treats the metric and affine connection as independent variables. In the presence of non-minimal scalar-curvature couplings, which are central to induced gravity scenarios, this choice modifies the dynamics after transformation to the Einstein frame. Unlike the metric case, the Palatini formulation does not generate additional kinetic contributions proportional to derivatives of the non-minimal coupling function under conformal transformations. This difference typically results in a significantly flatter effective inflaton potential, thereby suppressing the tensor-to-scalar ratio $r$ and allowing compatibility with observational constraints in regions of parameter space that would otherwise be excluded. Moreover, this effective flattening can raise the perturbative unitarity cutoff associated with large non-minimal couplings, improving the ultraviolet (UV) behavior of the theory and making the Palatini framework particularly attractive for constructing phenomenologically consistent inflationary models~\cite{Bauer:2010jg,Rasanen:2018fom,Enckell_2019,Tenkanen:2020dge,McDonald:2020lpz, Bostan:2024idr, Bostan:2025vkt, Bostan:2025zdt}.

Complementing this gravitational perspective, induced gravity models, motivated by Brans-Dicke theory and the desire to dynamically generate the fundamental Planck mass from the vacuum expectation value of a scalar field, provide a highly compelling and natural framework for early Universe cosmology~\cite{Zee:1978wi,Spokoiny:1984bd,Fakir:1990iu,Cervantes-Cota:1994qci, Gialamas:2020snr}. Within this context, the non-minimal coupling not only drives the symmetry breaking that establishes the gravitational scale but also inherently modifies the Einstein-frame potential. Crucially, this setup naturally accommodates \emph{multi-phase inflation}, wherein the Universe undergoes successive and distinct inflationary stages driven by a single scalar degree of freedom~\cite{Silk:1986vc,Polarski:1992dq, ADAMS1997405, Burgess:2005sb, Artymowski:2016ikw}. For instance, the dynamics can seamlessly transition from an initial large-field linear or polynomial regime, responsible for setting the observable large-scale CMB perturbations, into a small-field plateau or quadratic phase near the minimum of the potential~\cite{Kannike:2015kda, Artymowski:2016dlz}. This multi-phase evolution is particularly advantageous because it partially decouples the high-energy physics governing the horizon exit from the low-energy dynamics dictating the end of inflation. Consequently, it yields distinctive primordial observational signatures. It allows for a significantly richer post-inflationary phenomenology, facilitating non-trivial reheating mechanisms and resonant particle production scenarios without requiring severe fine-tuning of the underlying parameters~\cite{Cerioni:2009kn, Lalak:2016idb,  Lozanov:2016hid, Racioppi:2017spw, Garcia:2020eof,Drees:2021wgd, Bernal:2021qrl,  Ghoshal:2023jhh, Ghoshal:2024ycp, Drees:2024hok}.

The post-inflationary reheating epoch bridges the cold, scalar-dominated Universe at the end of inflation and the hot, radiation-dominated plasma of the early Universe. In many scenarios, this transition is driven by the decay of the inflaton condensate. The existence of such a non-baryonic DM component has now firmly been verified by a multitude of independent astrophysical and cosmological observations, ranging from galactic rotation curves and gravitational lensing to the precise anisotropies of the CMB and the formation of LSS~\cite{1980ApJ...238..471R, Clowe:2006eq, Planck:2018vyg}. Despite constituting approximately \(26\%\) of the Universe's total energy density, its fundamental microscopic nature remains entirely unknown, strongly motivating the search for a production mechanism beyond the Standard Model (SM)~\cite{Bertone:2004pz}. If the inflaton couples directly to dark-sector particles, these decays can non-thermally produce DM~\cite{Chung:2004nh,Allahverdi:2010xz,Bernal:2017kxu,Mambrini:2021zpp,Bernal:2021qrl}, establishing a direct connection between the inflationary energy scale and the DM relic abundance.

Viable non-thermal DM candidates can span a wide mass range, from keV to Planck scales, depending on the reheating temperature and relevant coupling strengths. Equally compelling is the necessity to explain the observed baryon asymmetry of the Universe (BAU). Precision measurements from Big Bang Nucleosynthesis (BBN) and the CMB dictate a striking imbalance between matter and antimatter, quantified by a baryon-to-entropy ratio of \(Y_B \simeq 8.7\times10^{-11}\)~\cite{Planck:2018vyg, ParticleDataGroup:2022pth}. Since the SM of particle physics fails to fully satisfy the Sakharov conditions required to dynamically generate this asymmetry, lacking both sufficient CP violation and a strong first-order phase transition, new physics is unequivocally required~\cite{Sakharov:1967dj, Canetti:2012zc}.

Moreover, the inflaton may also decay into RHNs, whose presence in the beyond Standard Model theory is well motivated from the requirement of tiny masses of SM neutrinos from neutrino oscillation experiments~\cite{Super-Kamiokande:1998kpq,  SNO:2001kpb}. The mass generation is elegantly achieved by introducing heavy right-handed, SM-singlet neutrinos, which naturally generate the minuscule active neutrino masses via the Type-I seesaw mechanism~\cite{Minkowski:1977sc, Gell-Mann:1979vob, Mohapatra:1979ia}. While alternative neutrino mass generation schemes exist, such as Type-II~\cite{MAGG198061} and Type-III~\cite{Foot:1988aq} seesaws or radiative mass models~\cite{Zee:1980ai}, we adopt the Type-I framework as a representative benchmark. This choice allows us to illustrate how the underlying gravitational formulation, the metric-Palatini distinction, and the specific shape of the inflationary potential directly affect the viable phenomenological parameter space of the BSM theory. If the inflaton also decays into these right-handed states, their subsequent out-of-equilibrium, CP-violating decays can simultaneously generate the observed baryon asymmetry of the Universe via non-thermal leptogenesis~\cite{Asaka:1999yd,Fukugita:1986hr,Giudice:2003jh,Hahn-Woernle:2008tsk,Mambrini:2021zpp}. In this way, inflation, neutrino mass generation, DM production, and baryogenesis can be consistently incorporated within a unified cosmological framework.

A comparative study of these cosmological sequences within both metric and Palatini formulations of gravity is, therefore, well motivated. The two formalisms can predict different effective inflaton mass scales, couplings, and reheating dynamics, which in turn affect the resulting DM abundances and the efficiency of leptogenesis. Understanding these differences is crucial both for interpreting forthcoming cosmological observations and for evaluating the theoretical viability of particle physics models associated with inflation.

Ultimately, the central objective of this work is to establish a unified cosmological framework that seamlessly connects the dynamics of inflation, DM production, active neutrino mass generation, and the origin of the matter-antimatter asymmetry. By doing so, we aim to uncover explicit theoretical correlations between primordial CMB observables, such as \(n_s\) and \(r\), and the permissible parameter spaces for post-inflationary phenomenology, demonstrating how the underlying geometry of gravity explicitly dictates the Universe's thermal history.

\emph{In this work}, we present a comprehensive analysis of induced multi-phase inflation within both the metric and Palatini formulations of gravity. Following a review of the Lagrangian density and the inflationary slow-roll approximation in Subsections~\ref{Sec:Lagrangian} and~\ref{Sec:Slow_roll}, we systematically investigate three classes of Einstein-frame potentials arising from non-minimally coupled scalar fields. Specifically, we analyze linear (Subsec.~\ref{sec:linear_inflation}), Brans-Dicke-like (Subsec.~\ref{sec:BD_inflation}), and Higgs-like (Subsec.~\ref{sec:Higgs_like}) models, computing the inflationary observables;  \(r\), \(n_s\), and the running of the spectral index \(\alpha_s\), to identify parameter space regions consistent with current CMB constraints. In Sec.~\ref{Sec:reheating}, we examine the subsequent reheating dynamics, demonstrating that perturbative inflaton decays dominate while non-perturbative preheating effects remain subdominant due to radiative stability constraints and Higgs backreactions; within this regime, we evaluate the production of non-thermal fermionic DM. Finally, Sec.~\ref{sec:lepto} explores neutrino mass generation via the Type-I seesaw mechanism and non-thermal leptogenesis, contrasting the predictions of the metric and Palatini formalisms to highlight how the choice of gravitational formulation impacts the viable parameter space of the unified model. Supplementary results are presented in Apps.~\ref{app:Inflationaries}--~\ref{app:other_leptogenesis}. Throughout this work, we adopt natural units with the reduced Planck mass set to unity, \(M_{\rm Pl}\equiv 1\), and use the mostly-plus metric signature.
%%%%%%%%%%%%%%%%%%%%%%%%%%%%%%%%%%%%%%%%%%%%%%%%%%%%%%%%%%%%%%%%%%%%%%%%%%%%%%
%%%%%%%%%%%%%%%%%%%%%%%%%%%%%%%%%%%%%%%%%%%%%%%%%%%%%%%%%%%%%%%%%%%%%%%%%%%%%%
%%%%%%%%%%%%%%%%%%%%%%%%%%%%%%%%%%%%%%%%%%%%%%%%%%%%%%%%%%%%%%%%%%%%%%%%%%%%%%
\section{Inflationary Framework}
%%%%%%%%%%%%%%%%%%%%%%%%%%%%%%%%%%%%%%%%%%%%%%%%%%%%%%%%%%%%%%%%%%%%%%%%%%%%%%
\subsection{Lagrangian Density}
\label{Sec:Lagrangian}
The action for the inflaton in the Jordan frame is given by
%%%%%%%%%%%%%%%%%%%%%%%%%%%%%%%%%%%%%%%%%%%%%%%%%%%%%%%%%%%%%%%%%%%%%%%%%%%%%%
\begin{equation}\label{Eq:Jordan frame action}
S_{\rm J} = \int \mathrm{d}^4 x \,\sqrt{-g}
\left[
\frac{1}{2}\,f(\phi)\,\mathcal{R}
+ \frac{1}{2}\,g^{\mu\nu}\partial_\mu \phi\,\partial_\nu \phi
- V_{\rm J}(\phi)
\right],
\end{equation}
%%%%%%%%%%%%%%%%%%%%%%%%%%%%%%%%%%%%%%%%%%%%%%%%%%%%%%%%%%%%%%%%%%%%%%%%%%%%%%
where $g \equiv \det(g_{\mu\nu})$, $\mathcal{R}$ is the Ricci scalar constructed from the Jordan-frame metric, and $V_{\rm J}(\phi)$ denotes the potential in the Jordan-frame for the inflaton, \(\phi\). After performing a conformal transformation to the Einstein frame\footnote{The physical equivalence between the Jordan and Einstein frames is a well-established result at the classical level. While inflation is formulated here in the Jordan frame to define the non-minimal coupling to gravity explicitly, cosmological observables, such as the primordial spectra and the tensor-to-scalar ratio, remain invariant under conformal transformations. Consequently, evaluating the inflationary dynamics in the Einstein frame yields physically equivalent predictions for both the background evolution and curvature perturbations~\cite{Flanagan:2004bz, Faraoni:2006fx,Chiba:2013mha, Postma:2014vaa}.}
%%%%%%%%%%%%%%%%%%%%%%%%%%%%%%%%%%%%%%%%%%%%%%%%%%%%%%%%%%%%%%%%%%%%%%%%%%%%%%
\begin{equation}\label{eq:conformal}
\tilde{g}_{\mu\nu} = f(\phi)\, g_{\mu\nu},
\end{equation}
where a tilde denotes Einstein-frame quantities, the scalar potential transforms as
\begin{equation}\label{Eq:pot-Jordan_to_Einstein}
V_E(\phi) = \frac{V_{\rm J}(\phi)}{f^2(\phi)} \,.
\end{equation}
%%%%%%%%%%%%%%%%%%%%%%%%%%%%%%%%%%%%%%%%%%%%%%%%%%%%%%%%%%%%%%%%%%%%%%%%%%%%%%
Here, $V_E (\phi)$ is the Einstein-frame potential, $f(\phi)$ is the non-minimal coupling function. 

To obtain a canonical kinetic term for the inflaton in the Einstein frame, we redefine the field $\phi \mapsto \zeta$ as
%%%%%%%%%%%%%%%%%%%%%%%%%%%%%%%%%%%%%%%%%%%%%%%%%%%%%%%%%%%%%%%%%%%%%%%%%%%%%%
\begin{equation}\label{Eq:conversion-inflaton-EToJ}
\frac{\mathrm{d}\zeta}{\mathrm{d}\phi} = \sqrt{\Pi(\phi)} \,,
\end{equation}
%%%%%%%%%%%%%%%%%%%%%%%%%%%%%%%%%%%%%%%%%%%%%%%%%%%%%%%%%%%%%%%%%%%%%%%%%%%%%%
where the function $\Pi(\phi)$ captures the effects of the non-minimal coupling to gravity and is given by
%%%%%%%%%%%%%%%%%%%%%%%%%%%%%%%%%%%%%%%%%%%%%%%%%%%%%%%%%%%%%%%%%%%%%%%%%%%%%%
\begin{equation}\label{eq:Pi-general}
\Pi(\phi) = \frac{1}{f(\phi)} + \kappa\,\frac{3}{2} \left(\frac{f'(\phi)}{f(\phi)}\right)^2 , \qquad
\kappa = \begin{cases}
1 & \text{(metric formalism)}, \\[2mm]
0 & \text{(Palatini formalism)} .
\end{cases}
\end{equation}
%%%%%%%%%%%%%%%%%%%%%%%%%%%%%%%%%%%%%%%%%%%%%%%%%%%%%%%%%%%%%%%%%%%%%%%%%%%%%%
Here, primes denote derivatives with respect to the scalar field $\phi$, and the explicit form of the non-minimal coupling function $f(\phi)$ will be specified in Section~\ref{induced}, according to the models considered in this work.\footnote{We employ distinct non-minimal coupling functions, $f(\phi)$, tailored to each specific Jordan-frame potential, $V_J(\phi)$, rather than assuming a single universal coupling across all models. This tailored approach is physically motivated as transforming to the Einstein frame yields $V_E = V_J / f^2$, so the precise functional interplay between $f(\phi)$ and $V_J(\phi)$ is required to generate the asymptotically flat plateaus necessary to satisfy the stringent CMB constraints on inflationary observables. Furthermore, within the context of induced gravity, this specific pairing ensures that the fundamental Planck scale is dynamically and consistently generated at the minimum of the respective potentials without disrupting the required slow-roll dynamics~\cite{Galante:2014ifa}.}

After this field redefinition, the Einstein-frame action takes the form
%%%%%%%%%%%%%%%%%%%%%%%%%%%%%%%%%%%%%%%%%%%%%%%%%%%%%%%%%%%%%%%%%%%%%%%%%%%%%%
\begin{equation}
S_{\rm E}= \int \mathrm{d}^4 x\,\sqrt{-\tilde{g}}
\left(
\frac{1}{2} \mathcal{\tilde{R}} + \frac{1}{2} \Pi (\phi)\,(\partial\phi)^2 - V_E(\phi)
\right)\,,
\end{equation}
%%%%%%%%%%%%%%%%%%%%%%%%%%%%%%%%%%%%%%%%%%%%%%%%%%%%%%%%%%%%%%%%%%%%%%%%%%%%%%
where \(V_E(\phi)\) is the potential given in the Einstein-frame, further details related to different potentials are discussed in Sec.~\ref{induced}. This form of the action is then used to analyze the inflationary dynamics and compute the slow-roll parameters in the following subsection.

%%%%%%%%%%%%%%%%%%%%%%%%%%%%%%%%%%%%%%%%%%%%%%%%%%%%%%%%%%%%%%%%%%%%%%%%%%%%%%
%%%%%%%%%%%%%%%%%%%%%%%%%%%%%%%%%%%%%%%%%%%%%%%%%%%%%%%%%%%%%%%%%%%%%%%%%%%%%%
%%%%%%%%%%%%%%%%%%%%%%%%%%%%%%%%%%%%%%%%%%%%%%%%%%%%%%%%%%%%%%%%%%%%%%%%%%%%%%
\subsection{Slow-roll approximation in inflation}
\label{Sec:Slow_roll}
In this section, we present a general discussion of the slow-roll parameters employed throughout the various scenarios considered in this work. The inflationary observables can then be expressed in terms of these slow-roll parameters, following the standard formalism reviewed in Ref.~\cite{Lyth:2009zz}
%%%%%%%%%%%%%%%%%%%%%%%%%%%%%%%%%%%%%%%%%%%%%%%%%%%%%%%%%%%%%%%%%%%%%%%%%%%%%%
\begin{equation}\label{slowroll1}
\epsilon =\frac{M_{\rm Pl}^2}{2}\left( \frac{V_{\zeta} }{V}\right) ^{2}\,, \quad
\eta = M_{\rm Pl}^2 \frac{V_{\zeta \zeta} }{V}  \,, \quad
\kappa ^{2} = M_{\rm Pl}^4 \frac{V_{\zeta} V_{\zeta\zeta\zeta} }{V^{2}}\,.
\end{equation}
%%%%%%%%%%%%%%%%%%%%%%%%%%%%%%%%%%%%%%%%%%%%%%%%%%%%%%%%%%%%%%%%%%%%%%%%%%%%%%
Here, the subscript $\zeta$ denotes derivatives with respect to the canonically normalized scalar field $\zeta$. Within the slow-roll approximation, the inflationary observables, namely the scalar spectral index $n_s$, the tensor-to-scalar ratio $r$, and the running of the spectral index $\alpha_s$, are given by
%%%%%%%%%%%%%%%%%%%%%%%%%%%%%%%%%%%%%%%%%%%%%%%%%%%%%%%%%%%%%%%%%%%%%%%%%%%%%%
\begin{align}
    n_s &= 1 - 6 \epsilon_\zeta + 2 \eta_\zeta, \\
    r &= 16 \epsilon_\zeta, \\
    \alpha_s &\equiv \frac{\mathrm{d} n_s}{\mathrm{d} \ln k} = 16 \epsilon_\zeta \eta_\zeta - 24 \epsilon_\zeta^2 - 2 \kappa_\zeta^2,
\end{align}
%%%%%%%%%%%%%%%%%%%%%%%%%%%%%%%%%%%%%%%%%%%%%%%%%%%%%%%%%%%%%%%%%%%%%%%%%%%%%%
where the slow-roll parameters are defined as
%%%%%%%%%%%%%%%%%%%%%%%%%%%%%%%%%%%%%%%%%%%%%%%%%%%%%%%%%%%%%%%%%%%%%%%%%%%%%%
\begin{align}
    \epsilon_\zeta &\equiv \frac{1}{2} \left( \frac{V_\zeta'}{V} \right)^2, &
    \eta_\zeta &\equiv \frac{V_\zeta''}{V}, &
    \kappa_\zeta^2 &\equiv \frac{V_\zeta' V_\zeta'''}{V^2}.
\end{align}
%%%%%%%%%%%%%%%%%%%%%%%%%%%%%%%%%%%%%%%%%%%%%%%%%%%%%%%%%
The number of \(e\)-folds $N_*$ in the slow-roll approximation is given by\footnote{It is worth noting that the raw definition of the number of \(e\)-folds exhibits frame dependence due to the conformal rescaling of the scale factor. This discrepancy becomes remarkably pronounced in the Palatini formulation at large non-minimal couplings~\cite{Racioppi:2021jai}. However, formulating the expansion in terms of a frame-invariant normalized physical distance resolves this ambiguity, and its mathematical outcome identically matches the standard Einstein-frame computation~\cite{Racioppi:2021jai}. Therefore, evaluating $N_*$ using the Einstein-frame variable intrinsically provides a frame-invariant and physically consistent measure of the inflationary expansion.}
%%%%%%%%%%%%%%%%%%%%%%%%%%%%%%%%%%%%%%%%%%%%%%%%%%%%%%%%%%%%%%%%%%%%%%%%%%%%%%
\begin{equation} \label{efold1}
N_*=\int^{\zeta_*}_{\zeta_e}\frac{\rm{d}\zeta}{\sqrt{2\epsilon_\zeta}}\,. \end{equation}
%%%%%%%%%%%%%%%%%%%%%%%%%%%%%%%%%%%%%%%%%%%%%%%%%%%%%%%%%%%%%%%%%%%%%%%%%%%%%%
Here, the subscript ``$*$'' denotes quantities evaluated at the horizon exit of the pivot scale, while $\zeta_e$ denotes the value of the inflaton at the end of inflation. Inflation is assumed to end when the slow-roll approximation breaks down, i.e., when $\max(\epsilon,|\eta|)=1$. In the models considered here, the slow-roll parameter $\epsilon$ typically increases more rapidly than $|\eta|$ near the end of inflation, so that the condition $\epsilon(\zeta_e)=1$ provides a robust and sufficient criterion for determining $\zeta_e$. We have explicitly verified that imposing the alternative condition $|\eta|=1$ yields nearly identical values of $\zeta_e$, with differences negligible for the precision required in our subsequent analyses, including reheating dynamics and cosmological observables. Our results are therefore insensitive to the specific choice of slow-roll termination criterion given as
%%%%%%%%%%%%%%%%%%%%%%%%%%%%%%%%%%%%%%%%%%%%%%%%%%%%%%%%%%%%%%%%%%%%%%%%%%%%%%
\begin{equation}
\epsilon(\zeta_e)=1\,.
\end{equation}
%%%%%%%%%%%%%%%%%%%%%%%%%%%%%%%%%%%%%%%%%%%%%%%%%%%%%%%%%%%%%%%%%%%%%%%%%%%%%%
In the slow-roll approximation, the amplitude of the primordial curvature perturbation is expressed as
%%%%%%%%%%%%%%%%%%%%%%%%%%%%%%%%%%%%%%%%%%%%%%%%%%%%%%%%%%%%%%%%%%%%%%%%%%%%%%
\begin{equation}
\Delta_{\mathcal R}^2
= \left. \frac{V_E(\zeta)}{24\pi^2\,\epsilon(\zeta)} \right|_{\zeta=\zeta_*}
= \frac{V_{E,*}}{24\pi^2\,\epsilon_*}\,.
\label{perturb1}
\end{equation}
%%%%%%%%%%%%%%%%%%%%%%%%%%%%%%%%%%%%%%%%%%%%%%%%%%%%%%%%%%%%%%%%%%%%%%%%%%%%%%
The best-fit value of the pivot scale is $k_* = 0.05~\mathrm{Mpc}^{-1}$, for which the amplitude of the primordial curvature perturbation is
%%%%%%%%%%%%%%%%%%%%%%%%%%%%%%%%%%%%%%%%%%%%%%%%%%%%%%%%%%%%%%%%%%%%%%%%%%%%%%
\begin{equation}
    \Delta_\mathcal{R}^2 \approx 2.1\times 10^{-9},
\end{equation}
%%%%%%%%%%%%%%%%%%%%%%%%%%%%%%%%%%%%%%%%%%%%%%%%%%%%%%%%%%%%%%%%%%%%%%%%%%%%%%
as determined by the \emph{Planck} measurements~\cite{Planck:2018vyg}.

However, obtaining an analytic expression for an inflationary potential originally defined as $V_J(\phi)$ in terms of the canonically normalized field, $V_E(\zeta)$, is not always straightforward, and in some cases may be impossible, depending on the functional form of the potential. In such situations, it is convenient to perform numerical computations of the inflationary dynamics directly in terms of the original scalar field $\phi$, without explicitly transforming to the canonical field $\zeta$. Accordingly, the slow-roll parameters must be expressed in terms of $\phi$, and take the following form~\cite{Linde_2011}
%%%%%%%%%%%%%%%%%%%%%%%%%%%%%%%%%%%%%%%%%%%%%%%%%%%%%%%%%%%%%%%%%%%%%%%%%%%%%%
\begin{equation}
    \epsilon_\phi = \frac{1}{2} \left( \frac{V_\phi}{V} \right)^2, \qquad
    \eta_\phi = \frac{V_{\phi\phi}}{V}, \qquad
    \kappa_\phi^2 = \frac{V_\phi V_{\phi\phi\phi}}{V^2},
\end{equation}
%%%%%%%%%%%%%%%%%%%%%%%%%%%%%%%%%%%%%%%%%%%%%%%%%%%%%%%%%%%%%%%%%%%%%%%%%%%%%%
where subscripts denote derivative with respect to the scalar field \(\phi\).

Then, the slow-roll parameters can be written as
%%%%%%%%%%%%%%%%%%%%%%%%%%%%%%%%%%%%%%%%%%%%%%%%%%%%%%%%%%%%%%%%%%%%%%%%%%%%%%
\begin{eqnarray}
\epsilon=Z\epsilon_{\phi}\,, \ \
\eta=Z\eta_{\phi}+{\rm sgn}(V')Z'\sqrt{\frac{\epsilon_{\phi}}{2}}\,.
\end{eqnarray}
%%%%%%%%%%%%%%%%%%%%%%%%%%%%%%%%%%%%%%%%%%%%%%%%%%%%%%%%%%%%%%%%%%%%%%%%%%%%%%
Here, \( Z(\phi) = \left(\Pi(\phi)\right)^{-1} \) denotes the field-space metric, with \(\Pi(\phi)\) defined in Eq.~\eqref{eq:Pi-general}. The number of $e$-folds between the horizon exit at $\phi_*$ and the end of inflation at $\phi_e$ is then given by
%%%%%%%%%%%%%%%%%%%%%%%%%%%%%%%%%%%%%%%%%%%%%%%%%%%%%%%%%%%%%%%%%%%%%%%%%%%%%%
\begin{equation}
    N_* = \int_{\phi_e}^{\phi_*} \frac{\, V(\phi)}{ Z(\phi)\,V'(\phi)}\, \mathrm{d}\phi = \rm{sgn}(V')\int^{\phi_*}_{\phi_e}\frac{\mathrm{d}\phi}{Z(\phi)\,\sqrt{2\epsilon_{\phi}}}\, .
\end{equation}
%%%%%%%%%%%%%%%%%%%%%%%%%%%%%%%%%%%%%%%%%%%%%%%%%%%%%%%%%%%%%%%%%%%%%%%%%%%%%%
The amplitude of the curvature perturbation evaluated at horizon exit reads
%%%%%%%%%%%%%%%%%%%%%%%%%%%%%%%%%%%%%%%%%%%%%%%%%%%%%%%%%%%%%%%%%%%%%%%%%%%%%%
\begin{equation}
    \Delta_\mathcal{R}^2
    = \frac{1}{12\pi^2}
    \frac{V^3(\phi_*)}{\left[V'(\phi_*)\right]^2 Z(\phi_*)}
    = \frac{1}{24\pi^2}\frac{V(\phi_*)}{\epsilon_\phi(\phi_*)},
\end{equation}
%%%%%%%%%%%%%%%%%%%%%%%%%%%%%%%%%%%%%%%%%%%%%%%%%%%%%%%%%%%%%%%%%%%%%%%%%%%%%%
where all quantities are evaluated at $\phi=\phi_*$.

To compute the numerical values of the inflationary observables, such as the scalar spectral index $n_s$ and the tensor-to-scalar ratio $r$, it is necessary to determine the number of $e$-folds, $N_*$, corresponding to the pivot scale. Assuming a standard thermal history after inflation, the number of $e$-folds associated with the pivot scale $k_* = 0.05~\mathrm{Mpc}^{-1}$ can be approximated as~\cite{Liddle_2003}
%%%%%%%%%%%%%%%%%%%%%%%%%%%%%%%%%%%%%%%%%%%%%%%%%%%%%%%%%%%%%%%%%%%%%%%%%%%%%%
\begin{equation} \label{efolds}
N_* \simeq 61.5
+ \frac{1}{2}\ln\!\left(\frac{\rho_*}{M_{\rm Pl}^4}\right)
- \frac{1}{3(1+\omega_r)}\ln\!\left(\frac{\rho_e}{M_{\rm Pl}^4}\right)
+ \left[\frac{1}{3(1+\omega_r)} - \frac{1}{4}\right]
\ln\!\left(\frac{\rho_r}{M_{\rm Pl}^4}\right).
\end{equation}
%%%%%%%%%%%%%%%%%%%%%%%%%%%%%%%%%%%%%%%%%%%%%%%%%%%%%%%%%%%%%%%%%%%%%%%%%%%%%%
Here, $\rho_e = V(\phi_e)$ denotes the energy density at the end of inflation, $\rho_r$ denotes the energy density at the end of reheating, and $\rho_* \approx V(\phi_*)$ corresponds to the energy density when the scale associated with $k_*$ exits the horizon. $\omega_r$ denotes the equation-of-state parameter during reheating.

The energy densities $\rho_r$ and $\rho_*$ can be expressed as
%%%%%%%%%%%%%%%%%%%%%%%%%%%%%%%%%%%%%%%%%%%%%%%%%%%%%%%%%%%%%%%%%%%%%%%%%%%%%%
\begin{align}
    \rho_r &= \frac{\pi^2}{30} g_\mathrm{*}\, T_\mathrm{rh}^4, \\
    \rho_* &\approx V(\phi_*).
\end{align}
%%%%%%%%%%%%%%%%%%%%%%%%%%%%%%%%%%%%%%%%%%%%%%%%%%%%%%%%%%%%%%%%%%%%%%%%%%%%%%
Here, $g_\mathrm{*} = 106.75$ denotes the number of relativistic degrees of freedom at reheating, and $T_\mathrm{rh}$ is the reheat temperature, defined as the temperature at which the Universe reaches thermal equilibrium and the energy density becomes dominated by radiation.

%%%%%%%%%%%%%%%%%%%%%%%%%%%%%%%%%%%%%%%%%%%%%%%%%%%%%%%%%%%%%%%%%%%%%%%%%%%%%%
%%%%%%%%%%%%%%%%%%%%%%%%%%%%%%%%%%%%%%%%%%%%%%%%%%%%%%%%%%%%%%%%%%%%%%%%%%%%%%
%%%%%%%%%%%%%%%%%%%%%%%%%%%%%%%%%%%%%%%%%%%%%%%%%%%%%%%%%%%%%%%%%%%%%%%%%%%%%%
\section{Induced Gravity Multi-phase Inflation}
\label{induced}
In the following subsections, we investigate explicit realizations of induced multi-phase inflation. We begin with linear inflation models, which are characterized by approximately linear Einstein-frame potentials at large field values\footnote{For extensive discussions on how linear inflation successfully emerges as an attractor solution in the strong coupling limit, either from loop corrections to a quartic potential or within generalized scalar-tensor theories, see Refs.~\cite{Kannike:2015kda, Artymowski:2016dlz}.}, and then proceed to Brans-Dicke-like and Higgs-like scenarios. For each model class, we derive analytical approximations for key inflationary observables and present representative numerical benchmarks that are consistent with current CMB constraints. We also systematically compare the predictions obtained in the metric and Palatini formalisms, 
emphasizing the tendency of the Palatini formulation to yield flatter effective potentials and, consequently, suppressed tensor-to-scalar ratios. For clarity, the main text displays the principal correlations between \(r\) and \(n_s\), as well as between \(r\) and the non-minimal coupling \(\xi\) 
for each model, while additional results, including the running of the scalar spectral index 
\(\alpha_s\) and the dependence of the initial field value \(\phi_i\) on \(\xi\), are deferred to the appendices.

%%%%%%%%%%%%%%%%%%%%%%%%%%%%%%%%%%%%%%%%%%%%%%%%%%%%%%%%%%%%%%%%%%%%%%%%%%%%%%
%%%%%%%%%%%%%%%%%%%%%%%%%%%%%%%%%%%%%%%%%%%%%%%%%%%%%%%%%%%%%%%%%%%%%%%%%%%%%%
%%%%%%%%%%%%%%%%%%%%%%%%%%%%%%%%%%%%%%%%%%%%%%%%%%%%%%%%%%%%%%%%%%%%%%%%%%%%%%
\subsection{Linear Inflation}
\label{sec:linear_inflation}
In the Palatini formulation, the field-space metric reduces to $\Pi(\phi) = 1/f(\phi)$, leading to a significantly flatter Einstein-frame potential at large field values compared to the metric case. We consider two Jordan-frame potentials which, after the conformal transformation and canonical normalization, yield an approximately \emph{linear} Einstein-frame potential in the large-field regime
%%%%%%%%%%%%%%%%%%%%%%%%%%%%%%%%%%%%%%%%%%%%%%%%%%%%%%%%%%%%%%%%%%%%%%%%%%%%%%
\begin{align}
  V_J^{(1)}(\phi) &= M^4\, f^2(\phi)\,\ln\!\bigl[f(\phi)\bigr], \label{eq:U1_Palatini}\\
  V_J^{(2)}(\phi) &= M^4\,\bigl[f(\phi)-1\bigr]^2 \ln\!\bigl[f(\phi)\bigr]. \label{eq:U2_Palatini}
\end{align}
%%%%%%%%%%%%%%%%%%%%%%%%%%%%%%%%%%%%%%%%%%%%%%%%%%%%%%%%%%%%%%%%%%%%%%%%%%%%%%
The non-minimal coupling is taken to have a power-law form
%%%%%%%%%%%%%%%%%%%%%%%%%%%%%%%%%%%%%%%%%%%%%%%%%%%%%%%%%%%%%%%%%%%%%%%%%%%%%%
\begin{equation}
f(\phi) = 1 + \xi\phi^{n}, \qquad \xi > 0, \quad n \geq \tfrac{1}{2}.
\end{equation}
%%%%%%%%%%%%%%%%%%%%%%%%%%%%%%%%%%%%%%%%%%%%%%%%%%%%%%%%%%%%%%%%%%%%%%%%%%%%%%
In the regime $\phi \gg \xi^{-1/n}$, the logarithmic term dominates. As a result, the Einstein-frame potential is approximately linear in the canonically normalized field $\zeta$ in the metric case, whereas it becomes logarithmically flatter in the Palatini formulation.
%%%%%%%%%%%%%%%%%%%%%%%%%%%%%%%%%%%%%%%%%%%%%%%%%%%%%%%%%%%%%%%%%%%%%%%%%%%%%%

\begin{table}[t!]
\centering
\renewcommand{\arraystretch}{1.3}
\resizebox{\textwidth}{!}{
\begin{tabular}{|c|c|c|c|c|c|c|c|c||c|c|c|c|c|c|c|c|c|}
\hline
\multicolumn{9}{|c||}{\textbf{Metric}} & \multicolumn{9}{c|}{\textbf{Palatini}} \\
\hline
$n$ & $\xi$ & $M\times10^{-3}$ & $\phi_i/M_{\rm Pl}$ & $\phi_e/M_{\rm Pl}$ & $n_s$ & $r$ & $-\alpha_s\times 10^{-4}$ & $N$ & $n$ & $\xi$ & $M\times10^{-3}$ & $\phi_i/M_{\rm Pl}$ & $\phi_e/M_{\rm Pl}$ & $n_s$ & $r$ & $-\alpha_s\times 10^{-4}$ & $N$\\
\hline
\multirow{4}{*}{\rotatebox[origin=c]{90}{0.5}} & 0.1 & 7.94 & 7.91 & 0.35 & 0.978 & 0.032 & 3.83 & 56.3 & \multirow{4}{*}{\rotatebox[origin=c]{90}{0.5}} & 0.1 & 7.94 & 7.94 & 0.35 & 0.978 & 0.032 & 3.83 & 56.8 \\
& 1 & 4.57 & 10.54 & 0.25 & 0.978 & 0.021 & 3.75 & 56.7 & & 1 & 4.57 & 10.67 & 0.36 & 0.978 & 0.021 & 3.75 & 56.5 \\
& 100 & 2.88 & 44.84 & $6.1\times10^{-5}$ & 0.964 & 0.014 & 3.36 & 56.2 & & 100 & 2.34 & 75.55 & 0.73 & 0.981 & 0.007 & 3.36 & 56.3 \\
& 1000 & 3.63 & 32.57 & $6.1\times 10^{-7}$ & 0.931 & 0.047 & 3.28 & 57.9 & & 1000 & 2.00 & 269.49 & 1.80 & 0.981 & 0.005 & 3.28 & 55.3 \\
\hline
\multirow{3}{*}{\rotatebox[origin=c]{90}{1}} & 0.1 & 6.76 & 12.46 & 0.70 & 0.974 & 0.054 & 4.56 & 57.4 & \multirow{3}{*}{\rotatebox[origin=c]{90}{1}} & 0.1 & 6.76 & 12.49 & 0.71 & 0.974 & 0.054 & 4.56 & 57.2 \\
& 1 & 4.27 & 22.16 & 0.50 & 0.975 & 0.034 & 4.50 & 56.1 & & 1 & 4.17 & 24.07 & 0.72 & 0.976 & 0.031 & 4.24 & 56.5 \\
& 100 & 3.80 & 48.82 & $7.8\times 10^{-3}$ & 0.965 & 0.056 & 2.54 & 57.5 & & 100 & 2.40 & 570.65 & 1.83 & 0.980 & 0.012 & 3.59 & 56.7 \\
\hline
\multirow{2}{*}{\rotatebox[origin=c]{90}{2}} & 0.1 & 4.57 & 41.90 & 1.35 & 0.972 & 0.074 & 5.06 & 57.1 & \multirow{2}{*}{\rotatebox[origin=c]{90}{1.5}} & 0.1 & 5.37 & 21.39 & 1.06 & 0.972 & 0.062 & 5.13 & 57.1 \\
& 1 & 4.07 & 56.92 & 0.79 & 0.974 & 0.070 & 4.63 & 57.3 & & 1 & 3.63 & 98.89 & 1.09 & 0.975 & 0.038 & 4.47 & 56.1 \\
\hline
\end{tabular}%
}
\caption{\it Comparison of the parameter values for linear inflation, as given in Eq.~\eqref{eq:U1_Palatini}, between the metric and Palatini formulations.}
\label{tab:infl-para_1}
\end{table}
%%%%%%%%%%%%%%%%%%%%%%%%%%%%%%%%%%%%%%%%%%%%%%%%%%%%%%%%%%%%%%%%%%%%%%%%%%%%%%
In the large-field regime, analytical approximations for the primary inflationary observables, derived from the slow-roll parameters, are given by
%%%%%%%%%%%%%%%%%%%%%%%%%%%%%%%%%%%%%%%%%%%%%%%%%%%%%%%%%%%%%%%%%%%%%%%%%%%%%%
\begin{align}
    \label{eq:ns_1}
n_s^{(1)} &\approx
\begin{cases}
1 - \dfrac{2n\big(n+2+\xi\phi_i^n\big)}{n\big(n+2+\xi\phi_i^n\big)L^2 + 2\phi_i^2}\,,~~
L=\ln(1+\xi\phi_i^n) & \text{(metric)}, \\[8pt]
1 - n\xi\phi_i^{n-2}\Bigl(\dfrac{3n}{L^2}-\dfrac{n-2}{L}\Bigr)\,,~~L=\ln(1+\xi\phi_i^n) & \text{(Palatini).}
\end{cases}\\[12pt]
\label{eq:ns_2}
n_s^{(2)} &\approx
\begin{cases}
1-\dfrac{3}{2N_*} & \text{(metric)},\\[6pt]
1-\dfrac{9n^2+6n}{\phi_i^2} & \text{(Palatini).}
\end{cases}
\end{align}
%%%%%%%%%%%%%%%%%%%%%%%%%%%%%%%%%%%%%%%%%%%%%%%%%%%%%%%%%%%%%%%%%%%%%%%%%%%%%%
\begin{align}
    \label{eq:r_1}
    r^{(1)} &\approx
    \begin{cases}
    \dfrac{16n^2}{2\phi_i^2+3n^2 L^2}\,,~~L=\ln(1+\xi\phi_i^n) & \text{(metric)},\\[8pt]
    \dfrac{8n^2}{\phi_i^2}\,\dfrac{\bigl(1+\frac{u}{2}\bigr)^2}{1+u}\,,~~u=\xi\phi_i^n & \text{(Palatini).}
    \end{cases}\\[12pt]
    \label{eq:r_2}
    r^{(2)} &\approx
    \begin{cases}
    \dfrac{144n^2}{27n^2 L^2 + 2\phi_i^2\!\left(1+\frac{2}{3}u\,e^{-u}\right)}\,,~~u=\xi\phi_i^n,\,L=\ln(1+u) & \text{(metric)},\\[8pt]
    \dfrac{8n^2}{\phi_i^2}\,\dfrac{(u+2L)^2}{(1+u)L^2}\,,~~u=\xi\phi_i^n,\,L=\ln(1+u) & \text{(Palatini).}
    \end{cases}
\end{align}
%%%%%%%%%%%%%%%%%%%%%%%%%%%%%%%%%%%%%%%%%%%%%%%%%%%%%%%%%%%%%%%%%%%%%%%%%%%%%%
 These expressions capture the asymptotic behavior of the model in the large-field regime, where the non-minimal coupling $\xi$ and the power $n$ determine the degree of flattening of the effective Einstein-frame potential. In particular, for the Palatini formulation with the potential $V_J^{(1)}$, Eq.~\eqref{eq:ns_1} shows that the scalar spectral index $n_s$ approaches unity as $\xi$ increases, reflecting the increasingly flat potential. Meanwhile, the tensor-to-scalar ratio $r$, given in Eq.~\eqref{eq:r_1}, scales inversely with the square of the initial field value $\phi_i$, leading to a strong suppression of tensor modes.
 %%%%%%%%%%%%%%%%%%%%%%%%%%%%%%%%%%%%%%%%%%%%%%%%%%%%%%%%%%%%%%%%%%%%%%%%%%%%%%

\begin{figure}[t!]
\centering
\includegraphics[width=0.45\linewidth]{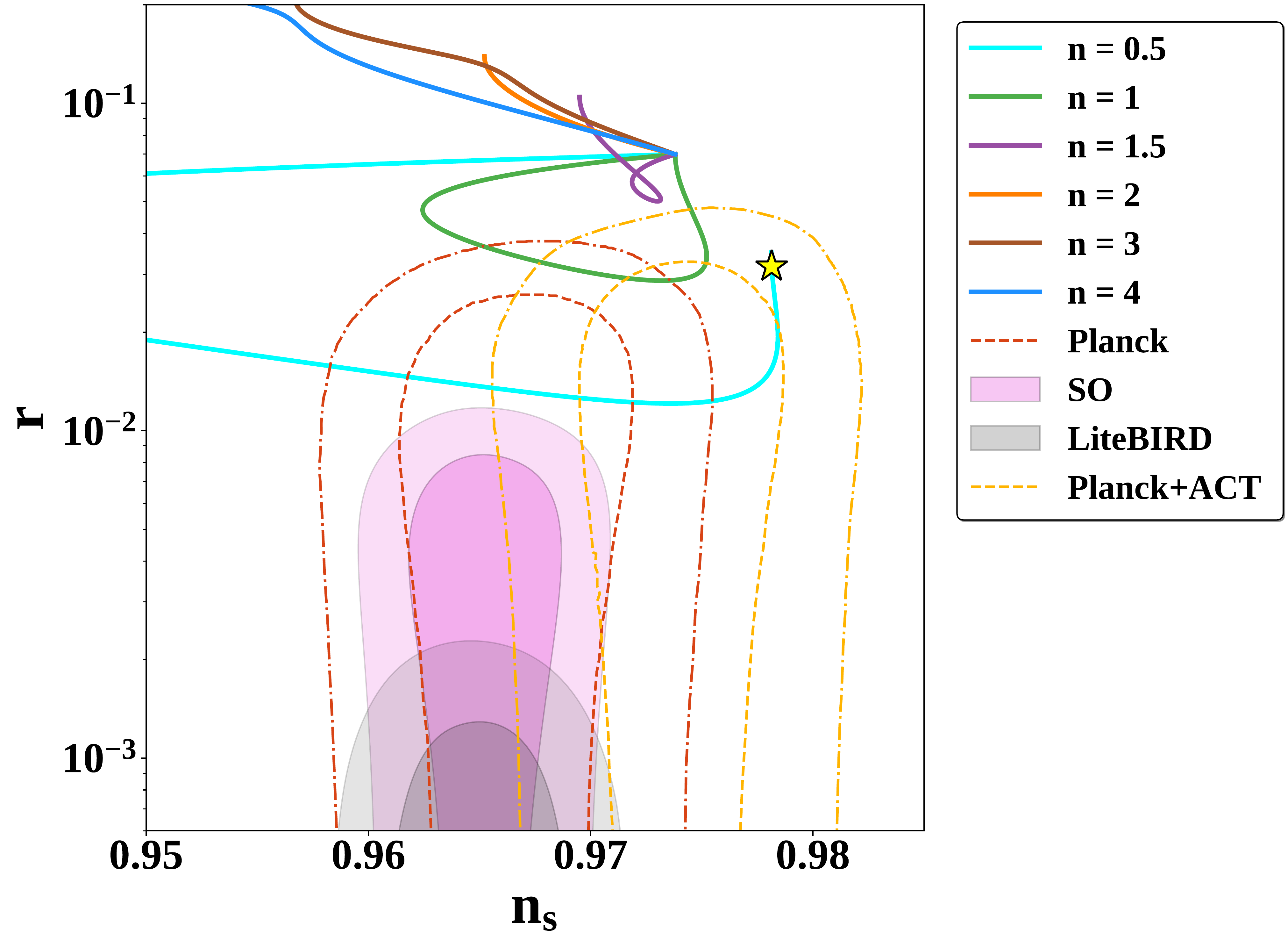}
\includegraphics[width=0.52\linewidth]{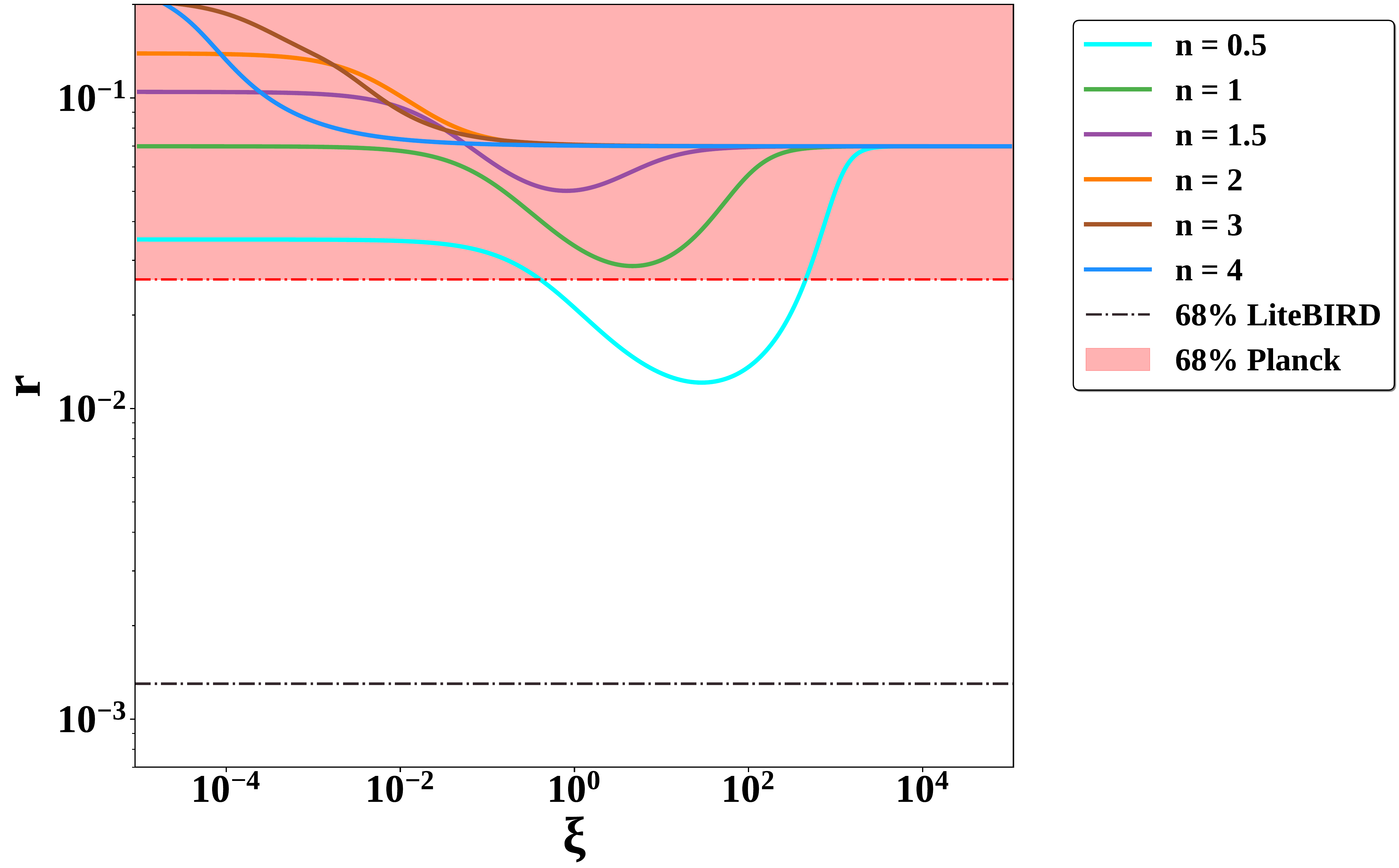}
\caption{\it Inflationary observables for the linear inflation model defined by Eq. \eqref{eq:U1_Palatini} in \textbf{metric} formalism.}
\label{fig:r_ns_1_m}
\end{figure}
%%%%%%%%%%%%%%%%%%%%%%%%%%%%%%%%%%%%%%%%%%%%%%%%%%%%%%%%%%%%%%%%%%%%%%%%%%%%%%
\begin{figure}[t!]
\centering
\includegraphics[width=0.45\linewidth]{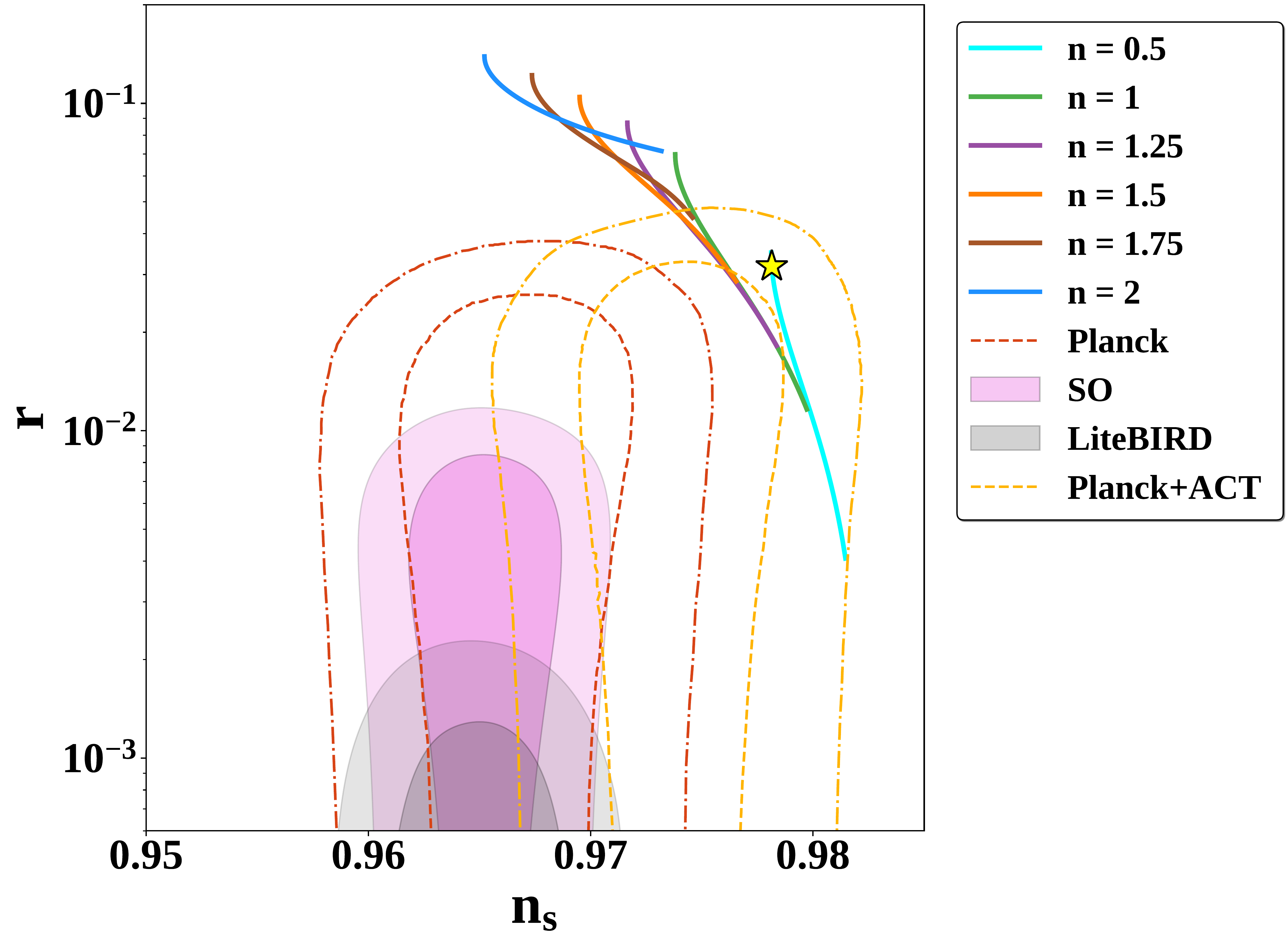}\hfill
\includegraphics[width=0.52\linewidth]{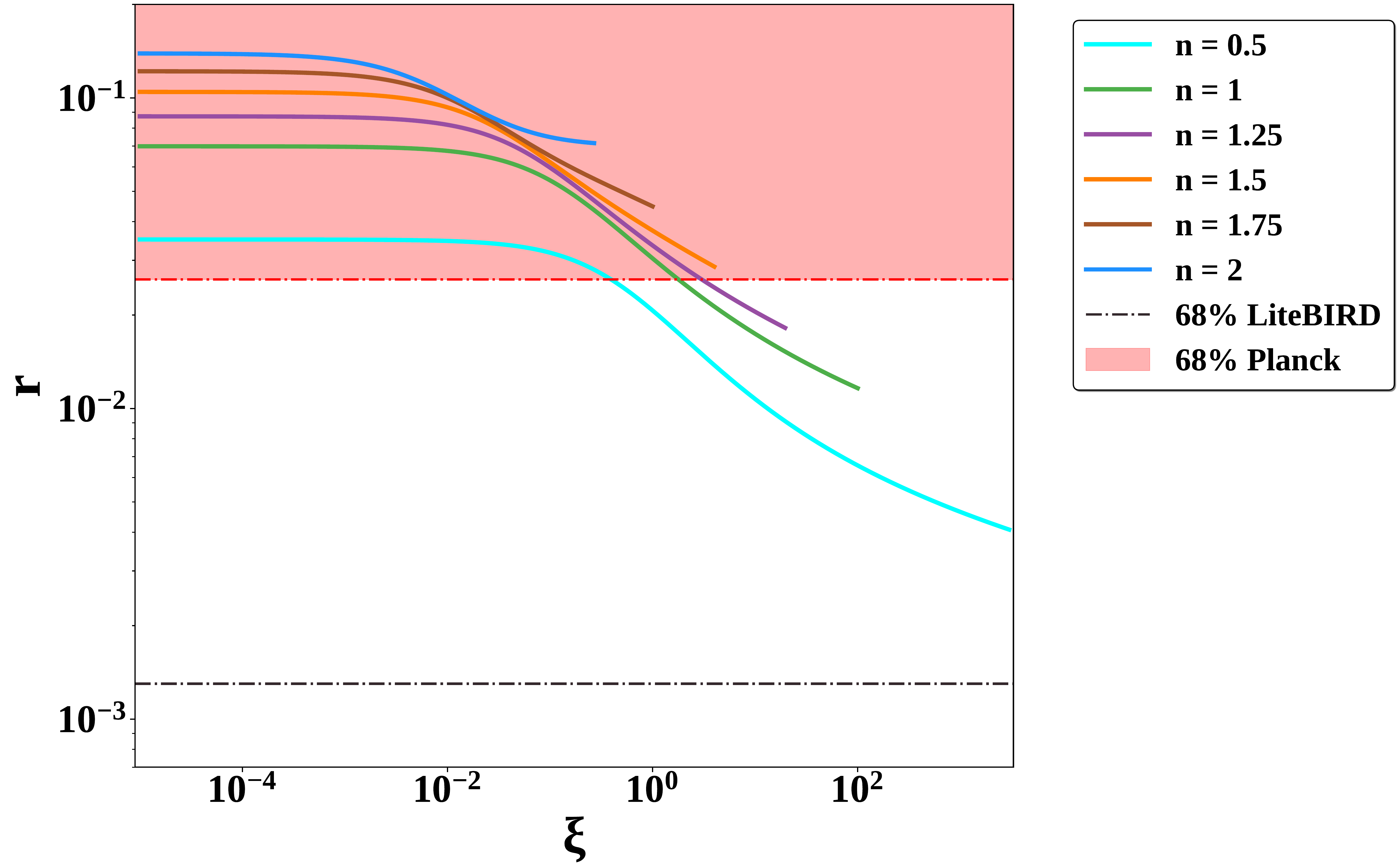}\hfill
\caption{\it Same as Fig.~\ref{fig:r_ns_1_m} but in the \textbf{Palatini} formalism.}
\label{fig:r_ns_1_P}
\end{figure}
\begin{figure}[t!]
\centering
\includegraphics[width=0.45\linewidth]{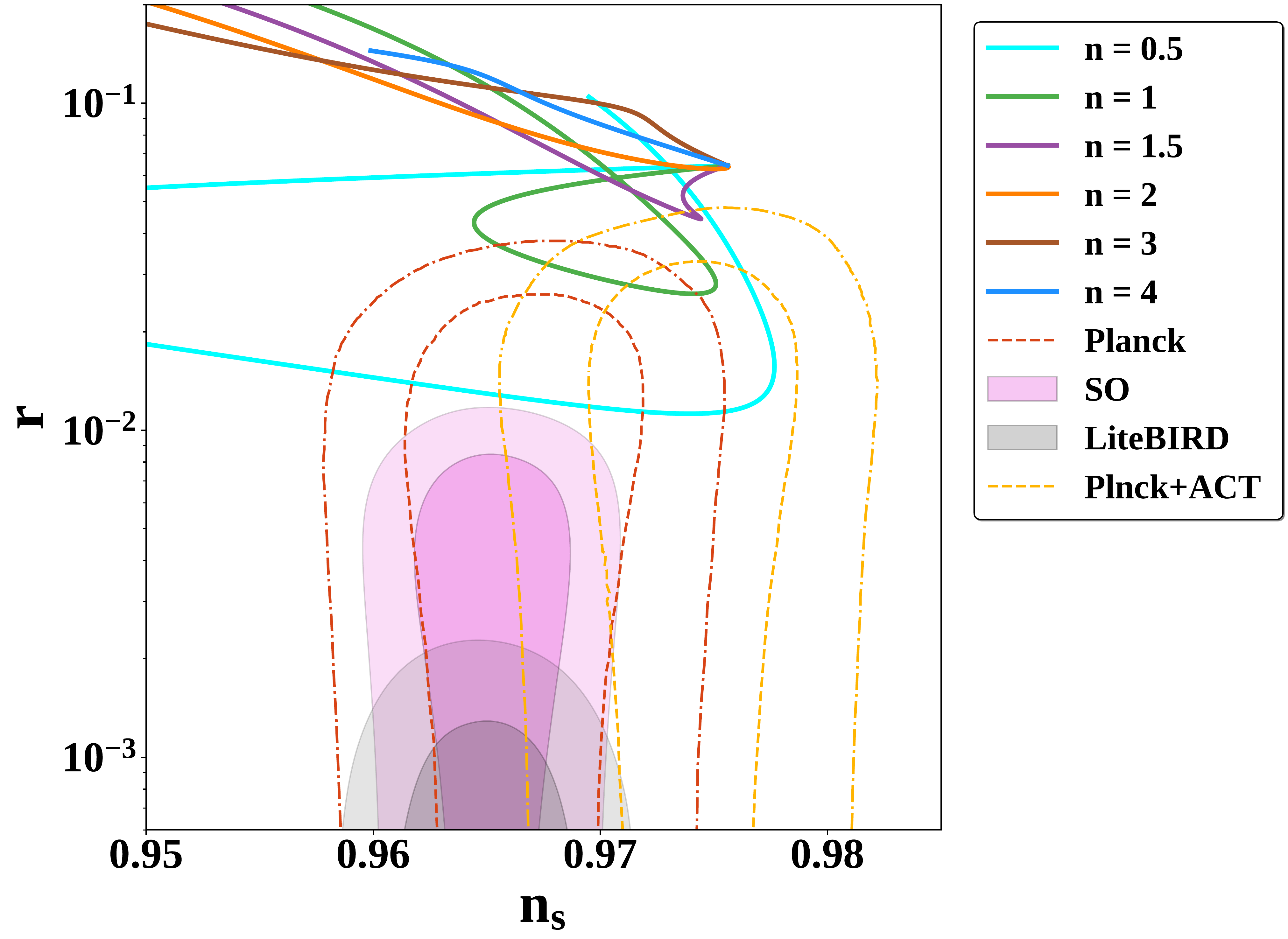}\hfill
\includegraphics[width=0.52\linewidth]{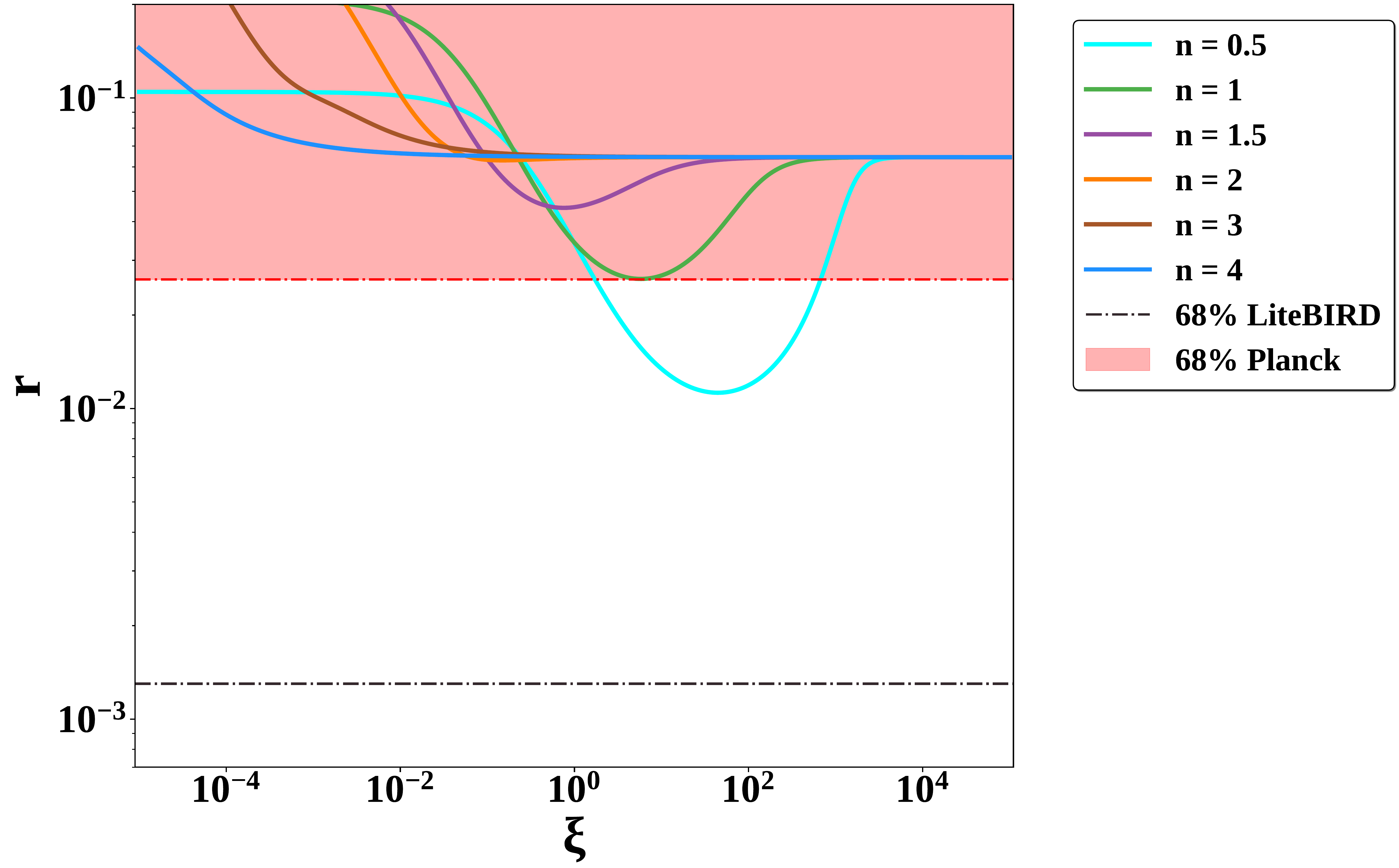}\hfill
\caption{\it Inflationary observables for the linear inflation model defined by Eq. \eqref{eq:U2_Palatini} in \textbf{metric} formalism.}
\label{fig:r_ns_2_m}
\end{figure}
\begin{figure}[t!]
\centering
\includegraphics[width=0.45\linewidth]{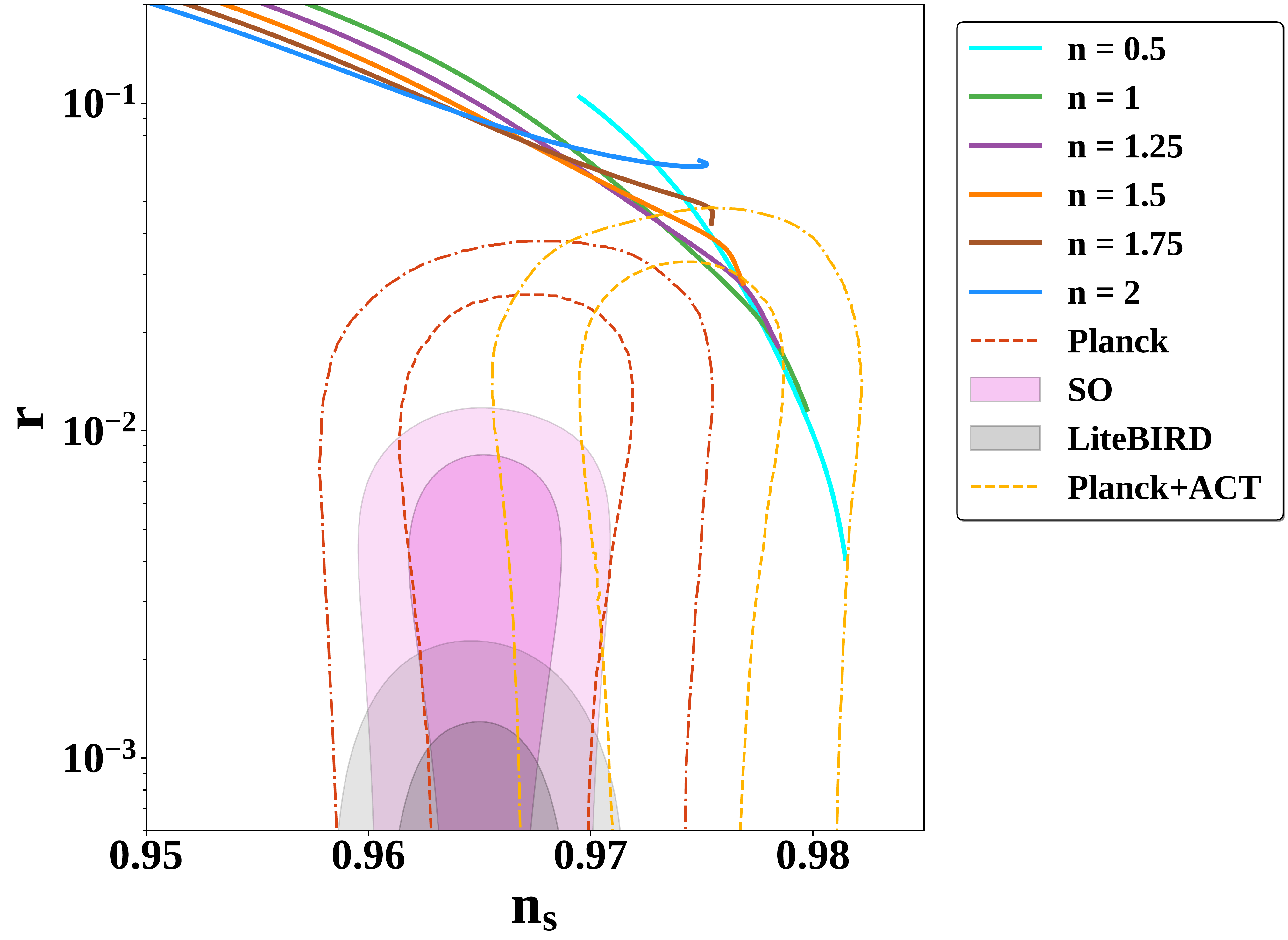}\hfill
\includegraphics[width=0.52\linewidth]{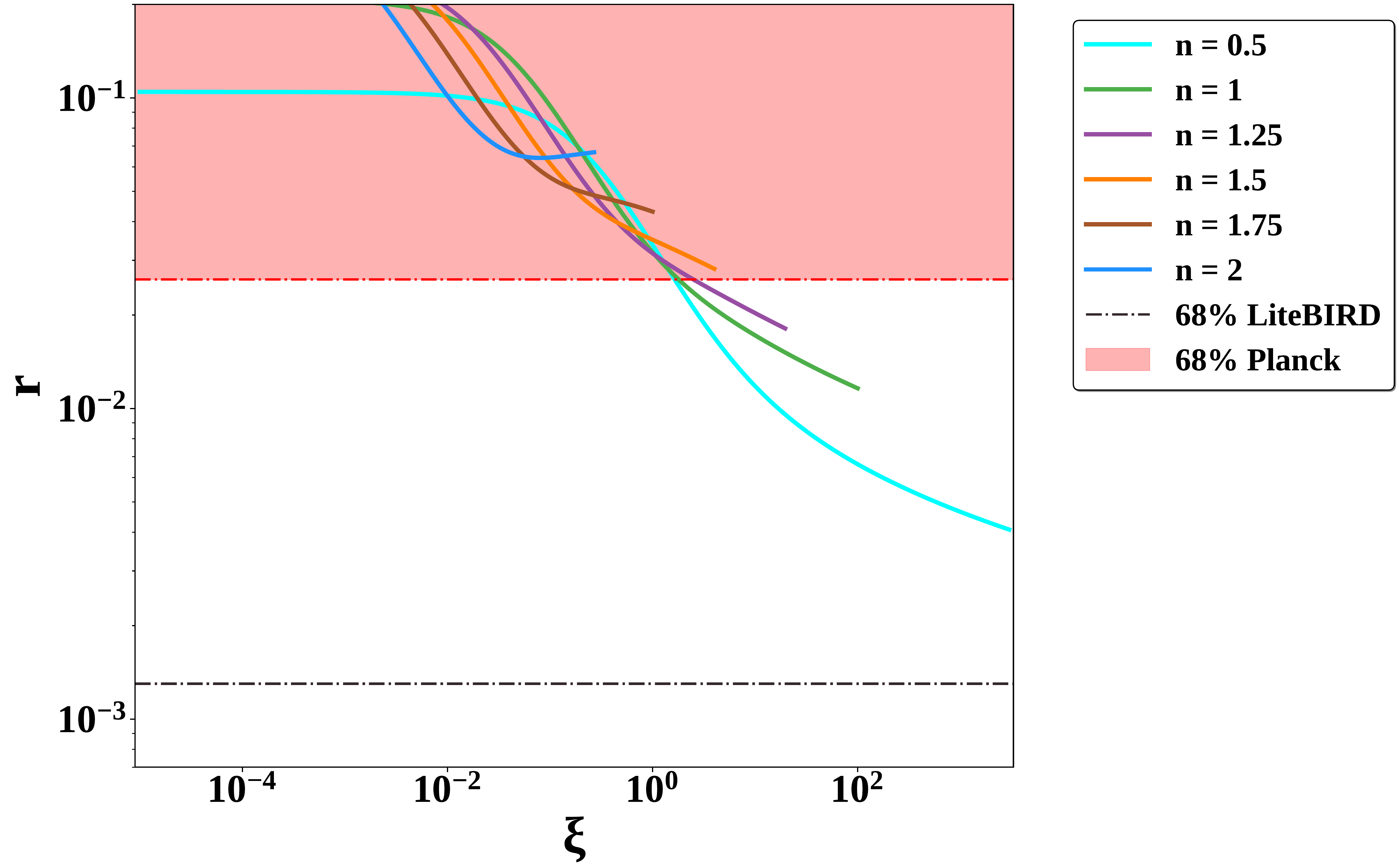}\hfill
\caption{\it Same as Fig.~\ref{fig:r_ns_2_m} but in the \textbf{Palatini} formalism.}
\label{fig:r_ns_2_P}
\end{figure}

%%%%%%%%%%%%%%%%%%%%%%%%%%%%%%%%%%%%%%%%%%%%%%%%%%%%%%%%%%%%%%%%%%%%%%%%%%%%%%
Numerical results for representative choices of $n$ and $\xi$, yielding $N_* \simeq 55$-$58$ $e$-folds and reproducing the observed amplitude of curvature perturbations, $\Delta_\mathcal{R}^2 \simeq 2.1 \times 10^{-9}$, are summarized in Tabs.~\ref{tab:infl-para_1} and~\ref{tab:infl-para_2}. These tables list the initial and end-of-inflation field values, $\phi_i$ and $\phi_e$, together with the inflationary observables $n_s$, $r$, $\alpha_s$, and $N_*$. The corresponding trajectories in the $(n_s, r)$ plane, illustrating the dependence on the non-minimal coupling $\xi$, are shown in Figs.~\ref{fig:r_ns_1_m}--\ref{fig:r_ns_2_P}, alongside current constraints from \emph{Planck}~\cite{Planck:2018jri}, combined \emph{Planck}+ACT data~\cite{ACT:2025tim}, and projected sensitivities of forthcoming CMB experiments, LiteBIRD~\cite{Hazumi_2020} and SO~\cite{SimonsObservatory:2018koc}.

To explicitly demonstrate the unified nature of our cosmological framework, bridging the inflationary analyses with the subsequent thermal history, we highlight a specific benchmark point in the $r$ vs.\ $n_s$ plots in both Fig.~\ref{fig:r_ns_1_m} and Fig.~\ref{fig:r_ns_1_P}, denoted by yellow stars. This point corresponds to the parameter choice $n = 0.5$ and $\xi = 0.1$. An identical methodology is rigorously applied to the other models, for which we similarly select representative, observationally viable benchmarks to drive post-inflationary phenomenological computations. The explicit parameter values for these benchmarks, along with their corresponding inflationary predictions, are detailed in Tabs.~\ref{tab:infl-para_1}--\ref{tab:infl-para_4}. We emphasize that these specific benchmarks are chosen primarily for illustrative purposes to provide a clear narrative for the reader; naturally, numerous other viable parameter combinations within the allowed regions could also be considered.
%%%%%%%%%%%%%%%%%%%%%%%%%%%%%%%%%%%%%%%%%%%%%%%%%%%%%%%%%%%%%%%%%%%%%%%%%%%%%%

\begin{table}[t!]
\centering
\renewcommand{\arraystretch}{1.3} 
\resizebox{\textwidth}{!}{
\begin{tabular}{|c|c|c|c|c|c|c|c|c||c|c|c|c|c|c|c|c|c|}
\hline
\multicolumn{9}{|c||}{\textbf{Metric}} & \multicolumn{9}{c|}{\textbf{Palatini}} \\
\hline
$n$ & $\xi$ & $M\times10^{-3}$ & $\phi_i/M_{\rm Pl}$ & $\phi_e/M_{\rm Pl}$ & $n_s$ & $r$ & $-\alpha_s \times10^{-4}$ & $N$ & $n$ & $\xi$ & $M\times10^{-3}$ & $\phi_i/M_{\rm Pl}$ & $\phi_e/M_{\rm Pl}$ & $n_s$ & $r$ & $-\alpha_s \times10^{-4}$ & $N$ \\
\hline
\multirow{4}{*}{\rotatebox[origin=c]{90}{0.5}} & 0.1 & 18.20 & 13.47 & 1.03 & 0.972 & 0.081 & 5.05 & 57.4 & \multirow{4}{*}{\rotatebox[origin=c]{90}{0.5}} & 0.1 & 18.20 & 13.47 & 1.03 & 0.972 & 0.081 & 5.01 & 57.4 \\
& 1 & 3.63 & 15.32 & 0.78 & 0.976 & 0.034 & 4.32 & 56.6 & & 1 & 5.62 & 15.48 & 0.87 & 0.976 & 0.034 & 4.27 & 56.9 \\
& 100 & 2.75 & 50.54 & $5.1 \times 10^{-3}$ & 0.969 & 0.012 & 8.01 & 56.4 & & 100 & 2.34 & 77.09 & 0.81 & 0.981 & 0.007 & 3.38 & 56.9 \\
& 1000 & 3.31 & 50.77 & $5.1\times10^{-6}$ & 0.930 & 0.035 & 5.94 & 57.8 & & 1000 & 1.20 & 269.24 & 1.82 & 0.981 & 0.005 & 3.28 & 55.1 \\
\hline
\multirow{3}{*}{\rotatebox[origin=c]{90}{1}} & 0.1 & 8.71 & 20.29 & 1.99 & 0.967 & 0.094 & 6.13 & 57.5 & \multirow{3}{*}{\rotatebox[origin=c]{90}{1}} & 0.1 & 8.71 & 20.45 & 2.00 & 0.967 & 0.093 & 6.13 & 57.9 \\
& 1 & 4.27 & 28.53 & 1.29 & 0.974 & 0.032 & 4.77 & 56.7 & & 1 & 4.17 & 30.37 & 1.61 & 0.975 & 0.032 & 4.54 & 56.3 \\
& 100 & 3.63 & 65.55 & $22.58\times10^{-3}$ & 0.966 & 0.048 & 2.86 & 57.8 & & 100 & 2.40 & 571.01 & 1.97 & 0.980 & 0.012 & 3.58 & 56.7 \\
\hline
\multirow{2}{*}{\rotatebox[origin=c]{90}{2}} & 0.1 & 4.27 & 58.97 & 3.06 & 0.975 & 0.062 & 4.91 & 57.9 & \multirow{2}{*}{\rotatebox[origin=c]{90}{1.5}} & 0.1 & 5.13 & 32.21 & 2.82 & 0.970 & 0.060 & 5.90 & 58.0 \\
& 1 & 3.89 & 70.85 & 1.40 & 0.976 & 0.063 & 4.16 & 57.6 & & 1 & 3.55 & 110.47 & 2.19 & 0.976 & 0.035 & 4.23 & 56.6 \\
\hline
\end{tabular}%
}
\caption{\it Same as Table~\ref{tab:infl-para_1}, but for the model defined in Eq.~\eqref{eq:U2_Palatini}.}
\label{tab:infl-para_2}
\end{table}
%%%%%%%%%%%%%%%%%%%%%%%%%%%%%%%%%%%%%%%%%%%%%%%%%%%%%%%%%%%%%%%%%%%%%%%%%%%%%%
In the metric formalism, the inflaton undergoes moderate field excursions, typically in the range $\phi_i \sim 8$-$70\, M_{\rm Pl}$, as shown in Table~\ref{tab:infl-para_1} (for instance, $\phi_i \simeq 57$ for $n=2$ and $\xi=1$). This behavior arises from the $\kappa = 1$ contribution in Eq.~\eqref{eq:Pi-general}, which enhances the effective kinetic term and thus limits the displacement of the canonically normalized field.

As the non-minimal coupling $\xi$ increases, the tensor-to-scalar ratio $r$ initially decreases, reaching values of order $r \sim 10^{-2}$ (see Fig.~\ref{fig:r_ns_1_m}), due to the flattening of the Einstein-frame potential. However, for sufficiently large couplings, $\xi \gtrsim 10^3$, this trend reverses: higher-order contributions become dominant, resulting in a steeper effective potential, a redder scalar spectrum, and a corresponding increase in $r$. This behavior is illustrated, for example, in the case $n=0.5$ and $\xi = 1000$, for which Tab.~\ref{tab:infl-para_1} yields $n_s = 0.931$.
%%%%%%%%%%%%%%%%%%%%%%%%%%%%%%%%%%%%%%%%%%%%%%%%%%%%%%%%%%%%%%%%%%%%%%%%%%%%%%

%%%%%%%%%%%%%%%%%%%%%%%%%%%%%%%%%%%%%%%%%%%%%%%%%%%%%%%%%%%%%%%%%%%%%%%%%%%%%%
%%%%%%%%%%%%%%%%%%%%%%%%%%%%%%%%%%%%%%%%%%%%%%%%%%%%%%%%%%%%%%%%%%%%%%%%%%%%%%
By contrast, in the Palatini formalism, corresponding to the choice $\kappa = 0$ in Eq.~\eqref{eq:Pi-general}, the inflaton can undergo substantially larger field excursions for large values of the non-minimal coupling $\xi$, often reaching several hundred Planck units (for example, $\phi_i \simeq 571$ for $n = 1$ and $\xi = 100$, as shown in Tab.~\ref{tab:infl-para_2}). This behavior results in an ultra-flat Einstein-frame potential, producing a slightly blue-tilted scalar spectrum with $n_s \simeq 0.980$-$0.981$ and a strongly suppressed tensor-to-scalar ratio, $r \lesssim 5 \times 10^{-3}$, for sufficiently large couplings, in agreement with the analytical estimate in Eq.~\eqref{eq:r_1}. Consequently, the corresponding primordial gravitational-wave signal is expected to lie well below the projected sensitivity of forthcoming experiments.

A comparison of the two Jordan-frame potentials defined in Eqs.~\eqref{eq:U1_Palatini} and~\eqref{eq:U2_Palatini} shows that the additional $(f-1)^2$ factor in $V_J^{(2)}$ systematically increases the tensor-to-scalar ratio $r$ and decreases the scalar spectral index $n_s$ relative to $V_J^{(1)}$ for fixed values of $(n,\xi)$, particularly in the low-to-moderate $\xi$ regime (for instance, $r = 0.081$ versus $r = 0.032$ for $n = 0.5$ and $\xi = 0.1$, as presented in Tabs.~\ref{tab:infl-para_1} and~\ref{tab:infl-para_2}). From a physical perspective, this modification introduces an approximately quadratic minimum at small field values, enabling a secondary hilltop-like inflationary phase. Such a phase can amplify curvature perturbations and may have significant implications for post-inflationary phenomena, including non-thermal DM production, as illustrated by the comparison of Figs.~\ref{fig:dm_1} and~\ref{fig:dm_2}, as we will discuss in detail in the following sections.

Across all scenarios, the number of \(e\)-folds at horizon exit lies in the range 
\(N_* \simeq 55\text{--}58\), consistent with standard assumptions about reheating. 
The running of the scalar spectral index is mildly negative, with 
\(|\alpha_s| \sim (3\text{-}6)\times 10^{-4}\). 
All predicted observables are compatible with the latest 
\emph{Planck}+ACT constraints~\cite{Planck:2018jri,ACT:2025tim}.

Overall, Palatini formulations exhibit attractor-like behavior, similar to linear inflation, characterized by an essentially negligible tensor-to-scalar ratio. In contrast, the metric formulations span a broader and experimentally testable parameter space, with $r \sim 0.02$-$0.09$, potentially within reach of future CMB experiments.

%%%%%%%%%%%%%%%%%%%%%%%%%%%%%%%%%%%%%%%%%%%%%%%%%%%%%%%%%%%%%%%%%%%%%%%%%%%%%%
%%%%%%%%%%%%%%%%%%%%%%%%%%%%%%%%%%%%%%%%%%%%%%%%%%%%%%%%%%%%%%%%%%%%%%%%%%%%%%
%%%%%%%%%%%%%%%%%%%%%%%%%%%%%%%%%%%%%%%%%%%%%%%%%%%%%%%%%%%%%%%%%%%%%%%%%%%%%%
\subsection{Brans-Dicke-like induced inflation}
\label{sec:BD_inflation}
We now consider a Brans-Dicke-like model with the Jordan-frame potential
%%%%%%%%%%%%%%%%%%%%%%%%%%%%%%%%%%%%%%%%%%%%%%%%%%%%%%%%%%%%%%%%%%%%%%%%%%%%%%
\begin{equation}
\label{eq:3_potential}
V_J(\phi) = M^2\!\left(\frac{n}{\xi}\,\phi^{\frac{1}{1-n}} - 1 \right)^2 ,
\end{equation}
%%%%%%%%%%%%%%%%%%%%%%%%%%%%%%%%%%%%%%%%%%%%%%%%%%%%%%%%%%%%%%%%%%%%%%%%%%%%%%
together with a shifted non-minimal coupling of the form
%%%%%%%%%%%%%%%%%%%%%%%%%%%%%%%%%%%%%%%%%%%%%%%%%%%%%%%%%%%%%%%%%%%%%%%%%%%%%%
\begin{equation}
f(\phi) = \frac{\xi}{n}\bigl[1 + (\phi - 1)^n\bigr].
\end{equation}
%%%%%%%%%%%%%%%%%%%%%%%%%%%%%%%%%%%%%%%%%%%%%%%%%%%%%%%%%%%%%%%%%%%%%%%%%%%%%%
The function \( f(\phi) \) possesses a minimum at \( \phi \simeq 1 \), giving rise 
to a small-field, approximately quadratic potential in the Einstein frame. 
This potential becomes further flattened for large values of \( n \) or \( \xi \), 
thereby supporting slow-roll inflation with strictly sub-Planckian field excursions.
%%%%%%%%%%%%%%%%%%%%%%%%%%%%%%%%%%%%%%%%%%%%%%%%%%%%%%%%%%%%%%%%%%%%%%%%%%%%%%

\begin{table}[t!]
\centering
\renewcommand{\arraystretch}{1.3} 
\resizebox{\textwidth}{!}{
\begin{tabular}{|c|c|c|c|c|c|c|c|c||c|c|c|c|c|c|c|c|c|}
\hline
\multicolumn{9}{|c||}{\textbf{Metric}} & \multicolumn{9}{c|}{\textbf{Palatini}} \\
\hline
$n$ & $\xi$ & $M\!\times\!10^{-9}$ & $\phi_i/M_{\rm Pl}$ & $\phi_e/M_{\rm Pl}$ & $n_s$ & $r$ & $-\alpha_s\!\times\!10^{-4}$ & $N$ & $n$ & $\xi$ & $M\!\times\!10^{-9}$ & $\phi_i/M_{\rm Pl}$ & $\phi_e/M_{\rm Pl}$ & $n_s$ & $r$ & $-\alpha_s\!\times\!10^{-4}$ & $N$ \\
\hline
\multirow{3}{*}{\rotatebox[origin=c]{90}{5}} & 0.1 & 2.76 & 1.31 & 2.63 & 0.951 & 0.001 & 8.76 & 56.8 & \multirow{3}{*}{\rotatebox[origin=c]{90}{5}} & 0.1 & 2.55 & 1.31 & 2.15 & 0.954 & 0.001 & 7.41 & 56.1 \\
& 1 & 49.42 & 1.15 & 1.96 & 0.951 & $3.14\!\times\!10^{-5}$ & 8.94 & 54.9 & & 1 & 48.52 & 1.14 & 1.77 & 0.952 & $3.03\!\times\!10^{-5}$ & 7.75 & 54.7 \\
& 4 & 1004.51 & 1.09 & 1.77 & 0.951 & $3.17\!\times\!10^{-6}$ & 9.13 & 54.1 & & 4 & 1001.23 & 1.09 & 1.64 & 0.951 & $3.15\!\times\!10^{-6}$ & 7.62 & 54.1 \\
\hline
\multirow{3}{*}{\rotatebox[origin=c]{90}{35}} & 0.1 & 0.04 & 1.84 & 2.00 & 0.963 & 0.001 & 6.76 & 55.7 & \multirow{3}{*}{\rotatebox[origin=c]{90}{35}} & 0.1 & 0.04 & 1.84 & 1.98 & 0.963 & 0.001 & 6.28 & 55.5 \\
& 1 & 1.08 & 1.78 & 1.96 & 0.962 & $5.31\!\times\!10^{-5}$ & 6.88 & 54.8 & & 1 & 1.08 & 1.78 & 1.94 & 0.962 & $5.31\!\times\!10^{-5}$ & 6.32 & 54.8 \\
& 34 & 6700.84 & 1.70 & 1.91 & 0.961 & $1.32\!\times\!10^{-6}$ & 7.12 & 53.8 & & 34 & 6620 & 1.70 & 1.89 & 0.962 & $1.29\!\times\!10^{-6}$ & 6.34 & 53.8 \\
\hline
\multirow{3}{*}{\rotatebox[origin=c]{90}{55}} & 0.1 & 0.01 & 1.89 & 2.00 & 0.963 & $4\!\times\!10^{-4}$ & 6.70 & 55.5 & \multirow{3}{*}{\rotatebox[origin=c]{90}{55}} & 0.1 & 0.01 & 1.89 & 1.98 & 0.963 & $4\!\times\!10^{-4}$ & 6.22 & 55.4 \\
& 1 & 0.37 & 1.85 & 1.97 & 0.963 & $3.82\!\times\!10^{-5}$ & 6.83 & 54.7 & & 1 & 0.37 & 1.85 & 1.96 & 0.963 & $3.82\!\times\!10^{-4}$ & 6.25 & 54.7 \\
& 54 & 7427 & 1.79 & 1.93 & 0.962 & $6.31\!\times\!10^{-7}$ & 7.09 & 53.6 & & 54 & 7427 & 1.79 & 1.92 & 0.962 & $6.31\!\times\!10^{-7}$ & 6.27 & 53.6 \\
\hline
\end{tabular}%
}
\caption{\it  Comparison of parameter values for the Brans-Dicke-like inflation model expressed in Eq.~\eqref{eq:3_potential} across metric and Palatini formulations.}
\label{tab:infl-para_3}
\end{table}
%%%%%%%%%%%%%%%%%%%%%%%%%%%%%%%%%%%%%%%%%%%%%%%%%%%%%%%%%%%%%%%%%%%%%%%%%%%%%%
The large-field analytical approximations (which also remain accurate near the minimum) are given by
%%%%%%%%%%%%%%%%%%%%%%%%%%%%%%%%%%%%%%%%%%%%%%%%%%%%%%%%%%%%%%%%%%%%%%%%%%%%%%
\begin{align}
\label{eq:ns_3}
n_s &\approx
\begin{cases}
1 - \dfrac{2n\!\left(n+2+\xi (\phi_i-1)^n\right)}{n\!\left(n+2+\xi (\phi_i-1)^n\right)L^2 + 2(\phi_i-1)^2}\,,~~
L = \ln\!\left[1 + \xi (\phi_i-1)^n\right] & \text{(metric)},\\[10pt]
1 - n\xi (\phi_i-1)^{n-2}\left(\dfrac{3n}{L^2} - \dfrac{n-2}{L}\right)\,,~~
L = \ln\!\left[1 + \xi (\phi_i-1)^n\right] & \text{(Palatini)}.
\end{cases}\\[15pt]
\label{eq:r_3}
r &\approx
\begin{cases}
\dfrac{16n^2}{2(\phi_i-1)^2 + 3n^2 L^2}\,,~~
L = \ln\!\left[1 + \xi (\phi_i-1)^n\right] & \text{(metric)},\\[10pt]
\dfrac{8n^2}{(\phi_i-1)^2}\,\dfrac{\left(1 + \dfrac{u}{2}\right)^2}{1+u}\,,~~
u = \xi (\phi_i-1)^n & \text{(Palatini).}
\end{cases}
\end{align}
%%%%%%%%%%%%%%%%%%%%%%%%%%%%%%%%%%%%%%%%%%%%%%%%%%%%%%%%%%%%%%%%%%%%%%%%%%%%%%
In the small-field limit (\( u \ll 1 \), with \( L \simeq u \)), both formalisms 
reduce to quadratic inflation, yielding
%%%%%%%%%%%%%%%%%%%%%%%%%%%%%%%%%%%%%%%%%%%%%%%%%%%%%%%%%%%%%%%%%%%%%%%%%%%%%%
\begin{equation}
n_s \simeq 1 - \frac{2}{N_*}
\qquad \text{and} \qquad
r \simeq \frac{8}{N_*}.
\end{equation}
%%%%%%%%%%%%%%%%%%%%%%%%%%%%%%%%%%%%%%%%%%%%%%%%%%%%%%%%%%%%%%%%%%%%%%%%%%%%%%
The analytical expressions in Eqs.~\eqref{eq:ns_3} and~\eqref{eq:r_3} clarify how the 
non-minimal coupling \( \xi \) and the exponent \( n \) control the curvature of the 
inflationary potential. In particular, in the metric formulation, the tensor-to-scalar 
ratio \( r \) in Eq.~\eqref{eq:r_3} decreases inversely with the squared shifted field, 
\( (\phi_i - 1)^2 \), leading to an efficient suppression of tensor perturbations.
%%%%%%%%%%%%%%%%%%%%%%%%%%%%%%%%%%%%%%%%%%%%%%%%%%%%%%%%%%%%%%%%%%%%%%%%%%%%%%

 \begin{figure}[t!]
\centering
\includegraphics[width=0.45\linewidth]{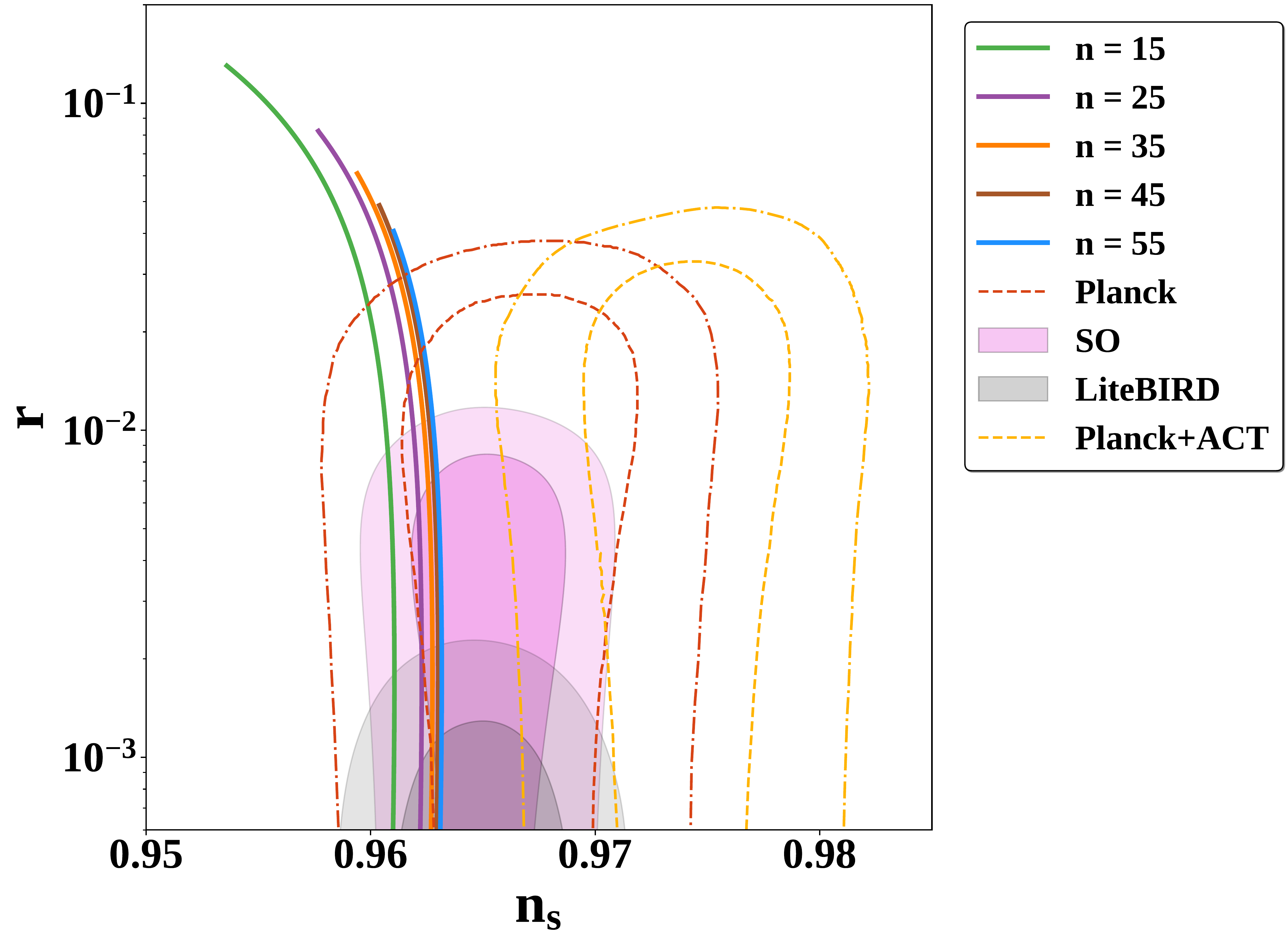}\hfill
\includegraphics[width=0.52\linewidth]{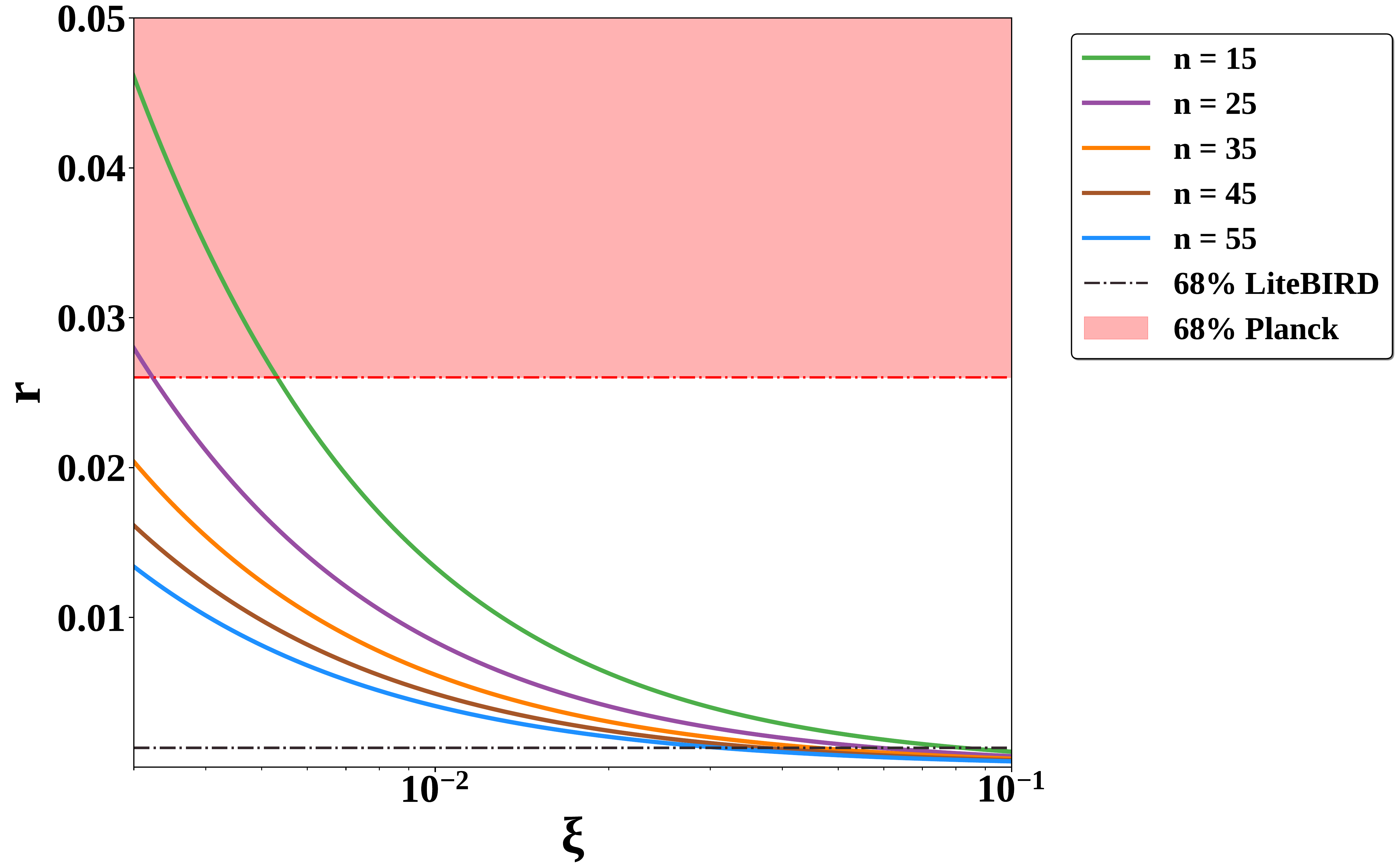}\hfill
\caption{\it Inflationary observables for the Brans-Dicke-like inflation model defined by Eq. \eqref{eq:3_potential} in \textbf{metric} formalism.}
\label{fig:r_ns_3_m}
\end{figure}
%%%%%%%%%%%%%%%%%%%%%%%%%%%%%%%%%%%%%%%%%%%%%%%%%%%%%%%%%%%%%%%%%%%%%%%%%%%%%%
\begin{figure}[t!]
\centering
\includegraphics[width=0.45\linewidth]{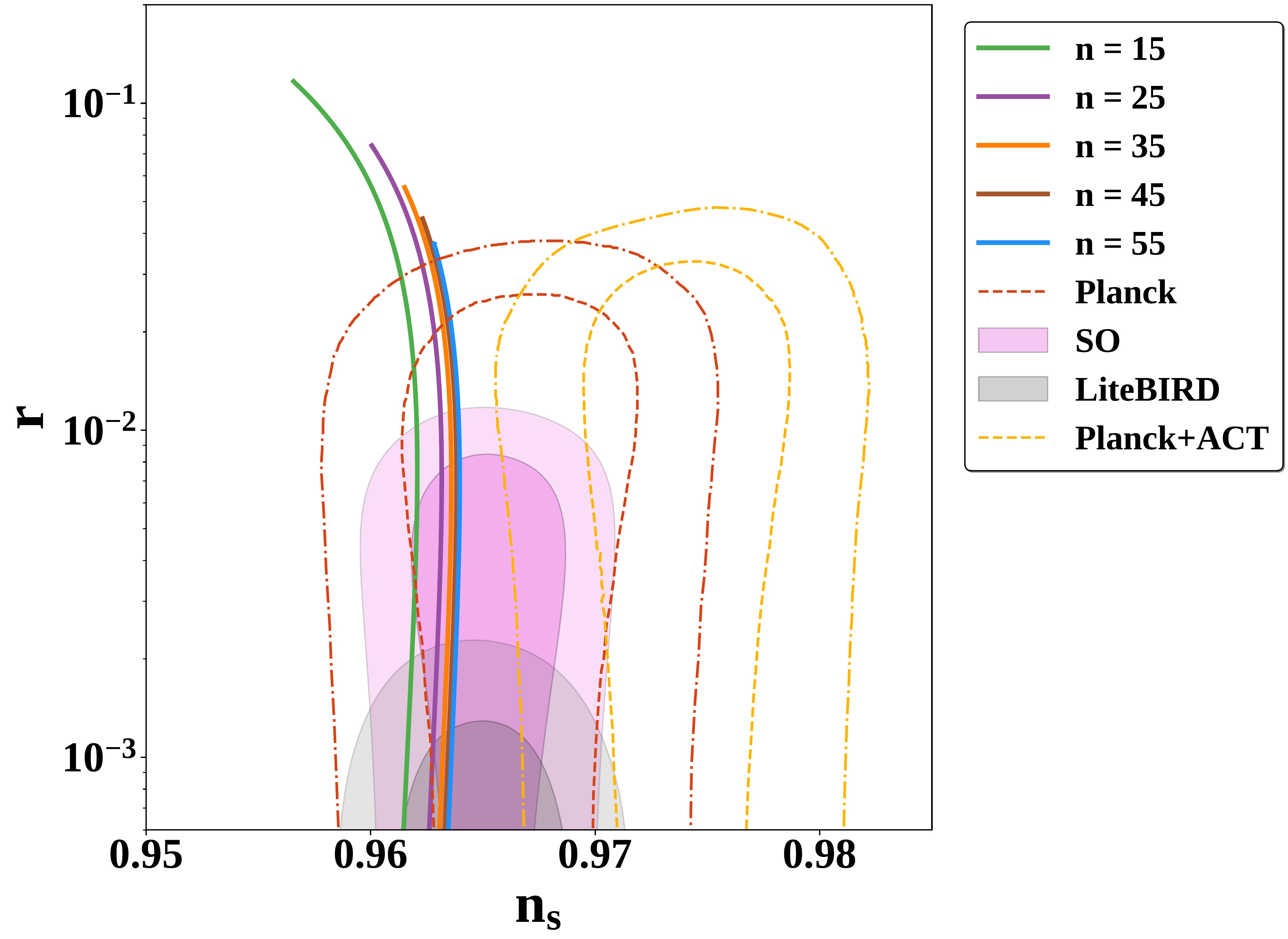}\hfill
\includegraphics[width=0.52\linewidth]{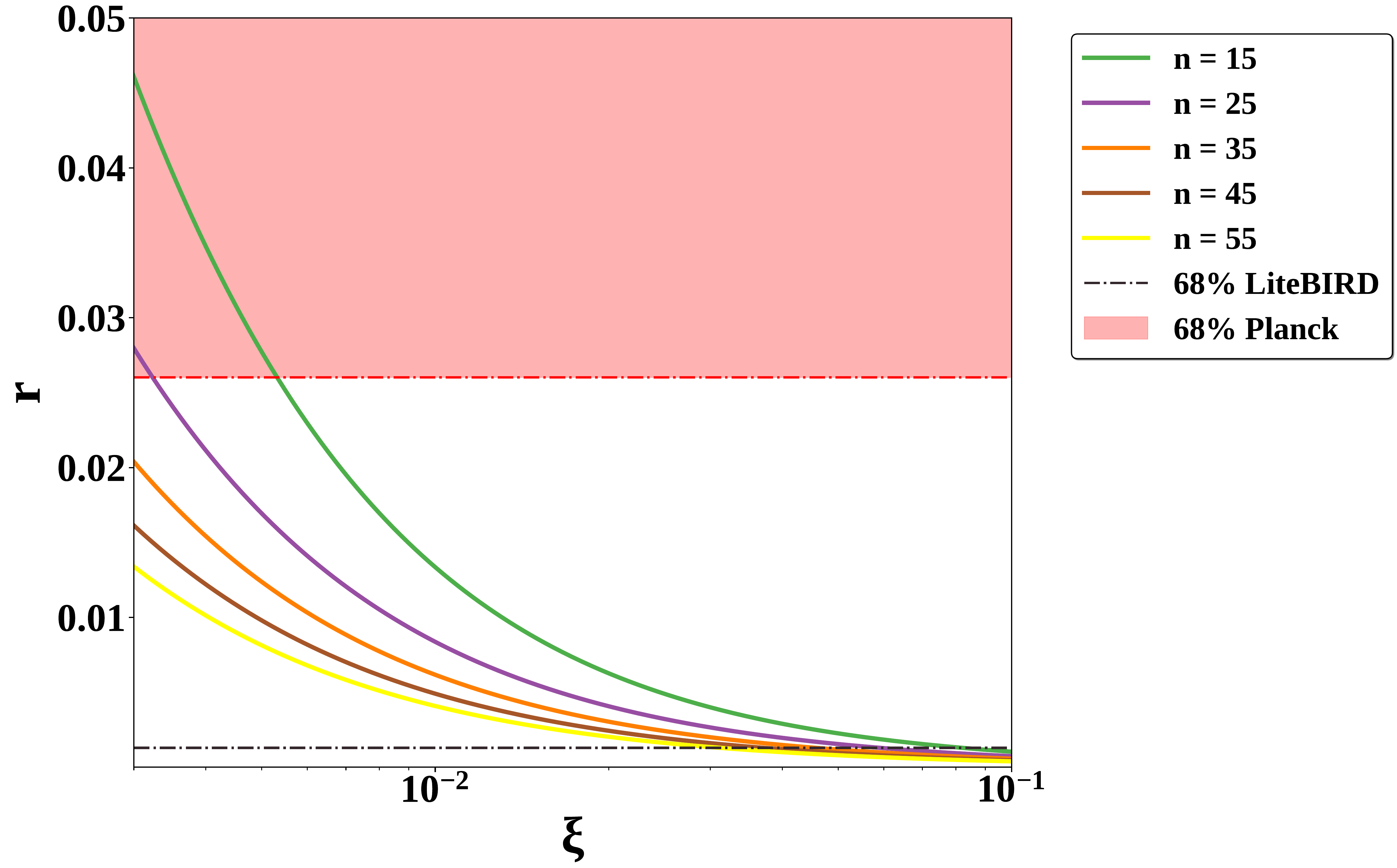}\hfill
\caption{\it Same as Fig.~\ref{fig:r_ns_3_P} but in the \textbf{Palatini} formalism.}
\label{fig:r_ns_3_P}
\end{figure}
%%%%%%%%%%%%%%%%%%%%%%%%%%%%%%%%%%%%%%%%%%%%%%%%%%%%%%%%%%%%%%%%%%%%%%%%%%%%%%
Benchmark calculations are performed for \( n = 5, 35, 55 \) and selected values of \( \xi \), which are calibrated to yield \( N_* \simeq 54\text{--}57 \) and reproduce the observed amplitude of scalar perturbations. The numerical results are summarized in Tab.~\ref{tab:infl-para_3}. The corresponding trajectories of these observables as functions of \( \xi \) 
are shown in Figs.~\ref{fig:r_ns_3_m} (metric) and~\ref{fig:r_ns_3_P} (Palatini), 
confronted with the current and forthcoming CMB constraints~%
\cite{Planck:2018jri,SimonsObservatory:2018koc,Hazumi_2020,ACT:2025tim}.

 Field displacements remain firmly sub-Planckian, with
$\Delta \phi \lesssim 2\,M_{\rm Pl}$ (for example, $\phi_i \simeq 1.89$ 
for $n = 55$ and $\xi = 0.1$ in the metric case), in good agreement 
with swampland distance conjectures that disfavor trans-Planckian field excursions in effective theories coupled to quantum gravity~\cite{Ooguri:2006in,Obied:2018sgi}.

In contrast to linear inflationary models, where the predictions of metric 
and Palatini gravity diverge significantly, the two formalisms yield closely 
aligned results here. This convergence arises because inflation proceeds in 
the small-field regime, specifically where \( \phi \ll 1/\sqrt{\xi} \). In this 
limit, the non-minimal coupling correction to the kinetic term is subdominant, 
rendering the difference between the metric (\( \kappa = 1 \)) and Palatini 
(\( \kappa = 0 \)) kinetic terms in Eq.~\eqref{eq:Pi-general} negligible. 
Consequently, the field-space metric approaches unity in both formalisms, 
leading to indistinguishable predictions for the inflationary observables.

The scalar spectral index lies in the range
\( n_s \simeq 0.951\text{--}0.963 \), increasing with larger values of \( n \)
(see Fig.~\ref{fig:r_ns_3_m}), and is consistent with current \emph{Planck}
constraints~\cite{Planck:2018jri} as well as the projected sensitivities of
forthcoming experiments such as the Simons Observatory~\cite{SimonsObservatory:2018koc}
and LiteBIRD~\cite{Hazumi_2020}. Furthermore, tensor perturbations are strongly suppressed,
with \( r \lesssim 10^{-3} \), decreasing to \( \mathcal{O}(10^{-7}) \) at large
\( \xi \), in agreement with the analytical expectation in
Eq.~\eqref{eq:r_3}.

%%%%%%%%%%%%%%%%%%%%%%%%%%%%%%%%%%%%%%%%%%%%%%%%%%%%%%%%%%%%%%%%%%%%%%%%%%%%%%
%%%%%%%%%%%%%%%%%%%%%%%%%%%%%%%%%%%%%%%%%%%%%%%%%%%%%%%%%%%%%%%%%%%%%%%%%%%%%%
%%%%%%%%%%%%%%%%%%%%%%%%%%%%%%%%%%%%%%%%%%%%%%%%%%%%%%%%%%%%%%%%%%%%%%%%%%%%%%
\subsection{Higgs inflation-like induced inflation}
\label{sec:Higgs_like}
We finally consider a generalized Higgs-inflation-like model characterized by
a constant Jordan-frame potential
%%%%%%%%%%%%%%%%%%%%%%%%%%%%%%%%%%%%%%%%%%%%%%%%%%%%%%%%%%%%%%%%%%%%%%%%%%%%%%
\begin{equation}
\label{eq:4_potential}
V_J(\phi) = M^2 
\left(
\frac{\xi}{\xi + n\bigl(n\lambda\bigr)^{1/(1-n)}}
\right)^2,
\qquad 
\lambda = \frac{1}{n},
\end{equation}
%%%%%%%%%%%%%%%%%%%%%%%%%%%%%%%%%%%%%%%%%%%%%%%%%%%%%%%%%%%%%%%%%%%%%%%%%%%%%%
together with a shifted non-minimal coupling
%%%%%%%%%%%%%%%%%%%%%%%%%%%%%%%%%%%%%%%%%%%%%%%%%%%%%%%%%%%%%%%%%%%%%%%%%%%%%%
\begin{equation}
f(\phi) = 1 + \frac{\xi}{n}\bigl[1 + (\phi - 1)^n\bigr].
\end{equation}
%%%%%%%%%%%%%%%%%%%%%%%%%%%%%%%%%%%%%%%%%%%%%%%%%%%%%%%%%%%%%%%%%%%%%%%%%%%%%%
This setup generates a plateau-like Einstein-frame potential at large field
values, with the degree of flattening controlled by the parameters \( n \) and
\( \xi \), while maintaining sub-Planckian field excursions.

In the large-field regime, analytical approximations for the main observables
take the form
%%%%%%%%%%%%%%%%%%%%%%%%%%%%%%%%%%%%%%%%%%%%%%%%%%%%%%%%%%%%%%%%%%%%%%%%%%%%%%
\begin{align}
\label{eq:ns_4}
n_s &\approx
\begin{cases}
1 - \dfrac{16|a|}{\bigl[1 + W_{-1}(z)\bigr]^2},
\qquad
z = -\bigl(1+e^{-8|a|N_*}\bigr)e^{-8|a|N_*},
\qquad
e^{2\sqrt{|a|}\phi_i} = -W_{-1}(z),
& \text{(metric)}. \\[12pt]
1 - \dfrac{6}{N_*\!\left(1+8\xi N_*\right)}
+ \dfrac{3+8\xi N_*}{N_*\sqrt{1+8\xi N_*}},
& \text{(Palatini)}.
\end{cases}
\\[15pt]
\label{eq:r_4}
r &\approx
\begin{cases}
\dfrac{128|a|}{\bigl[1 + W_{-1}(z)\bigr]^2},
& \text{(metric)}, \\[8pt]
\dfrac{16}{N_*\!\left(1+8\xi N_*\right)},
& \text{(Palatini)}.
\end{cases}
\end{align}
%%%%%%%%%%%%%%%%%%%%%%%%%%%%%%%%%%%%%%%%%%%%%%%%%%%%%%%%%%%%%%%%%%%%%%%%%%%%%%
\begin{table}[t!]
\centering
\renewcommand{\arraystretch}{1.3}
\resizebox{\textwidth}{!}{
\begin{tabular}{|c|c|c|c|c|c|c|c|c||c|c|c|c|c|c|c|c|c|}
\hline
\multicolumn{9}{|c||}{\textbf{Metric}} & \multicolumn{9}{c|}{\textbf{Palatini}} \\
\hline
$n$ & $\xi$ & $M\!\times\!10^{-4}$ & $\phi_i/M_{\rm Pl}$ & $\phi_e/M_{\rm Pl}$ & $n_s$ & $r$ & $-\alpha_s\!\times\!10^{-4}$ & $N$ & $n$ & $\xi$ & $M\!\times\!10^{-4}$ & $\phi_i/M_{\rm Pl}$ & $\phi_e/M_{\rm Pl}$ & $n_s$ & $r$ & $-\alpha_s\!\times\!10^{-4}$ & $N$ \\
\hline
\multirow{3}{*}{\rotatebox[origin=c]{90}{5}} 
 & 0.1 & 0.49 & 1.31 & 3.09 & 0.951 & $2.85\!\times\!10^{-5}$ & 8.95 & 54.9 & \multirow{3}{*}{\rotatebox[origin=c]{90}{5}} & 0.1 & 0.48 & 1.31 & 2.68 & 0.952 & $2.77\!\times\!10^{-4}$ & 8.80 & 54.6 \\
 & 1   & 0.03 & 1.15 & 2.16 & 0.951 & $5.22\!\times\!10^{-6}$ & 9.09 & 54.3 & & 1   & 0.03 & 1.14 & 1.95 & 0.951 & $5.18\!\times\!10^{-6}$ & 9.01 & 54.2 \\
 & 4   & 0.01 & 1.09 & 1.85 & 0.950 & $1.46\!\times\!10^{-6}$ & 9.20 & 53.9 & & 4   & 0.01 & 1.09 & 1.70 & 0.951 & $1.42\!\times\!10^{-6}$ & 9.16 & 53.9 \\
\hline
\multirow{3}{*}{\rotatebox[origin=c]{90}{35}} 
 & 0.1 & 0.82 & 1.84 & 2.08 & 0.962 & $1.77\!\times\!10^{-6}$ & 7.10 & 53.9 & \multirow{3}{*}{\rotatebox[origin=c]{90}{35}} & 0.1 & 0.83 & 1.84 & 2.06 & 0.961 & $1.80\!\times\!10^{-6}$ & 7.09 & 53.9 \\
 & 1   & 0.08 & 1.78 & 2.01 & 0.962 & $1.50\!\times\!10^{-6}$ & 7.11 & 53.8 & & 1   & 0.08 & 1.78 & 1.99 & 0.962 & $1.50\!\times\!10^{-6}$ & 7.10 & 56.8 \\
 & 34  & 0.01 & 1.70 & 1.92 & 0.962 & $6.37\!\times\!10^{-7}$ & 7.17 & 53.6 & & 34  & 0.01 & 1.70 & 1.90 & 0.962 & $6.37\!\times\!10^{-7}$ & 7.16 & 53.6 \\
\hline
\multirow{3}{*}{\rotatebox[origin=c]{90}{55}} 
 & 0.1 & 0.86 & 1.89 & 2.05 & 0.962 & $7.77\!\times\!10^{-7}$ & 7.08 & 53.7 & \multirow{3}{*}{\rotatebox[origin=c]{90}{55}} & 0.1 & 0.86 & 1.89 & 2.04 & 0.962 & $7.75\!\times\!10^{-7}$ & 7.07 & 53.7 \\
 & 1   & 0.08 & 1.85 & 2.00 & 0.962 & $7.05\!\times\!10^{-7}$ & 7.08 & 53.6 & & 1   & 0.08 & 1.85 & 1.99 & 0.962 & $7.00\!\times\!10^{-7}$ & 7.08 & 53.6 \\
 & 54  & 0.00394 & 1.79 & 1.94 & 0.962 & $3.12\!\times\!10^{-7}$ & 7.14 & 53.4 & & 54  & 0.00394 & 1.79 & 1.93 & 0.962 & $3.12\!\times\!10^{-7}$ & 7.13 & 53.4 \\
\hline
\end{tabular}%
}
\caption{\it Comparison of parameter values for the Higgs-like inflation model expressed in Eq.~\eqref{eq:4_potential} across metric and Palatini formulations.}
\label{tab:infl-para_4}
\end{table}
%%%%%%%%%%%%%%%%%%%%%%%%%%%%%%%%%%%%%%%%%%%%%%%%%%%%%%%%%%%%%%%%%%%%%%%%%%%%%%
where, in the metric formulation
%%%%%%%%%%%%%%%%%%%%%%%%%%%%%%%%%%%%%%%%%%%%%%%%%%%%%%%%%%%%%%%%%%%%%%%%%%%%%%
\begin{equation}
|a| = \frac{\xi}{\xi + \tfrac{3}{2} n^2 \xi}.
\end{equation}
%%%%%%%%%%%%%%%%%%%%%%%%%%%%%%%%%%%%%%%%%%%%%%%%%%%%%%%%%%%%%%%%%%%%%%%%%%%%%%
In the limit of large non-minimal coupling, both formulations approach the
well-known universal attractor predictions~\cite{Jarv:2017azx}
%%%%%%%%%%%%%%%%%%%%%%%%%%%%%%%%%%%%%%%%%%%%%%%%%%%%%%%%%%%%%%%%%%%%%%%%%%%%%%
\begin{equation}
n_s \simeq 1 - \frac{2}{N_*}, \qquad
r \simeq \frac{12}{N_*^2} \quad \text{(metric)}, \qquad
r \simeq \frac{2}{\xi N_*^2} \quad \text{(Palatini)}.
\end{equation}
%%%%%%%%%%%%%%%%%%%%%%%%%%%%%%%%%%%%%%%%%%%%%%%%%%%%%%%%%%%%%%%%%%%%%%%%%%%%%%
The analytical expressions for this model, given in 
Eqs.~\eqref{eq:ns_4} and~\eqref{eq:r_4}, elucidate the role of the non-minimal 
coupling \( \xi \) and the exponent \( n \) in shaping the inflationary plateau. 
In particular, within the Palatini approximation, Eq.~\eqref{eq:ns_4} shows that 
the scalar spectral index \( n_s \) rapidly approaches its attractor value as 
\( \xi \) increases, while the tensor-to-scalar ratio in 
Eq.~\eqref{eq:r_4} decreases quadratically with the number of \(e\)-folds 
\( N_* \), thereby strongly suppressing the primordial gravitational-wave signal.
%%%%%%%%%%%%%%%%%%%%%%%%%%%%%%%%%%%%%%%%%%%%%%%%%%%%%%%%%%%%%%%%%%%%%%%%%%%%%%

\begin{figure}[t!]
\centering
\includegraphics[width=0.5\linewidth]{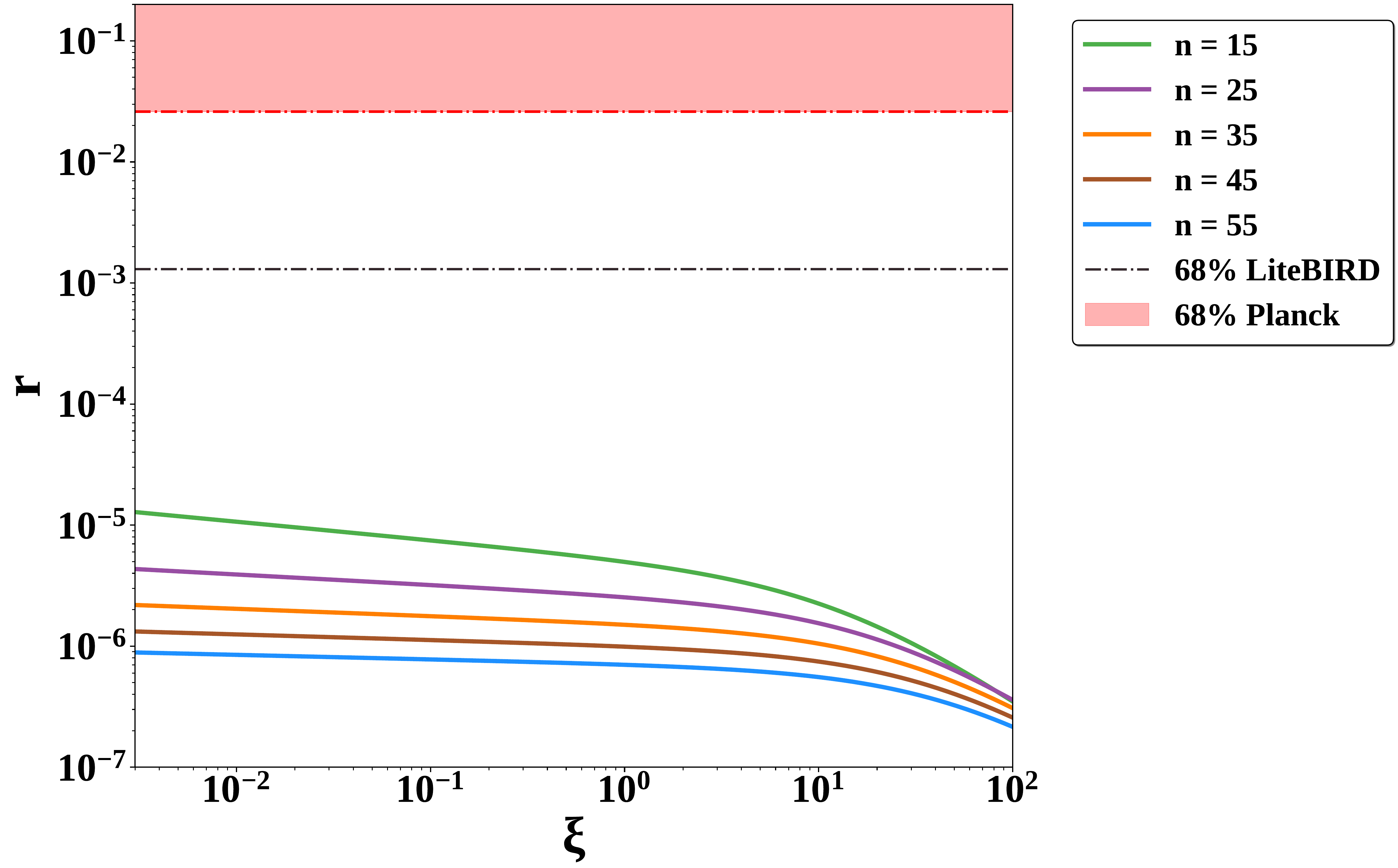}\hfill
\includegraphics[width=0.5\linewidth]{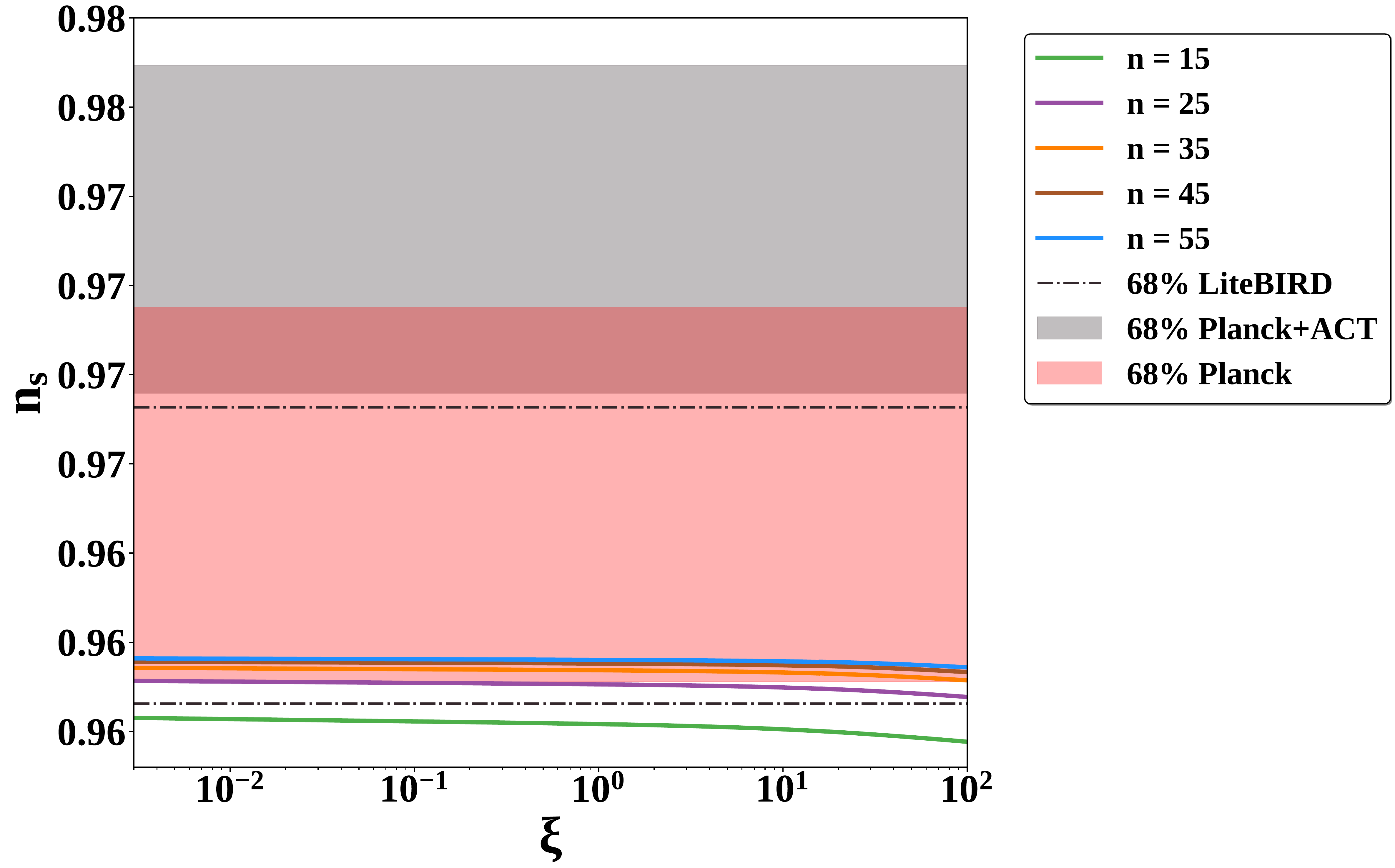}\\
\caption{\it Inflationary observables for the Higgs-like inflation defined by Eq. \eqref{eq:4_potential} in \textbf{metric} formalism.}
\label{fig:r_ns_4_m}
\end{figure}
%%%%%%%%%%%%%%%%%%%%%%%%%%%%%%%%%%%%%%%%%%%%%%%%%%%%%%%%%%%%%%%%%%%%%%%%%%%%%%
\begin{figure}[t!]
\centering
\includegraphics[width=0.5\linewidth]{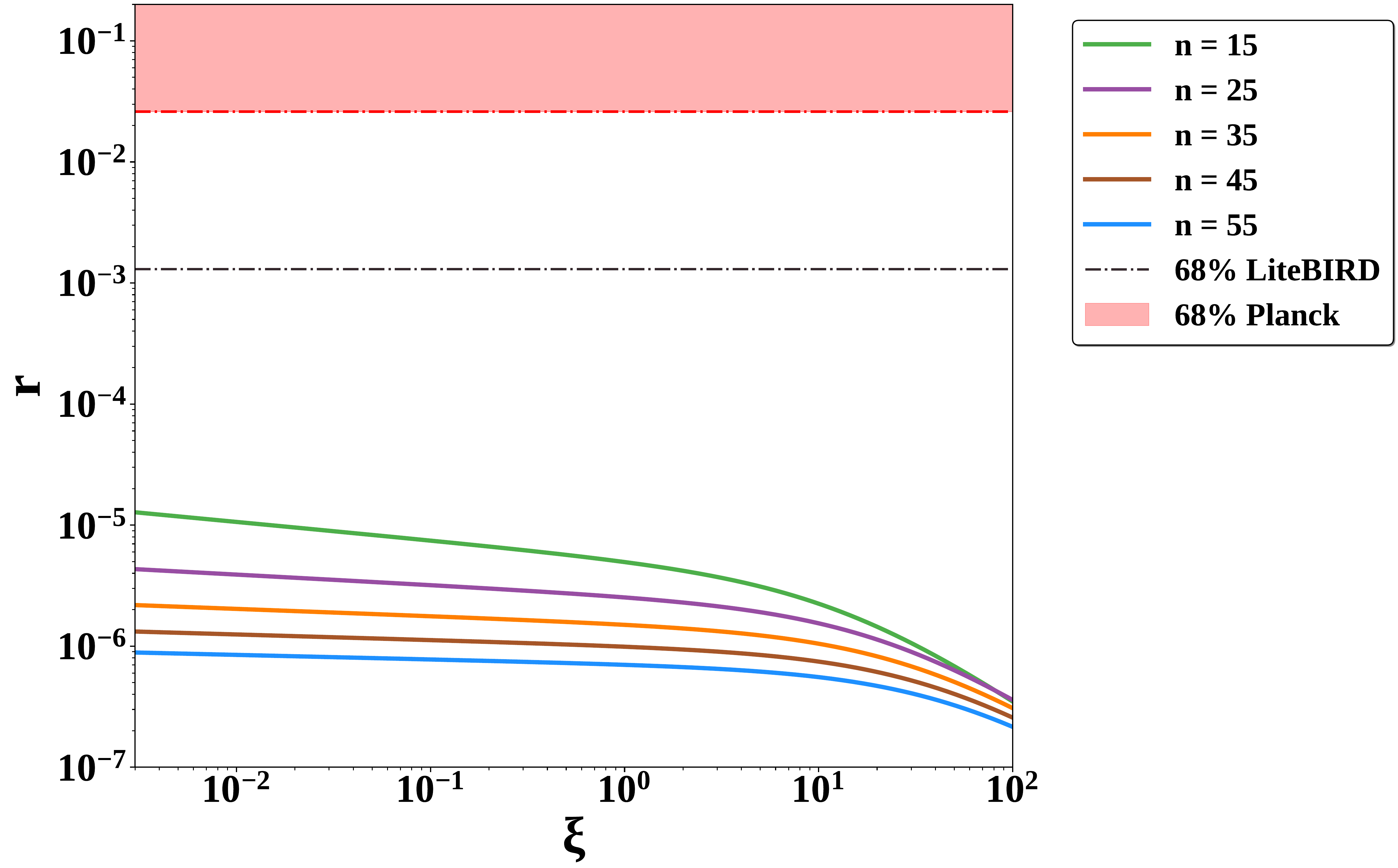}\hfill
\includegraphics[width=0.5\linewidth]{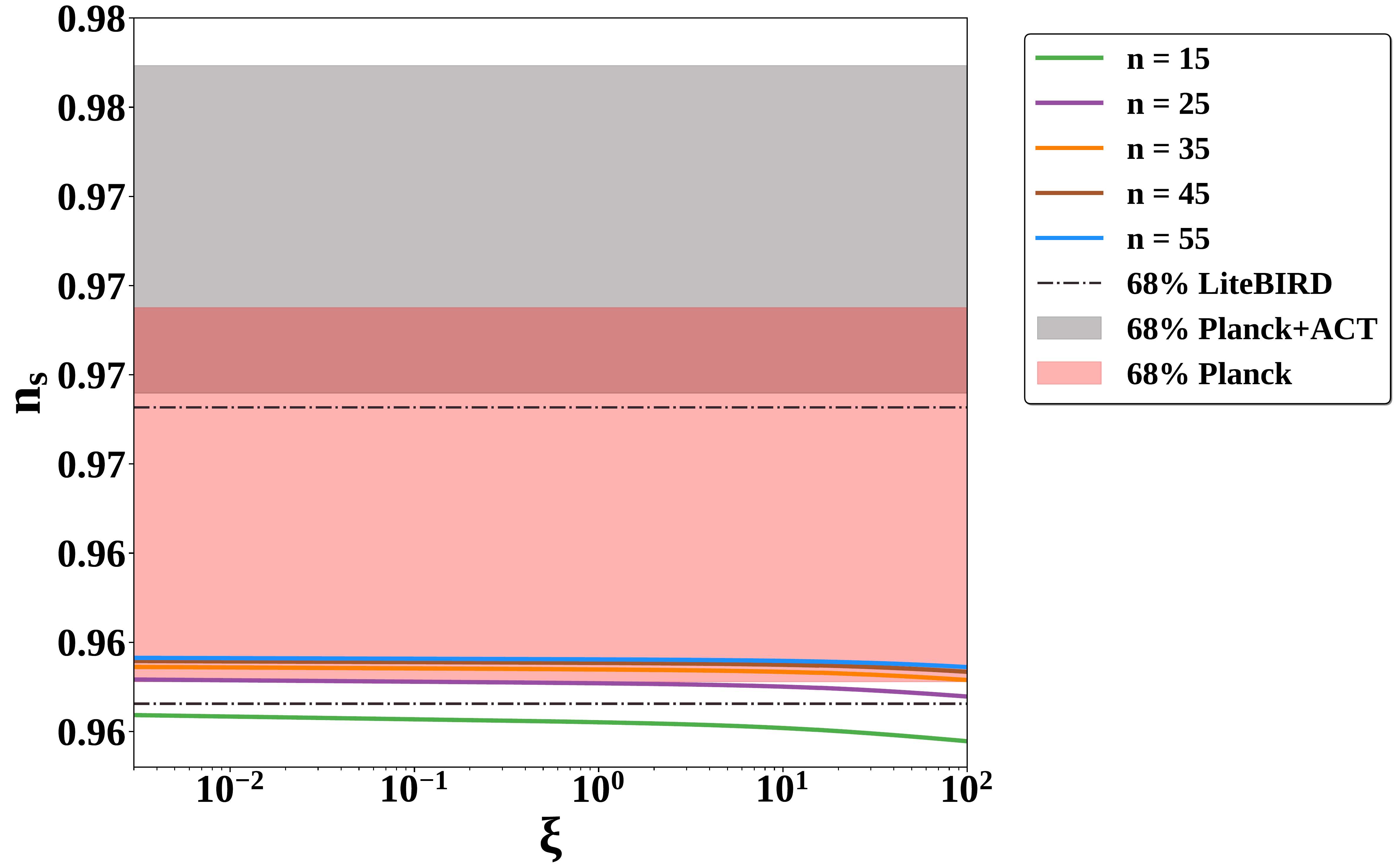}\\
\caption{\it Same as Fig.~\ref{fig:r_ns_4_m} but in the Palatini formalism.}
\label{fig:r_ns_4_P}
\end{figure}
%%%%%%%%%%%%%%%%%%%%%%%%%%%%%%%%%%%%%%%%%%%%%%%%%%%%%%%%%%%%%%%%%%%%%%%%%%%%%%
Representative numerical results for \( n = 5,\,35,\,55 \) and selected values of
\( \xi \), calibrated to yield \( N_* \simeq 53\text{--}57 \), are shown in
Tab.~\ref{tab:infl-para_4}, and the corresponding
trajectories as functions of \( \xi \) are displayed in
Figs.~\ref{fig:r_ns_4_m} (metric) and~\ref{fig:r_ns_4_P} (Palatini). In all cases, the field excursion remains safely sub-Planckian,
\( \Delta\phi \lesssim 3\,M_{\rm Pl} \) (for example,
\( \phi_i \simeq 1.89 \) for \( n = 55 \) and \( \xi = 0.1 \) in the metric formulation),
supporting the UV consistency of the effective field theory description.  Moreover, the predictions of the metric and Palatini formalisms
are closely aligned, reflecting the fact that inflation proceeds in a
small-field regime where the \( \kappa \)-dependent corrections to the canonical normalization play only a minor role.

The scalar spectral tilt lies in the range 
\( n_s \simeq 0.950\text{--}0.962 \), 
increasing mildly with larger values of \( n \) 
(see Fig.~\ref{fig:r_ns_4_m}), and is consistent with current \emph{Planck}
constraints~\cite{Planck:2018jri} as well as the projected sensitivities of
future observations such as LiteBIRD~\cite{Hazumi_2020}. Furthermore, tensor perturbations are strongly suppressed, with values in the range
\( r \sim 10^{-7}\text{--}10^{-4} \), in agreement with the analytical estimate in
Eq.~\eqref{eq:r_4}. However, the running of the scalar spectral index remains mildly
negative, \( |\alpha_s| \simeq (7\text{--}9)\times 10^{-4} \).

The inflaton undergoes upward field evolution (\( \phi_e > \phi_i \)), 
climbing over a mild Einstein-frame hump. This feature naturally supports 
multi-phase inflationary dynamics, which may leave distinctive imprints 
on late-time cosmology. Compared to the Brans-Dicke-like scenario, the model with a constant 
Jordan-frame potential generates a broader inflationary plateau, further 
suppressing the tensor-to-scalar ratio \( r \) and slightly reducing the 
total number of \(e\)-folds \( N_* \). In sharp contrast to the linear models \( V_J^{(1,2)} \), which typically involve 
trans-Planckian field excursions and predict \( r \gtrsim 5 \times 10^{-3} \), 
the present construction yields bounded field evolution with 
\( r \lesssim 10^{-4} \). This combination renders it the most empirically 
robust and theoretically well-controlled framework among those considered.

Having established the inflationary dynamics and observational predictions of the
induced multi-phase inflation scenarios, we now shift our focus to the post-inflationary epoch. 
Reheating plays a crucial role in connecting inflation to the hot Big Bang paradigm. In the following sections, we examine the dynamics of preheating, perturbative inflaton decays with radiative stability, dark-matter production from inflaton decay, 
neutrino-mass generation via the Type-I seesaw mechanism, and non-thermal leptogenesis. 
We illustrate how these processes are embedded within our gravitational framework 
and how they collectively address outstanding cosmological questions.

%%%%%%%%%%%%%%%%%%%%%%%%%%%%%%%%%%%%%%%%%%%%%%%%%%%%%%%%%%%%%%%%%%%%%%%%%%%%%%
%%%%%%%%%%%%%%%%%%%%%%%%%%%%%%%%%%%%%%%%%%%%%%%%%%%%%%%%%%%%%%%%%%%%%%%%%%%%%%
%%%%%%%%%%%%%%%%%%%%%%%%%%%%%%%%%%%%%%%%%%%%%%%%%%%%%%%%%%%%%%%%%%%%%%%%%%%%%%
\section{Reheating Dynamics} \label{Sec:reheating}
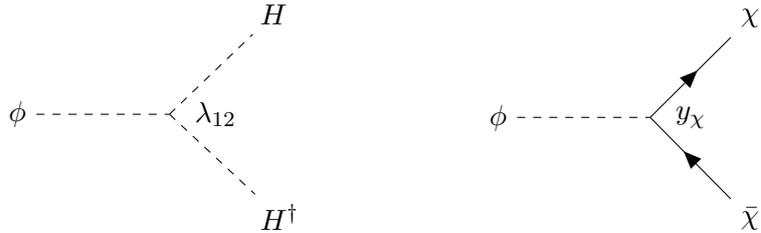
\begin{figure}[t!]
\centering
\begin{tikzpicture}
  \begin{feynman}
    \vertex (a) {\(\phi\)};
    \vertex [right=2cm of a] (b);
    \vertex [above right=1.5cm of b] (h1) {\(H\)};
    \vertex [below right=1.5cm of b] (h2) {\(H^\dagger\)};
    
    \diagram* {
      (a) -- [scalar] (b),
      (b) -- [scalar] (h1),
      (b) -- [scalar] (h2),
    };
    \node[right=0.2cm of b] {\(\lambda_{12}\)};
  \end{feynman}
\end{tikzpicture}
\hspace{2cm}
\begin{tikzpicture}
  \begin{feynman}
    \vertex (a) {\(\phi\)};
    \vertex [right=2cm of a] (b);
    \vertex [above right=1.5cm of b] (chi1) {\(\chi\)};
    \vertex [below right=1.5cm of b] (chi2) {\(\bar{\chi}\)};
    
    \diagram* {
      (a) -- [scalar] (b),
      (b) -- [fermion] (chi1),
      (chi2) -- [fermion] (b),
    };
    \node[right=0.2cm of b] {\(y_\chi\)};
  \end{feynman}
\end{tikzpicture}
 
\caption{\it Tree-level Feynman diagrams illustrating the perturbative decay of the inflaton, $\phi$. \textbf{Left:} Decay into SM Higgs doublets, governed by the trilinear coupling $\lambda_{12}$. \textbf{Right:} Decay into fermionic DM, $\chi$, governed by the Yukawa coupling $y_\chi$.}
\label{fig:feynman_decay}
\end{figure}

Following the end of inflation, the Universe enters the reheating epoch, a critical phase
during which the energy stored in the coherent oscillations of the inflaton field is
transferred to relativistic particles and subsequently thermalizes, giving rise to the
hot, radiation-dominated big bang Universe. In our framework, this process is primarily
mediated by interactions between the inflaton field, \( \phi \) and the SM Higgs boson, \( H \) as shown in Fig.~\ref{fig:feynman_decay}.

In this section, we present plots illustrating the reheating dynamics for the linear
inflation model defined in Eq.~\eqref{eq:U1_Palatini}. The corresponding figures for the
other scenarios considered in this work, including the second linear potential in
Eq.~\eqref{eq:U2_Palatini}, the Brans-Dicke-like model in
Eq.~\eqref{eq:3_potential}, and the Higgs-like model in Eq.~\eqref{eq:4_potential}, are
collected in App.~\ref{app:other_reheating}. An analogous convention is adopted for the figures related to dark-matter production 
from inflaton decay, which are presented in App.~\ref{app:other_DM}.

Before turning to a detailed discussion of the perturbative decay channels, we first
examine the possibility of non-perturbative particle production, or preheating, which
may occur immediately after the end of inflation~\cite{Kofman:1997yn}.
%%%%%%%%%%%%%%%%%%%%%%%%%%%%%%%%%%%%%%%%%%%%%%%%%%%%%%%%%%%%%%%%%%%%%%%%%%%%%%

%%%%%%%%%%%%%%%%%%%%%%%%%%%%%%%%%%%%%%%%%%%%%%%%%%%%%%%%%%%%%%%%%%%%%%%%%%%%%%
%%%%%%%%%%%%%%%%%%%%%%%%%%%%%%%%%%%%%%%%%%%%%%%%%%%%%%%%%%%%%%%%%%%%%%%%%%%%%%
%%%%%%%%%%%%%%%%%%%%%%%%%%%%%%%%%%%%%%%%%%%%%%%%%%%%%%%%%%%%%%%%%%%%%%%%%%%%%%
\subsection{Analysis of Preheating Mechanisms} \label{preheating}
Preheating is a non-perturbative process driven by parametric resonance, during which bosonic fields coupled to the coherently oscillating inflaton condensate can undergo exponential particle production~\cite{Kofman:1997yn,Lozanov:2019jxc}. The efficiency of this mechanism is highly sensitive to the structure and strength of the inflaton couplings appearing in the scalar potential.

In our setup, the interaction Lagrangian between the inflaton field \( \phi \) and the SM Higgs doublet \( H \) takes the form~\cite{Ghoshal:2024ycp}
%%%%%%%%%%%%%%%%%%%%%%%%%%%%%%%%%%%%%%%%%%%%%%%%%%%%%%%%%%%%%%%%%%%%%%%%%%%%%%
\begin{equation}
\mathcal{L}_{H\phi}
= - V(H,\phi)
= - \lambda_{12}\, \phi \, H^\dagger H
  - \frac{1}{2}\lambda_{22}\, \phi^2 \, H^\dagger H,
\end{equation}
%%%%%%%%%%%%%%%%%%%%%%%%%%%%%%%%%%%%%%%%%%%%%%%%%%%%%%%%%%%%%%%%%%%%%%%%%%%%%%
and contains both trilinear and quartic interaction terms.

The dynamics of preheating are governed by the post-inflationary oscillations of the inflaton field \( \phi \) about the minimum of its Einstein-frame potential \( V_E(\phi) \). The detailed form of \( V_E(\phi) \), which varies substantially among the four models considered here (Eqs.~\eqref{eq:U1_Palatini}, \eqref{eq:U2_Palatini}, \eqref{eq:3_potential}, and~\eqref{eq:4_potential}), as well as between the metric and Palatini formalisms, directly determines the oscillation frequency and amplitude of the inflaton condensate.

In particular, the large-field ``linear inflation'' scenarios defined by Eqs.~\eqref{eq:U1_Palatini} and~\eqref{eq:U2_Palatini} in the Palatini formulation exhibit extremely large field excursions (cf. Tabs.~\ref{tab:infl-para_1} and \ref{tab:infl-para_2}), resulting in high-frequency inflaton oscillations. This feature can significantly modify the structure of the resonance bands relative to the small-field models, namely the Brans-Dicke-like induced inflation model in Eq.~\eqref{eq:3_potential} and the Higgs-like scenario in Eq.~\eqref{eq:4_potential}.

We first consider preheating mediated by the trilinear interaction
%%%%%%%%%%%%%%%%%%%%%%%%%%%%%%%%%%%%%%%%%%%%%%%%%%%%%%%%%%%%%%%%%%%%%%%%%%%%%%
\begin{equation}
\lambda_{12}\,\phi\, H^\dagger H .
\end{equation}
%%%%%%%%%%%%%%%%%%%%%%%%%%%%%%%%%%%%%%%%%%%%%%%%%%%%%%%%%%%%%%%%%%%%%%%%%%%%%%
This coupling induces a time-dependent contribution to the Higgs effective mass~\cite{Kofman:1997yn}
%%%%%%%%%%%%%%%%%%%%%%%%%%%%%%%%%%%%%%%%%%%%%%%%%%%%%%%%%%%%%%%%%%%%%%%%%%%%%%
\begin{equation}
m_H^2(t) = m_{H,0}^2 + \lambda_{12}\,\phi(t) .
\end{equation}
%%%%%%%%%%%%%%%%%%%%%%%%%%%%%%%%%%%%%%%%%%%%%%%%%%%%%%%%%%%%%%%%%%%%%%%%%%%%%%
As the inflaton oscillates around the minimum of its potential, $\phi(t)$ periodically changes sign. During the intervals in which $\lambda_{12}\,\phi(t) < 0$ and dominates over the bare mass term, the Higgs mass-squared can become negative, triggering a tachyonic instability and potentially leading to explosive, non-perturbative Higgs production via tachyonic resonance~\cite{Dufaux:2006ee}.

However, the efficiency of this mechanism is strongly suppressed by the Higgs self-interaction term
%%%%%%%%%%%%%%%%%%%%%%%%%%%%%%%%%%%%%%%%%%%%%%%%%%%%%%%%%%%%%%%%%%%%%%%%%%%%%%
\begin{table}[t!]
\begin{center}
\begin{tabular}{|c|c||c|c|c|c|} 
 \hline
$n$ & $\xi$ & \multicolumn{2}{c|}{Stability for $y_
\chi$} & \multicolumn{2}{c|}{Stability for $\lambda_{12}$} \\ [0.5ex] 
\cline{3-6}
& & about $\mu=\phi_*$ & about $\mu=\phi_{\rm end}$ & about $\mu=\phi_*$ & about $\mu=\phi_{\rm end}$ \\ [0.5ex] 
\hline\hline
0.5 & 0.1 & $y_\chi<4.91\times 10^{-4}$ & $y_\chi<8.45\times 10^{-3}$ & $\lambda_{12}<4.90\times 10^{-6}$ & $\lambda_{12}<6.61\times 10^{-5}$ \\
\hline
1 & 0.1 & $y_\chi<3.63\times 10^{-4}$ & $y_\chi<2.50\times 10^{-3}$ & $\lambda_{12}<4.19\times 10^{-6}$ & $\lambda_{12}<1.14\times 10^{-5}$ \\ 
\hline
\end{tabular}
\caption{\it Allowed upper limits of $y_\chi$ and $\lambda_{12}$ for benchmark $(n, \xi)$ pairs in the linear inflation model of Eq.~\eqref{eq:U1_Palatini}, within \textbf{metric} gravity, as reported in Tab.~\ref{tab:infl-para_1}.}
\label{Table:stability0m}
\end{center}
\end{table}
%%%%%%%%%%%%%%%%%%%%%%%%%%%%%%%%%%%%%%%%%%%%%%%%%%%%%%%%%%%%%%%%%%%%%%%%%%%%%%
\begin{equation}
\lambda_H |H^\dagger H|^2 .
\end{equation}
%%%%%%%%%%%%%%%%%%%%%%%%%%%%%%%%%%%%%%%%%%%%%%%%%%%%%%%%%%%%%%%%%%%%%%%%%%%%%%
As the Higgs fluctuations grow, they generate a positive contribution to the effective mass through backreaction
%%%%%%%%%%%%%%%%%%%%%%%%%%%%%%%%%%%%%%%%%%%%%%%%%%%%%%%%%%%%%%%%%%%%%%%%%%%%%%
\begin{equation}
\delta m_H^2 \sim \lambda_H \langle |H|^2 \rangle ,
\end{equation}
%%%%%%%%%%%%%%%%%%%%%%%%%%%%%%%%%%%%%%%%%%%%%%%%%%%%%%%%%%%%%%%%%%%%%%%%%%%%%%
which counteracts the negative tachyonic term. Once this induced mass becomes comparable to or larger than $|\lambda_{12}\phi|$, the instability band closes and the tachyonic amplification shuts off. Consequently, the backreaction of the produced Higgs quanta efficiently terminates the resonance and prevents sustained energy transfer from the inflaton condensate. As a result, although trilinear couplings can accelerate the onset of
thermalization~\cite{Dufaux:2006ee}, the total fraction of energy transferred during this stage typically remains subdominant compared to the energy stored in the inflaton condensate.
%%%%%%%%%%%%%%%%%%%%%%%%%%%%%%%%%%%%%%%%%%%%%%%%%%%%%%%%%%%%%%%%%%%%%%%%%%%%%%

\begin{table}[t!]
\begin{center}
\begin{tabular}{|c|c||c|c|c|c|} 
 \hline
$n$ & $\xi$ & \multicolumn{2}{c|}{Stability for $y_
\chi$} & \multicolumn{2}{c|}{Stability for $\lambda_{12}$} \\ [0.5ex] 
\cline{3-6}
& & about $\mu=\phi_*$ & about $\mu=\phi_{\rm end}$ & about $\mu=\phi_*$ & about $\mu=\phi_{\rm end}$ \\ [0.5ex] 
\hline\hline
0.5 & 0.1 & $y_\chi<4.58\times 10^{-4}$ & $y_\chi<8.43\times 10^{-3}$ & $\lambda_{12}<4.25\times 10^{-6}$ & $\lambda_{12}<6.60\times 10^{-5}$ \\ 
\hline
1 & 0.1 & $y_\chi<3.79\times 10^{-4}$ & $y_\chi<2.49\times 10^{-3}$ & $\lambda_{12}<4.58\times 10^{-6}$ & $\lambda_{12}<1.14\times 10^{-5}$ \\ 
\hline
\end{tabular}
\caption{\it Same as Tab.~\ref{Table:stability0m} but for the \textbf{Palatini} framework.}
\label{Table:stability0}
\end{center}
\end{table}
%%%%%%%%%%%%%%%%%%%%%%%%%%%%%%%%%%%%%%%%%%%%%%%%%%%%%%%%%%%%%%%%%%%%%%%%%%%%%%
The efficiency of this channel can, in principle, exhibit a mild dependence on
the underlying gravitational formulation. The canonical normalization of the
inflaton field (Eq.~\eqref{Eq:conversion-inflaton-EToJ}) differs between the
metric and Palatini cases, which induces a corresponding rescaling of the
effective coupling $\lambda_{12}$ in the Einstein frame. Nevertheless, this
quantitative difference is insufficient to overcome the dominant suppression
arising from the Higgs backreaction.

Alternatively, preheating may proceed via the quartic interaction
%%%%%%%%%%%%%%%%%%%%%%%%%%%%%%%%%%%%%%%%%%%%%%%%%%%%%%%%%%%%%%%%%%%%%%%%%%%%%%
\begin{equation}
\lambda_{22}\,\phi^2\,H^\dagger H \, .
\end{equation}
%%%%%%%%%%%%%%%%%%%%%%%%%%%%%%%%%%%%%%%%%%%%%%%%%%%%%%%%%%%%%%%%%%%%%%%%%%%%%%
In this scenario, efficient parametric resonance typically requires a
sufficiently large coupling
%%%%%%%%%%%%%%%%%%%%%%%%%%%%%%%%%%%%%%%%%%%%%%%%%%%%%%%%%%%%%%%%%%%%%%%%%%%%%%
\begin{equation}
\lambda_{22} \gtrsim \mathcal{O}(10^{-8}) ,
\end{equation}
%%%%%%%%%%%%%%%%%%%%%%%%%%%%%%%%%%%%%%%%%%%%%%%%%%%%%%%%%%%%%%%%%%%%%%%%%%%%%%
as indicated in Ref.~\cite{Kofman:1997yn}. Such values, however, are incompatible
with the radiative stability of the inflaton potential. In particular, avoiding
sizeable loop corrections that would spoil the flatness of the inflationary
potential, a central requirement of our multi-phase inflationary
framework, imposes a much stronger constraint~\cite{Drees:2021wgd}
%%%%%%%%%%%%%%%%%%%%%%%%%%%%%%%%%%%%%%%%%%%%%%%%%%%%%%%%%%%%%%%%%%%%%%%%%%%%%%
\begin{equation}
\lambda_{22} \lesssim \mathcal{O}(10^{-10}) \,.
\end{equation}
%%%%%%%%%%%%%%%%%%%%%%%%%%%%%%%%%%%%%%%%%%%%%%%%%%%%%%%%%%%%%%%%%%%%%%%%%%%%%%
 Consequently, for the parameter space consistent with inflationary dynamics, preheating through the quartic inflaton-Higgs interaction remains inefficient in both the metric and Palatini formulations
considered here.

Based on the arguments presented above, we conclude that non-perturbative preheating effects remain subdominant in the post-inflationary dynamics of our
models. The Higgs backreaction efficiently suppresses the trilinear channel, whereas the requirement of radiative stability of the inflaton potential severely constrains the quartic channel.

This conclusion holds for all inflationary potentials considered in this work,
Eqs.~\eqref{eq:U1_Palatini},~\eqref{eq:U2_Palatini},
\eqref{eq:3_potential}, and~\eqref{eq:4_potential}, and for both the metric and
Palatini formalisms. It follows from the general features of the inflaton-Higgs
interactions and from the stability requirements of the inflationary potential,
rather than from model-specific details.
Accordingly, the dominant reheating mechanism in our framework is the
perturbative decay of the inflaton, to which we now turn.
%%%%%%%%%%%%%%%%%%%%%%%%%%%%%%%%%%%%%%%%%%%%%%%%%%%%%%%%%%%%%%%%%%%%%%%%%%%%%%

 \begin{figure}[t!]
     \centering
     \includegraphics[width=0.7\linewidth]{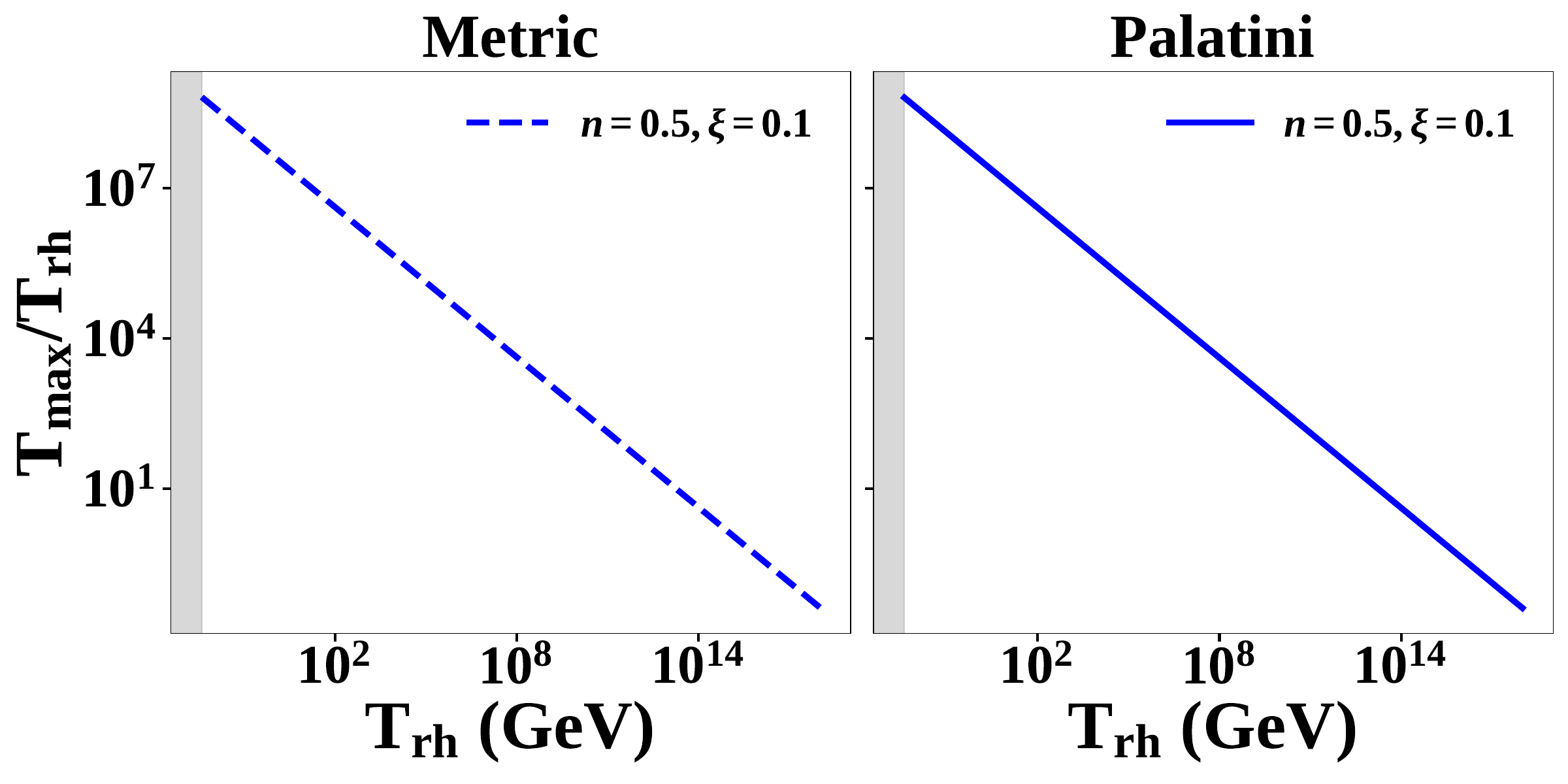}
     \caption{\it Variation of $T_{\rm max}/T_{\rm rh}$ as a function of $T_{\rm rh}$ for the benchmark values in Tables~\ref{Table:stability0m} and~\ref{Table:stability0}. The gray-shaded vertical band on the left of the plot indicates the lower bound on the reheat temperature, $T_{\rm rh} \ge 4~{\rm MeV}$.}
     \label{fig:tmax_1}
 \end{figure}
 %%%%%%%%%%%%%%%%%%%%%%%%%%%%%%%%%%%%%%%%%%%%%%%%%%%%%%%%%%%%%%%%%%%%%%%%%%%%%%
 %%%%%%%%%%%%%%%%%%%%%%%%%%%%%%%%%%%%%%%%%%%%%%%%%%%%%%%%%%%%%%%%%%%%%%%%%%%%%%
 %%%%%%%%%%%%%%%%%%%%%%%%%%%%%%%%%%%%%%%%%%%%%%%%%%%%%%%%%%%%%%%%%%%%%%%%%%%%%%
\subsection{Perturbative Decay of the Inflaton}
\label{Subsec:pert_decay}
The inflaton decays perturbatively into SM Higgs bosons and DM 
particles. For $m_\phi > 2 m_H$ and $m_\phi > 2 m_\chi$, the tree-level decay
rates are given by~\cite{Drees:2024hok}
%%%%%%%%%%%%%%%%%%%%%%%%%%%%%%%%%%%%%%%%%%%%%%%%%%%%%%%%%%%%%%%%%%%%%%%%%%%%%%
\begin{align}
\Gamma_{\phi \to H^{\dagger} H}
&= \frac{\lambda_{12}^2}{8\pi\, m_\phi}
\sqrt{1 - \frac{4 m_H^2}{m_\phi^2}} \,, \label{eq:decay_H} \\
\Gamma_{\phi \to \bar\chi \chi}
&= \frac{y_\chi^2\, m_\phi}{8\pi}
\left(1 - \frac{4 m_\chi^2}{m_\phi^2}\right)^{3/2} .
\label{eq:decay_chi}
\end{align}
%%%%%%%%%%%%%%%%%%%%%%%%%%%%%%%%%%%%%%%%%%%%%%%%%%%%%%%%%%%%%%%%%%%%%%%%%%%%%%
To incorporate a fermionic DM candidate $\chi$, we extend the
interaction Lagrangian by a Yukawa coupling
%%%%%%%%%%%%%%%%%%%%%%%%%%%%%%%%%%%%%%%%%%%%%%%%%%%%%%%%%%%%%%%%%%%%%%%%%%%%%%
\begin{equation}
\mathcal{L}_{\phi\chi} = - y_\chi \, \phi \, \bar{\chi} \chi.
\end{equation}
%%%%%%%%%%%%%%%%%%%%%%%%%%%%%%%%%%%%%%%%%%%%%%%%%%%%%%%%%%%%%%%%%%%%%%%%%%%%%%
Here, $y_\chi$ denotes the coupling between the inflaton, $\phi$, and the Dirac fermion, $\chi$, assumed to be a stable beyond-the-Standard-Model (BSM) particle constituting DM. Together with the inflaton–Higgs interaction, this coupling allows perturbative inflaton decays into both SM Higgs bosons and DM.
%%%%%%%%%%%%%%%%%%%%%%%%%%%%%%%%%%%%%%%%%%%%%%%%%%%%%%%%%%%%%%%%%%%%%%%%%%%%%%

\begin{table}[t!]
\begin{center}
\begin{tabular}{|c|c||c|c|c|c|} 
 \hline
$n$ & $\xi$ & \multicolumn{2}{c|}{Stability for $y_
\chi$} & \multicolumn{2}{c|}{Stability for $\lambda_{12}$} \\ [0.5ex] 
\cline{3-6}
& & about $\mu=\phi_*$ & about $\mu=\phi_{\rm end}$ & about $\mu=\phi_*$ & about $\mu=\phi_{\rm end}$ \\ [0.5ex] 
\hline\hline
0.5 & 0.1 & $y_\chi<3.08\times 10^{-4}$ & $y_\chi<2.54\times 10^{-3}$ & $\lambda_{12}<3.27\times 10^{-6}$ & $\lambda_{12}<1.72\times 10^{-5}$ \\ 
\hline
%\hline
1 & 0.1 & $y_\chi<2.19\times 10^{-4}$ & $y_\chi<1.62\times 10^{-3}$ & $\lambda_{12}<2.48\times 10^{-6}$ & $\lambda_{12}<1.06\times 10^{-5}$ \\ 
\hline
\end{tabular}
\caption{\it Allowed upper limits of $y_\chi$ and $\lambda_{12}$ for benchmark $(n, \xi)$ pairs in the linear inflation model of Eq.~\eqref{eq:U2_Palatini}, within \textbf{metric} gravity, as reported in Tab.~\ref{tab:infl-para_2}.}
\label{Table:stability1m}
\end{center}
\end{table}
%%%%%%%%%%%%%%%%%%%%%%%%%%%%%%%%%%%%%%%%%%%%%%%%%%%%%%%%%%%%%%%%%%%%%%%%%%%%%%
The inflaton mass $m_\phi$ is determined by the curvature of the Einstein-frame
potential at its minimum. The total decay width is
%%%%%%%%%%%%%%%%%%%%%%%%%%%%%%%%%%%%%%%%%%%%%%%%%%%%%%%%%%%%%%%%%%%%%%%%%%%%%%
\begin{equation}
\Gamma \equiv \Gamma_{\phi \to H^{\dagger} H}
+ \Gamma_{\phi \to \bar\chi \chi} .
\end{equation}
%%%%%%%%%%%%%%%%%%%%%%%%%%%%%%%%%%%%%%%%%%%%%%%%%%%%%%%%%%%%%%%%%%%%%%%%%%%%%%
In the parameter region where
\(
\Gamma_{\phi \to H^{\dagger} H}
\gg
\Gamma_{\phi \to \bar\chi \chi}
\),
the total width is dominated by decays into SM states~\cite{Ghoshal:2024ycp}
%%%%%%%%%%%%%%%%%%%%%%%%%%%%%%%%%%%%%%%%%%%%%%%%%%%%%%%%%%%%%%%%%%%%%%%%%%%%%%
\begin{equation}
\Gamma \simeq \Gamma_{\phi \to H^{\dagger} H} .
\end{equation}
The branching ratio into DM is then~\cite{Ghoshal:2024ycp}
\begin{equation}
\mathrm{Br}(\phi \to \bar\chi \chi)
=
\frac{\Gamma_{\phi \to \bar{\chi}\chi}}
{\Gamma_{\phi \to \bar{\chi}\chi}
+ \Gamma_{\phi \to H^{\dagger} H}} ,
\end{equation}
%%%%%%%%%%%%%%%%%%%%%%%%%%%%%%%%%%%%%%%%%%%%%%%%%%%%%%%%%%%%%%%%%%%%%%%%%%%%%%
which, in the limit $m_H, m_\chi \ll m_\phi$, simplifies to
%%%%%%%%%%%%%%%%%%%%%%%%%%%%%%%%%%%%%%%%%%%%%%%%%%%%%%%%%%%%%%%%%%%%%%%%%%%%%%
\begin{equation}\label{eq:Br}
\mathrm{Br}
\simeq
\frac{y_\chi^2 m_\phi^2}{\lambda_{12}^2 + y_\chi^2 m_\phi^2}
\simeq
\frac{y_\chi^2 m_\phi^2}{\lambda_{12}^2}
\quad
\text{for } y_\chi m_\phi \ll \lambda_{12}.
\end{equation}\\
%%%%%%%%%%%%%%%%%%%%%%%%%%%%%%%%%%%%%%%%%%%%%%%%%%%%%%%%%%%%%%%%%%%%%%%%%%%%%%
%%%%%%%%%%%%%%%%%%%%%%%%%%%%%%%%%%%%%%%%%%%%%%%%%%%%%%%%%%%%%%%%%%%%%%%%%%%%%%
%%%%%%%%%%%%%%%%%%%%%%%%%%%%%%%%%%%%%%%%%%%%%%%%%%%%%%%%%%%%%%%%%%%%%%%%%%%%%%
\textbf{Radiative Stability Conditions}

To preserve the flatness of the inflationary potential against quantum corrections, we impose radiative stability constraints using the one-loop 
Coleman-Weinberg (CW) correction~\cite{PhysRevD.7.1888}.~\footnote{For a comprehensive analysis of the CW effective potential applied specifically to linear inflation within both the metric and Palatini formulations, see Ref.~\cite{Racioppi:2017spw}. Additionally, the specific reheating constraints imposed by these one-loop corrections within the Palatini framework are extensively explored in Ref.~\cite{Bostan:2024cw}.}
The corresponding one-loop effective potential is given by~\cite{PhysRevD.7.1888}
%%%%%%%%%%%%%%%%%%%%%%%%%%%%%%%%%%%%%%%%%%%%%%%%%%%%%%%%%%%%%%%%%%%%%%%%%%%%%%
\begin{equation}
    V_{\rm 1\text{-}loop} 
    = \sum_j \frac{n_j}{64\pi^2} (-1)^{2s_j} \widetilde{m}_j^{4}
    \left[
        \ln\!\left(\frac{\widetilde{m}_j^{2}}{\mu^{2}}\right) - c_j
    \right].
    \label{eq:correction1}
\end{equation}
%%%%%%%%%%%%%%%%%%%%%%%%%%%%%%%%%%%%%%%%%%%%%%%%%%%%%%%%%%%%%%%%%%%%%%%%%%%%%%

\begin{table}[t!]
\begin{center}
\begin{tabular}{|c|c||c|c|c|c|} 
 \hline
$n$ & $\xi$ & \multicolumn{2}{c|}{Stability for $y_
\chi$} & \multicolumn{2}{c|}{Stability for $\lambda_{12}$} \\ [0.5ex] 
\cline{3-6}
& & about $\mu=\phi_*$ & about $\mu=\phi_{\rm end}$ & about $\mu=\phi_*$ & about $\mu=\phi_{\rm end}$ \\ [0.5ex] 
\hline\hline
0.5 & 0.1 & $y_\chi<4.01\times10^{-4}$ & $y_\chi<2.54\times 10^{-3}$ & $\lambda_{12}<5.58\times 10^{-6}$ & $\lambda_{12}<1.71\times 10^{-5}$ \\ 
\hline
1 & 0.1 & $y_\chi<2.35\times 10^{-4}$ & $y_\chi<1.61\times 10^{-3}$ & $\lambda_{12}<2.88\times 10^{-6}$ & $\lambda_{12}<1.06\times 10^{-5}$ \\ 
\hline
\end{tabular}
\caption{\it Same as Tab.~\ref{Table:stability1m} but for the \textbf{Palatini} framework.}
\label{Table:stability1}
\end{center}
\end{table}
%%%%%%%%%%%%%%%%%%%%%%%%%%%%%%%%%%%%%%%%%%%%%%%%%%%%%%%%%%%%%%%%%%%%%%%%%%%%%%
For the DM fermion $\chi$ and the SM Higgs $H$, the field-dependent masses are
%%%%%%%%%%%%%%%%%%%%%%%%%%%%%%%%%%%%%%%%%%%%%%%%%%%%%%%%%%%%%%%%%%%%%%%%%%%%%%
\begin{equation}
\widetilde{m}_{\chi}^2 (\phi) = \left( m_\chi + y_\chi \phi \right)^2, \qquad
\widetilde{m}_{H}^2 (\phi) = m_H^2 + \lambda_{12} \phi,
\end{equation}
%%%%%%%%%%%%%%%%%%%%%%%%%%%%%%%%%%%%%%%%%%%%%%%%%%%%%%%%%%%%%%%%%%%%%%%%%%%%%%
with constants $c_j = 3/2$, degrees of freedom $n_{H,\chi} = 4$, and spins $s_H = 0$, $s_\chi = 1/2$. We consider two renormalization scales: $\mu = \phi_*$ (horizon exit) and $\mu = \phi_{\rm end}$ (end of inflation).

The first and second derivatives of the CW potential are~\cite{Drees_2021}
%%%%%%%%%%%%%%%%%%%%%%%%%%%%%%%%%%%%%%%%%%%%%%%%%%%%%%%%%%%%%%%%%%%%%%%%%%%%%%
\begin{align}
	 & V_{\rm 1-loop}^\prime = \sum\limits_{j} \frac{n_j}{32 \pi^2}  (-1)^{2 s_j} 
	\widetilde{m}_j^2 \, \left(\widetilde{m}_j^{2}\right)^{\prime}
	\left[ \ln \left(\frac{\widetilde{m}_j^2}{\mu^2} \right)
	- 1 \right]\,,  \label{Eq:first derivative test1}\\
	& V_{\rm 1-loop}^{\prime\prime} =  \sum\limits_{j} \frac{n_j}{32 \pi^2} (-1)^{2 s_j} 
	\left\{ \left[ \left(\left(\widetilde{m}_j^{2}\right)^{\prime} \right)^2
	+ \widetilde{m}_j^2 \left(\widetilde{m}_j^{2}\right)^{\prime\prime} \right]
	\ln \left(\frac{\widetilde{m}_j^2}{\mu^2} \right)
	- \widetilde{m}_j^2 \left(\widetilde{m}_j^{2}\right)^{\prime\prime}  \right\}\,.   \label{Eq:second derivative test1}
\end{align}
%%%%%%%%%%%%%%%%%%%%%%%%%%%%%%%%%%%%%%%%%%%%%%%%%%%%%%%%%%%%%%%%%%%%%%%%%%%%%%
For $m_\chi\approx m_H \approx 0$, the explicit forms become
%%%%%%%%%%%%%%%%%%%%%%%%%%%%%%%%%%%%%%%%%%%%%%%%%%%%%%%%%%%%%%%%%%%%%%%%%%%%%%
\begin{align}
    V_{{\rm 1-loop},\chi}' &= -\frac{1}{4\pi^2}\, y_\chi^4 \,\mu^3 \left\{\ln(y_\chi^2)-1\right\}\, ,\qquad
    V_{{\rm 1-loop},\chi}'' = -\frac{y_\chi^4\,\mu^2}{4\pi^2}\left\{3\ln(y_\chi^2)-1\right\} \,,\\
     V_{{\rm 1-loop},H}'&=\frac{1}{8\pi^2}\lambda_{12}^2\,\mu\left\{\ln\left(\frac{\lambda_{12}}{\mu}\right)-1\right\}\,,\qquad V_{{\rm 1-loop},H}'' = \frac{1}{8\pi^2}\lambda_{12}^2 \ln\left(\frac{\lambda_{12}}{\mu}\right)\,.
\end{align}
%%%%%%%%%%%%%%%%%%%%%%%%%%%%%%%%%%%%%%%%%%%%%%%%%%%%%%%%%%%%%%%%%%%%%%%%%%%%%%
The upper bounds on $y_\chi$ and $\lambda_{12}$ are determined by requiring that quantum corrections do not dominate the tree-level potential

\begin{table}[t!]
\begin{center}
\begin{tabular}{|c|c||c|c|c|c|} 
 \hline
$n$ & $\xi$ & \multicolumn{2}{c|}{Stability for $y_
\chi$} & \multicolumn{2}{c|}{Stability for $\lambda_{12}$} \\ [0.5ex] 
\cline{3-6}
& & about $\mu=\phi_*$ & about $\mu=\phi_{\rm end}$ & about $\mu=\phi_*$ & about $\mu=\phi_{\rm end}$ \\ [0.5ex] 
\hline\hline
5 & 0.1 & $y_\chi<7.72\times 10^{-4}$ & $y_\chi<3.68\times 10^{-4}$ & $\lambda_{12}<9.77\times 10^{-7}$ & $\lambda_{12}<3.43\times 10^{-7}$ \\ 
\hline
5 & 4 & $y_\chi<5.53\times 10^{-4}$ & $y_\chi<5.75\times 10^{-4}$ & $\lambda_{12}<8.54\times 10^{-7}$ & $\lambda_{12}<8.40\times 10^{-7}$ \\ 
\hline
55 & 0.1 & $y_\chi<9.72\times 10^{-4}$ & $y_\chi<2.91\times10^{-3}$ & $\lambda_{12}<2.55\times 10^{-6}$ & $\lambda_{12}<2.43\times10^{-5}$ \\ 
\hline
55 & 54 & $y_\chi< 2.44\times 10^{-4}$ & $y_\chi<2.48\times10^{-4}$ & $\lambda_{12}<1.53\times 10^{-7}$ & $\lambda_{12}<1.70\times10^{-7}$ \\ 
\hline
\end{tabular}
\caption{\it Allowed upper limits of $y_\chi$ and $\lambda_{12}$ for benchmark $(n, \xi)$ pairs in the Brans-Dicke-like inflation model of Eq.~\eqref{eq:3_potential}, within \textbf{metric} gravity, as reported in Tab.~\ref{tab:infl-para_3}.}
\label{Table:stability2m}
\end{center}
\end{table}
%%%%%%%%%%%%%%%%%%%%%%%%%%%%%%%%%%%%%%%%%%%%%%%%%%%%%%%%%%%%%%%%%%%%%%%%%%%%%%
\begin{equation}\label{eq:constraints}
    \left|V'_{\rm 1-loop,j} (\phi = \mu) \right| <  V_E'(\phi = \mu)\, ,\qquad \left|V''_{\rm 1-loop,j} (\phi = \mu) \right| <  V_E''(\phi = \mu)\,.
\end{equation}
%%%%%%%%%%%%%%%%%%%%%%%%%%%%%%%%%%%%%%%%%%%%%%%%%%%%%%%%%%%%%%%%%%%%%%%%%%%%%%
The reheat temperature $T_{\rm rh}$, which marks the onset of radiation domination, 
is determined by the condition $H(T_{\rm rh}) = \frac{2}{3}\,\Gamma$, yielding~\cite{Ghoshal:2024ycp}
%%%%%%%%%%%%%%%%%%%%%%%%%%%%%%%%%%%%%%%%%%%%%%%%%%%%%%%%%%%%%%%%%%%%%%%%%%%%%%
\begin{equation}
\label{tre_metric}
T_{\rm rh} = \sqrt{\frac{2}{\pi}} 
\left(\frac{10}{g_*}\right)^{1/4} 
\sqrt{M_{\rm Pl}\,\Gamma}\,,
\end{equation}
%%%%%%%%%%%%%%%%%%%%%%%%%%%%%%%%%%%%%%%%%%%%%%%%%%%%%%%%%%%%%%%%%%%%%%%%%%%%%%
where $g_*$ denotes the effective number of relativistic degrees of freedom. 
For temperatures above the top-quark mass, one has $g_* = 106.75$.

Big Bang Nucleosynthesis (BBN) requires $T_{\rm rh} \gtrsim 4~\text{MeV}$~\cite{Sarkar:1995dd, Kawasaki:2000en, Hannestad:2004px, DEBERNARDIS2008192, deSalas:2015glj}, 
while the inflationary bound on the Hubble scale, $H_I \le 2.5 \times 10^{-5}\,M_{\rm Pl}$~\cite{Akrami:2018odb}, 
imposes an upper limit of $T_{\rm rh} \lesssim 7 \times 10^{15}~\text{GeV}$.

The maximum temperature attained during reheating, $T_{\rm max}$, 
can exceed $T_{\rm rh}$~\cite{Giudice:2000ex}. 
Using the inflationary Hubble scale
%%%%%%%%%%%%%%%%%%%%%%%%%%%%%%%%%%%%%%%%%%%%%%%%%%%%%%%%%%%%%%%%%%%%%%%%%%%%%%
\begin{equation}
\label{eq:HI}
H_I \simeq \sqrt{\frac{V_E(\phi_e)}{3\, M_{\rm Pl}^2}}\,,
\end{equation}
%%%%%%%%%%%%%%%%%%%%%%%%%%%%%%%%%%%%%%%%%%%%%%%%%%%%%%%%%%%%%%%%%%%%%%%%%%%%%%
the ratio $T_{\rm max}/T_{\rm rh}$ can be written as~\cite{Ghoshal_2024, Bernal_2021, Giudice_2001, Chung_1999}
%%%%%%%%%%%%%%%%%%%%%%%%%%%%%%%%%%%%%%%%%%%%%%%%%%%%%%%%%%%%%%%%%%%%%%%%%%%%%%
\begin{equation}
\frac{T_{\rm max}}{T_{\rm rh}} \simeq
\begin{cases}
\displaystyle 
\left(\frac{3}{8}\right)^{2/5} \left(\frac{H_I}{\mathcal{H}\left(T_{\rm rh}\right)}\right)^{1/4}  &\qquad \text{(for metric formalism)\,,} \\
 \frac{3^{1/4}}{2} \left( \frac{H_I}{\mathcal{H}(T_{\rm rh})} \right)^{1/4} &\qquad \text{(for Palatini formalism)\,,}
\end{cases}
\end{equation}
%%%%%%%%%%%%%%%%%%%%%%%%%%%%%%%%%%%%%%%%%%%%%%%%%%%%%%%%%%%%%%%%%%%%%%%%%%%%%%
where~\cite{Ghoshal:2024ycp}
%%%%%%%%%%%%%%%%%%%%%%%%%%%%%%%%%%%%%%%%%%%%%%%%%%%%%%%%%%%%%%%%%%%%%%%%%%%%%%
\begin{equation}\label{eq:H_T_rh}
    \mathcal{H}(T_{\rm rh}) \simeq \frac{\pi}{3M_{\rm Pl}}\sqrt{\frac{g_*}{10}}T_{\rm rh}^2\,.
\end{equation}
%%%%%%%%%%%%%%%%%%%%%%%%%%%%%%%%%%%%%%%%%%%%%%%%%%%%%%%%%%%%%%%%%%%%%%%%%%%%%%

\begin{table}[t!]
\begin{center}
\begin{tabular}{|c|c||c|c|c|c|} 
 \hline
$n$ & $\xi$ & \multicolumn{2}{c|}{Stability for $y_
\chi$} & \multicolumn{2}{c|}{Stability for $\lambda_{12}$} \\ [0.5ex] 
\cline{3-6}
& & about $\mu=\phi_*$ & about $\mu=\phi_{\rm end}$ & about $\mu=\phi_*$ & about $\mu=\phi_{\rm end}$ \\ [0.5ex] 
\hline\hline
5 & 0.1 & $y_\chi<1.88\times 10^{-3}$ & $y_\chi<1.51\times 10^{-3}$ & $\lambda_{12}<8.72\times 10^{-6}$ & $\lambda_{12}<6.97\times 10^{-6}$ \\ 
\hline
5 & 4 & $y_\chi<7.70\times 10^{-4}$ & $y_\chi<4.36\times 10^{-4}$ & $\lambda_{12}<9.74\times 10^{-7}$ & $\lambda_{12}<4.45\times 10^{-7}$ \\ 
\hline
55 & 0.1 & $y_\chi<9.72\times 10^{-4}$ & $y_\chi< 2.97\times10^{-3}$ & $\lambda_{12}<2.55\times 10^{-6}$ & $\lambda_{12}<2.51\times10^{-5}$ \\ 
\hline
55 & 54 & $y_\chi<2.44\times 10^{-4}$ & $y_\chi<2.36\times10^{-4}$ & $\lambda_{12}<1.52\times 10^{-7}$ & $\lambda_{12}<1.53\times10^{-7}$ \\ 
\hline
\end{tabular}
\caption{\it Same as Tab.~\ref{Table:stability2m} but for the \textbf{Palatini} framework.}
\label{Table:stability2}
\end{center}
\end{table}
%%%%%%%%%%%%%%%%%%%%%%%%%%%%%%%%%%%%%%%%%%%%%%%%%%%%%%%%%%%%%%%%%%%%%%%%%%%%%%
The radiative stability conditions, imposed through the 
CW one-loop corrections 
(Eqs.~\eqref{Eq:first derivative test1},~\eqref{Eq:second derivative test1}, and~\eqref{eq:constraints}), 
lead to stringent upper bounds on the Yukawa coupling $y_\chi$ 
to the DM fermion $\chi$ and on the trilinear coupling 
$\lambda_{12}$ to the SM Higgs $H$. 
These bounds ensure that quantum corrections do not spoil the 
flatness of the Einstein-frame potential $V_E(\phi)$, which is 
essential for sustaining the slow-roll conditions and the 
multi-phase inflationary trajectories discussed in 
Sec.~\ref{Sec:Slow_roll}. 
The constraints are evaluated at the renormalization scales 
$\mu=\phi_*$ (horizon exit) and $\mu=\phi_{\rm end}$ (end of inflation), 
reflecting their sensitivity to the inflaton field value during the 
inflationary epoch.

In the large-field linear inflation models 
(Eqs.~\eqref{eq:U1_Palatini} and \eqref{eq:U2_Palatini}), 
the resulting bounds are significantly stronger than in the 
small-field Brans-Dicke-like 
(Eq.~\eqref{eq:3_potential}) and Higgs-like 
(Eq.~\eqref{eq:4_potential}) scenarios. 
For example, in the metric formulation for $V_J^{(1)}(\phi)$ with 
$n=0.5$ and $\xi=0.1$ (Tab.~\ref{Table:stability0m}), 
one finds $y_\chi \lesssim 4.91 \times 10^{-4}$ at 
$\mu=\phi_*$, 
whereas in the Brans-Dicke-like case with 
$n=5$ and $\xi=0.1$ (Tab.~\ref{Table:stability2m}) 
the bound relaxes to 
$y_\chi \lesssim 7.72 \times 10^{-4}$. 

This hierarchy has a clear physical origin. 
In the linear models, the inflaton undergoes super-Planckian 
excursions ($\phi_i \sim 10$--$50\,M_{\rm Pl}$ in the metric case 
and up to $\sim 570\,M_{\rm Pl}$ in the Palatini case; 
cf.~Tabs.~\ref{tab:infl-para_1}--\ref{tab:infl-para_2}), 
which enhance loop corrections through the field-dependent masses 
$\widetilde{m}_j^2(\phi)$ that scale with powers of $\phi$. 
By contrast, in the small-field scenarios, the excursions are 
restricted to $\phi_i \sim 1$--$2\,M_{\rm Pl}$ 
(Tabs.~\ref{tab:infl-para_3}--\ref{tab:infl-para_4}), 
thereby suppressing the magnitude of the loop corrections and 
allowing comparatively larger couplings without destabilizing 
the flatness of the potential.

Comparing the two linear potentials, \(V_J^{(2)}(\phi)\) generally yields slightly looser bounds than \(V_J^{(1)}(\phi)\) due to the additional \((f(\phi)-1)^2\) factor, which steepens the Jordan-frame potential and modifies the Einstein-frame asymptotics, leading to marginally smaller field excursions for comparable parameters (e.g., \(\phi_i = 13.47 \, M_{\rm Pl}\) for \(V_J^{(2)}\) vs. \(7.91 \, M_{\rm Pl}\) for \(V_J^{(1)}\) at \(n=0.5\), \(\xi=0.1\), metric). This results in reduced logarithmic enhancements in the CW derivatives, as seen in the explicit forms for negligible masses. Among the small-field models, the Higgs-like variant imposes the most restrictive bounds (e.g., \(y_\chi \lesssim 2.04 \times 10^{-4}\) at \(\mu = \phi_*\) for \(n=5\), \(\xi=0.1\), metric; Tab.~\ref{Table:stability3m}), attributable to its plateau-like structure that demands even finer control over quantum corrections to preserve the attractor behavior akin to standard Higgs inflation.

The distinction between metric and Palatini formalisms is pronounced in large-field models, where Palatini's independent treatment of the affine connection softens the field-space metric \(\Pi(\phi)\) (Eq.~\eqref{eq:Pi-general} with \(\kappa=0\)), yielding larger canonical field excursions and thus tighter stability bounds. For example, in \(V_J^{(1)}(\phi)\) with \(n=0.5\), \(\xi=0.1\), the Palatini bound is \(y_\chi \lesssim 4.58 \times 10^{-4}\) at \(\mu = \phi_*\) (Tab.~\ref{Table:stability0}), compared to \(4.91 \times 10^{-4}\) in metric, reflecting the enhanced \(\phi\)-dependence in loop integrals. This effect diminishes in small-field regimes, where \(\Pi(\phi)\) differences are subdominant, leading to nearly identical bounds across formalisms (e.g., Tabs.~\ref{Table:stability2m}--\ref{Table:stability3}).
%%%%%%%%%%%%%%%%%%%%%%%%%%%%%%%%%%%%%%%%%%%%%%%%%%%%%%%%%%%%%%%%%%%%%%%%%%%%%%

\begin{table}[t!]
\begin{center}
\begin{tabular}{|c|c||c|c|c|c|} 
 \hline
$n$ & $\xi$ & \multicolumn{2}{c|}{Stability for $y_
\chi$} & \multicolumn{2}{c|}{Stability for $\lambda_{12}$} \\ [0.5ex] 
\cline{3-6}
& & about $\mu=\phi_*$ & about $\mu=\phi_{\rm end}$ & about $\mu=\phi_*$ & about $\mu=\phi_{\rm end}$ \\ [0.5ex] 
\hline\hline
5 & 0.1 & $y_\chi<2.04\times10^{-4}$ & $y_\chi<4.65\times10^{-4}$ & $\lambda_{12}<7.82\times10^{-8}$ & $\lambda_{12}< 1.16\times10^{-6}$ \\ 
\hline
5 & 4 & $y_\chi<6.83\times 10^{-5}$ & $y_\chi<3.89\times 10^{-4}$ & $\lambda_{12}<7.26\times 10^{-9}$ & $\lambda_{12}<4.00\times 10^{-7}$ \\ 
\hline
\end{tabular}
\caption{\it Allowed upper limits of $y_\chi$ and $\lambda_{12}$ for benchmark $(n, \xi)$ pairs in the Higgs-like inflation model of Eq.~\eqref{eq:4_potential}, within \textbf{metric} gravity, as reported in Tab.~\ref{tab:infl-para_4}.}
\label{Table:stability3m}
\end{center}
\end{table}
%%%%%%%%%%%%%%%%%%%%%%%%%%%%%%%%%%%%%%%%%%%%%%%%%%%%%%%%%%%%%%%%%%%%%%%%%%%%%%
Physically, these constraints link directly to inflationary observables: tighter bounds in large-field models correlate with lower \(r\) and higher \(n_s\) in Palatini (as discussed in Sec.~\ref{Sec:Slow_roll}), since suppressed couplings minimize perturbations that could alter the potential slope and thus the tensor modes. In small-field models, looser bounds align with the minuscule \(r \lesssim 10^{-5}\) and stable \(n_s \approx 0.95-0.96\), where the potential minimum's curvature is less sensitive to radiative shifts.

Furthermore, the reheat temperature \(T_{\rm rh}\) (Eq.~\eqref{tre_metric}), governed by the inflaton decay width Eq.~\eqref{eq:decay_H}, spans from the BBN lower limit of \(4 \, {\rm MeV}\) to an upper bound of \(\sim 7 \times 10^{15} \, {\rm GeV}\) set by the inflationary scale. The maximum temperature \(T_{\rm max}\) during reheating, which can significantly exceed \(T_{\rm rh}\), is captured by the ratio \(T_{\rm max}/T_{\rm rh}\) (Fig.~\ref{fig:tmax_1}), exhibiting a  declining trend with increasing \(T_{\rm rh}\). This behavior stems from the dependence on the inflationary Hubble scale Eq.~\eqref{eq:HI}, where higher \(T_{\rm rh}\) (faster decays) implies earlier radiation domination, reducing the duration over which the inflaton's energy density dominates and thus lowering the peak temperature relative to \(T_{\rm rh}\).

Across models, linear potentials yield higher ratios of $T_{\rm max}/T_{\rm rh}$, 
$\sim 10^1 - 10^9$ for $T_{\rm rh} \sim 10^2 - 10^{14}~{\rm GeV}$ 
(e.g., Fig.~\ref{fig:tmax_1} for $V_J^{(1)}$ and Fig.~\ref{fig:tmax_2} for $V_J^{(2)}$), 
compared to small-field models (cf. Figs.~\ref{fig:tmax_3} and~\ref{fig:tmax_4}), 
which yield ratios of $\sim 10^0 - 10^6$. This is due to larger $V_E(\phi_e)$ and consequently 
elevated $\mathcal{H}_I$ from super-Planckian excursions. Comparing $V_J^{(1)}$ and $V_J^{(2)}$, 
the latter shows slightly lower ratios because the quadratic modulation softens the potential end, 
reducing $\mathcal{H}_I$. In Brans-Dicke-like and Higgs-like models, the confined field values 
result in smaller $\mathcal{H}_I$, leading to flatter declines and lower peak ratios, 
consistent with their suppressed $r$ and plateau-like features, which imply lower inflationary energy scales.
%%%%%%%%%%%%%%%%%%%%%%%%%%%%%%%%%%%%%%%%%%%%%%%%%%%%%%%%%%%%%%%%%%%%%%%%%%%%%%

\begin{table}[t!]
\begin{center}
\begin{tabular}{|c|c||c|c|c|c|} 
 \hline
$n$ & $\xi$ & \multicolumn{2}{c|}{Stability for $y_
\chi$} & \multicolumn{2}{c|}{Stability for $\lambda_{12}$} \\ [0.5ex] 
\cline{3-6}
& & about $\mu=\phi_*$ & about $\mu=\phi_{\rm end}$ & about $\mu=\phi_*$ & about $\mu=\phi_{\rm end}$ \\ [0.5ex] 
\hline\hline
5 & 0.1 & $y_\chi<1.98\times10^{-4}$ & $y_\chi<4.46\times10^{-4}$ & $\lambda_{12}<7.10\times10^{-9}$ & $\lambda_{12}<1.12\times10^{-6}$ \\ 
\hline
5 & 4 & $y_\chi<6.75\times 10^{-5}$ & $y_\chi<3.68\times 10^{-4}$ & $\lambda_{12}<5.51\times 10^{-7}$ & $\lambda_{12}<3.29\times 10^{-7}$ \\ 
\hline
\end{tabular}
\caption{\it Same as Tab.~\ref{Table:stability3m} but for the \textbf{Palatini} framework.}
\label{Table:stability3}
\end{center}
\end{table}
%%%%%%%%%%%%%%%%%%%%%%%%%%%%%%%%%%%%%%%%%%%%%%%%%%%%%%%%%%%%%%%%%%%%%%%%%%%%%%
The metric and Palatini formalisms exhibit subtle differences in the ratio expressions, with the Palatini formulation often showing slightly steeper declines due to its distinct canonical normalization. However, for small-field models, the ratios converge, mirroring the similarity in bounds and reflecting the reduced impact of \(\kappa\) in Eq.~\eqref{eq:Pi-general}. These high \(T_{\rm max}\) values, even at low \(T_{\rm rh}\), are crucial for mechanisms like thermal leptogenesis, which require temperatures \(\gtrsim 10^9 \, {\rm GeV}\) for sufficient CP violation, while respecting BBN constraints. The branching ratio to DM (Eq.~\eqref{eq:Br}) further ties reheating to particle phenomenology: in large-field models, tighter \(y_\chi/\lambda_{12}\) limits imply smaller Br, favoring SM-dominated reheating, whereas small-field models permit higher DM yields, potentially alleviating relic density tensions.

It is important to emphasize that the radiative stability constraints derived from the one-loop CW conditions (Eqs.~\eqref{Eq:first derivative test1},~\eqref{Eq:second derivative test1}, and~\eqref{eq:constraints}) depend strictly on the coupling strength of the fields running in the loops, independent of their specific phenomenological roles. Because both the DM candidate \(\chi\) and the RHN \(N_1\) are fermions, their respective Yukawa couplings, \(y_\chi\) and \(y_{N_1}\), generate functionally identical negative loop corrections to the effective potential. Concurrently, the stringent mathematical upper bounds are imposed on \(y_{N_1}\) during the subsequent non-thermal leptogenesis analysis. Concurrently, the trilinear portal coupling \(\lambda_{12}\) plays a dual role; it governs the reheating temperature \(T_{\rm rh}\), while also driving the positive scalar loop corrections. Therefore, the requirement to suppress these scalar loops rigidly caps the maximum permissible value of \(\lambda_{12}\), which physically enforces an absolute theoretical ceiling on the maximum viable reheating temperature across all evaluated models, as can be followed in the next subsection, specifically  Eq.~\eqref{eq:Trh_max}.

In summary, the reheating dynamics underscore the interplay between gravitational formalisms and potential structures, with Palatini enhancing field effects in large-field regimes and small-field models offering greater flexibility for couplings. These features not only ensure consistency with CMB observables but also open avenues for post-inflationary particle production, paving the way for integrated models of inflation, DM, and baryogenesis.
%%%%%%%%%%%%%%%%%%%%%%%%%%%%%%%%%%%%%%%%%%%%%%%%%%%%%%%%%%%%%%%%%%%%%%%%%%%%%%
%%%%%%%%%%%%%%%%%%%%%%%%%%%%%%%%%%%%%%%%%%%%%%%%%%%%%%%%%%%%%%%%%%%%%%%%%%%%%%
%%%%%%%%%%%%%%%%%%%%%%%%%%%%%%%%%%%%%%%%%%%%%%%%%%%%%%%%%%%%%%%%%%%%%%%%%%%%%%
\subsection{Production of Dark Matter through Inflaton Decay}
\label{subsec:dm_decay}
In this subsection, we explore the production of the fermionic DM candidate \(\chi\) via the decay of the inflaton \(\phi\) during reheating. We assume these \(\chi\) particles are stable and non-thermal, potentially accounting for the full observed cold DM density. This mechanism links early-universe inflation directly to the DM abundance, providing a unified framework for cosmology and particle physics.

The evolution of the comoving number density \(N_\chi\) is governed by~\cite{Ghoshal:2023jhh, Ghoshal:2024ycp}
%%%%%%%%%%%%%%%%%%%%%%%%%%%%%%%%%%%%%%%%%%%%%%%%%%%%%%%%%%%%%%%%%%%%%%%%%%%%%%
\begin{equation}\label{eq:Comoving dm number density}
    \frac{d N_\chi}{dt} = a^3 \, \gamma\,,
\end{equation}
%%%%%%%%%%%%%%%%%%%%%%%%%%%%%%%%%%%%%%%%%%%%%%%%%%%%%%%%%%%%%%%%%%%%%%%%%%%%%%
which represents the integrated Boltzmann equation in an expanding Universe, tracking DM production while accounting for cosmic dilution.

Here, the production rate density for \(\phi \to \bar{\chi} \chi\) is~\cite{Ghoshal:2023jhh, Ghoshal:2024ycp}
%%%%%%%%%%%%%%%%%%%%%%%%%%%%%%%%%%%%%%%%%%%%%%%%%%%%%%%%%%%%%%%%%%%%%%%%%%%%%%
\begin{equation}\label{eq:gamma defined}
    \gamma = 2 \, \text{Br} \, \Gamma_\phi \, \frac{\rho_\phi}{m_\phi}\,.
\end{equation}
%%%%%%%%%%%%%%%%%%%%%%%%%%%%%%%%%%%%%%%%%%%%%%%%%%%%%%%%%%%%%%%%%%%%%%%%%%%%%%
The inflaton mass \(m_\phi\) corresponds to the effective mass arising from the second derivative of the Einstein-frame potential \(V_E(\phi)\) evaluated at its minimum
%%%%%%%%%%%%%%%%%%%%%%%%%%%%%%%%%%%%%%%%%%%%%%%%%%%%%%%%%%%%%%%%%%%%%%%%%%%%%%
\begin{equation}
m_\phi = \sqrt{V_E''(\phi_{\rm min})}\,.    
\end{equation}
%%%%%%%%%%%%%%%%%%%%%%%%%%%%%%%%%%%%%%%%%%%%%%%%%%%%%%%%%%%%%%%%%%%%%%%%%%%%%%
In the models considered here (Eqs.~\eqref{eq:U1_Palatini}, \eqref{eq:U2_Palatini}, \eqref{eq:3_potential}, and~\eqref{eq:4_potential}), the bare potential vanishes or remains kinematically suppressed at the minimum \(\phi\) in both metric and Palatini formalisms, as the curvature vanishes there. To ensure a well-defined oscillatory phase during reheating and maintain slow-roll viability, we introduce a bare mass term \(m_b^2 \phi^2 / 2\) to the Jordan-frame potential, following the discussion in ref.~\cite{Ghoshal:2024ycp}, which shifts the effective mass without spoiling inflationary predictions. This term is introduced in Eqs.~\eqref{eq:baremass-metric} and~\eqref{eq:bar mass limit Trh} for both metric and Palatini formulations, respectively, and it is consistent with the upper limits derived from radiative stability (Eqs.~\eqref{Eq:first derivative test1},  \eqref{Eq:second derivative test1}, and~\eqref{eq:constraints}). The bare mass term results in an effective \(m_\phi \approx m_b\) at the minimum, and incorporating a suppression factor \(\varrho = 10\) and \(100\) to align with the observational bounds on the inflationary Hubble parameter \(\mathcal{H}_I\) (Eq.~\eqref{eq:HI}). These values ensure \(m_\phi\) remains consistent with the reheating temperature range \(4 \, \text{MeV} \lesssim T_{\rm rh} \lesssim 10^{15} \, \text{GeV}\) and the decay kinematics discussed below.
%%%%%%%%%%%%%%%%%%%%%%%%%%%%%%%%%%%%%%%%%%%%%%%%%%%%%%%%%%%%%%%%%%%%%%%%%%%%%%

\begin{figure}[t!]
     \centering
     \includegraphics[width=0.9\linewidth]{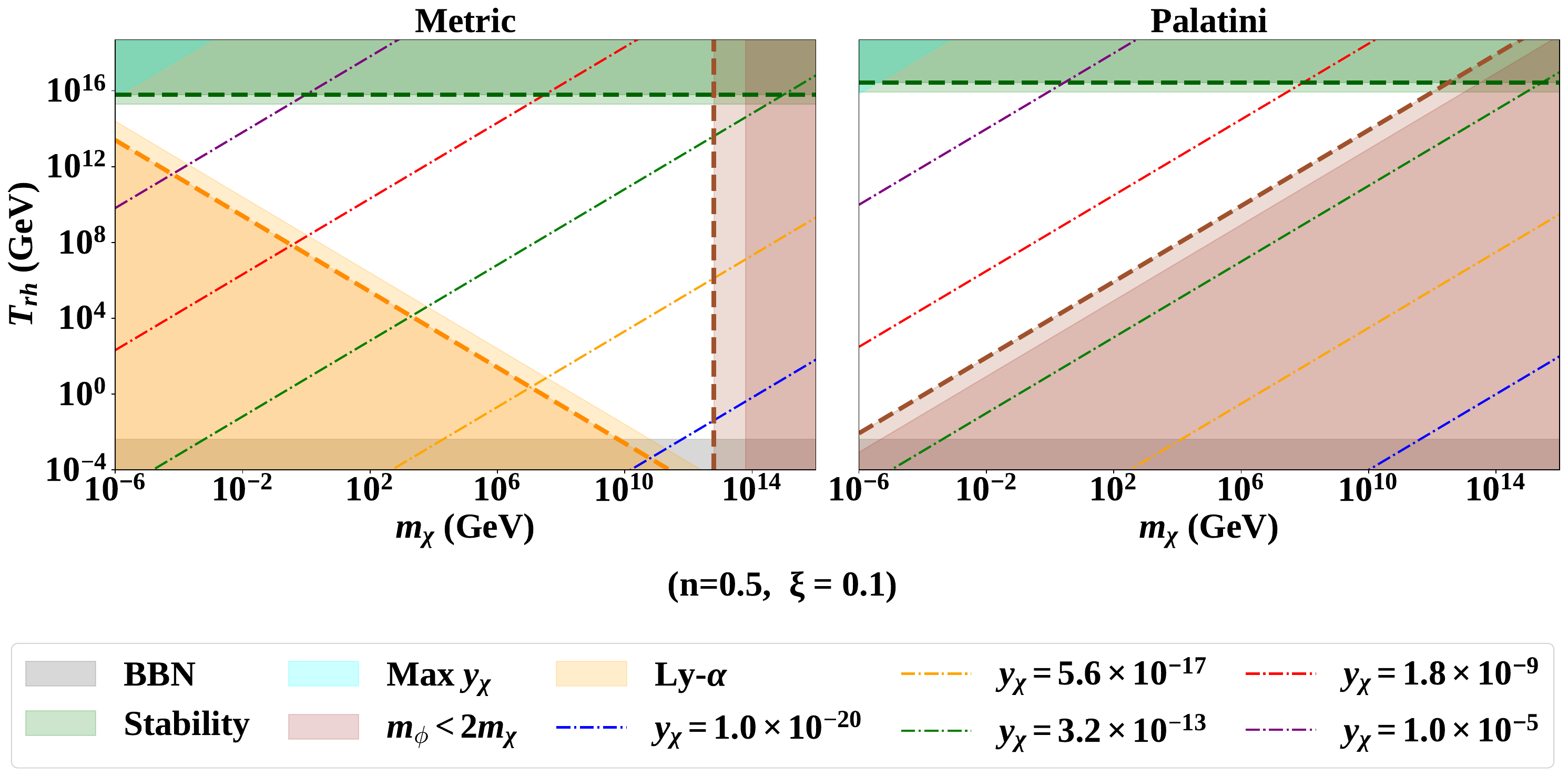}
   
     \caption{\it  The allowed regions (unshaded areas) in the $(T_{\rm rh}, m_\chi)$ plane for the linear inflation model expressed in Eq.~\eqref{eq:U1_Palatini} within \textbf{metric} (left) and \textbf{Palatini} (right) formalisms, along with the viable ranges of the Yukawa coupling $y_\chi$ (dotted lines) to account for the observed cold DM abundance. Shaded regions denote exclusion constraints described in the paper. In these regions, thick dashed lines indicate corresponding bounds for $\varrho=100$, while the continuous lines apply to $\varrho=10$. These values are as given in Tabs.~\ref{Table:stability0m} and~\ref{Table:stability0}, achieving viable inflationary parameters as detailed in Tab.~\ref{tab:infl-para_1}.}
     \label{fig:dm_1}
 \end{figure}
 %%%%%%%%%%%%%%%%%%%%%%%%%%%%%%%%%%%%%%%%%%%%%%%%%%%%%%%%%%%%%%%%%%%%%%%%%%%%%%
Physically, this rate quantifies how efficiently the inflaton's energy density \(\rho_\phi\) is converted into DM pairs, with the factor of 2 arising from the two fermions produced per decay. The branching fraction \(\text{Br}\) is approximately given by Eq.~\eqref{eq:Br}, and studied in detail within Refs. \cite{Ghoshal:2023jhh, Ghoshal:2023jvf}. This ratio highlights the competition between DM production and SM reheating channels, controlled by the relative strengths of the Yukawa \(y_\chi\) and trilinear \(\lambda_{12}\) couplings.

During reheating (\(T_{\max} > T > T_{\rm rh}\)), the Hubble parameter and inflaton energy density are~\cite{Bernal:2021qrl, Ghoshal_2024}
%%%%%%%%%%%%%%%%%%%%%%%%%%%%%%%%%%%%%%%%%%%%%%%%%%%%%%%%%%%%%%%%%%%%%%%%%%%%%%
\begin{equation}\label{Eq:Hubble parameter during reheating+Eq:rho_phi}
    \mathcal{H} \simeq \frac{\pi}{3} \sqrt{\frac{g_*}{10}} \frac{T^4}{M_P\, T_{\rm rh}^2}, \qquad 
    \rho_{\phi} \simeq \frac{ \pi^2  g_* }{30 } \frac{T^8}{ T_{\rm rh}^4 }\,.
\end{equation}
%%%%%%%%%%%%%%%%%%%%%%%%%%%%%%%%%%%%%%%%%%%%%%%%%%%%%%%%%%%%%%%%%%%%%%%%%%%%%%
These expressions capture the transition from inflaton domination to radiation, in which the Universe's expansion rate decelerates as energy is transferred to relativistic particles. Specifically, the non-standard temperature dependence, such as \(\mathcal{H} \propto T^4\) rather than \(T^2\) scaling of the radiation-dominated era as in Eq.~\eqref{eq:H_T_rh}, arises because the continuous entropy injection from the decaying, matter-like inflaton condensate modifies the standard adiabatic evolution of the scale factor to \(a\propto T^{-8/3}\).

The DM yield \(Y_\chi = n_\chi / s\) differs between formalisms due to the effective equation of state~\cite{Ghoshal:2024ycp}
%%%%%%%%%%%%%%%%%%%%%%%%%%%%%%%%%%%%%%%%%%%%%%%%%%%%%%%%%%%%%%%%%%%%%%%%%%%%%%
\begin{equation}\label{eq:Yield of DM}
Y_\chi \simeq 
\begin{cases}
   \frac{3}{\pi} \frac{g_*}{g_{*s}} \sqrt{\frac{10}{g_*}} \frac{M_P \, \Gamma_\phi}{m_{\phi}\, T_{\rm rh}} \text{Br} \,, & \text{(metric)},\\ 
 \frac{9}{2\pi} \frac{g_*}{g_{*s}} \sqrt{\frac{10}{g_*}} \frac{M_P \, \Gamma_\phi}{m_{\phi}\, T_{\rm rh}} \text{Br} \,, & \text{(Palatini)}.
\end{cases}
\end{equation}
%%%%%%%%%%%%%%%%%%%%%%%%%%%%%%%%%%%%%%%%%%%%%%%%%%%%%%%%%%%%%%%%%%%%%%%%%%%%%%
The factor-of-3/2 difference reflects distinct inflaton oscillation behaviors: matter-like in the metric (slower decay) versus radiation-like in the Palatini (faster energy dilution).

The present-day CDM yield is~\cite{Ghoshal:2024ycp}
%%%%%%%%%%%%%%%%%%%%%%%%%%%%%%%%%%%%%%%%%%%%%%%%%%%%%%%%%%%%%%%%%%%%%%%%%%%%%%
\begin{equation}\label{Eq:present day CDM yield}
 Y_{{\rm CDM},0} = \frac{4.3 \times 10^{-10}}{m_\chi / \text{GeV}} \,,
\end{equation}
%%%%%%%%%%%%%%%%%%%%%%%%%%%%%%%%%%%%%%%%%%%%%%%%%%%%%%%%%%%%%%%%%%%%%%%%%%%%%%
derived from observed relic density, setting the target for production mechanisms.

Setting \(Y_\chi \sim Y_{{\rm CDM},0}\) yields~\cite{Ghoshal:2024ycp}
%%%%%%%%%%%%%%%%%%%%%%%%%%%%%%%%%%%%%%%%%%%%%%%%%%%%%%%%%%%%%%%%%%%%%%%%%%%%%%
\begin{equation}\label{eq:Trh_mchi}
T_{\rm rh} \simeq
\begin{cases}
 6.49 \times 10^{25}\, y_\chi^2 \, m_\chi \,, & \text{(metric)} \,, \\%
 9.74 \times 10^{25}\, y_\chi^2 \, m_\chi \,, & \text{(Palatini)} \,.
\end{cases}
\end{equation}
%%%%%%%%%%%%%%%%%%%%%%%%%%%%%%%%%%%%%%%%%%%%%%%%%%%%%%%%%%%%%%%%%%%%%%%%%%%%%%
This expression serves as a relation for mapping the viable phenomenological parameter space in the (\(T_{\rm rh}. m_\chi\)) plane, as it explicitly connects the reheating temperature to the DM mass required to satisfy the observed relic density. The difference in the numerical coefficients between the two formulations originates directly from their distinct behavior during the reheating epoch, matter-like for the metric case and radiation-like for the Palatini case. Furthermore, the geometric boundaries of the allowed regions in the related figures, namely Fig.~\ref{fig:dm_1} and Figs.~\ref{fig:dm_2}--~\ref{fig:dm_4}, are fundamentally determined by the maximum permissible values of the Yukawa coupling, \(y_\chi\). Because these stringent \(y_\chi\) bounds are derived from the CW radiative stability conditions (Eqs.~\eqref{Eq:first derivative test1},~\eqref{Eq:second derivative test1}, and~\eqref{eq:constraints}), which are highly sensitive to the underlying canonical field excursions, they exhibit strong formalism dependence, as we show in the resulting plots across different models. Consequently, the combination of the scaling in Eq.~\eqref{eq:Trh_mchi} and the tighter Palatini-specific CW bounds is what ultimately shapes the different viable parameter spaces observed between the two gravitational frameworks.

To generalize across models, we introduce below the effective bare mass (inflaton oscillation mass) as~\cite{Ghoshal:2024ycp}
%%%%%%%%%%%%%%%%%%%%%%%%%%%%%%%%%%%%%%%%%%%%%%%%%%%%%%%%%%%%%%%%%%%%%%%%%%%%%%
\begin{align}
& m_{(m),\xi} = \frac{1}{\varrho} \, \sqrt{\frac{2}{3}\frac{\Lambda }{\xi^2}M_P^2}\,, \label{eq:baremass-metric}\\
& m_{{(P)},\xi} \simeq \frac{1}{\varrho}\, \sqrt{\Lambda} \left(\frac{\sqrt{3}\, M_P\, \lambda_{12}^2}{8\pi \,\Lambda} \right)^{1/3}
   = \frac{1}{\zeta} \, \left(\frac{3 \pi^2}{40 }\right)^{1/4} \, (g_* \, \Lambda)^{1/4} \, T_{\rm rh}\,, \label{eq:bar mass limit Trh}
\end{align}
%%%%%%%%%%%%%%%%%%%%%%%%%%%%%%%%%%%%%%%%%%%%%%%%%%%%%%%%%%%%%%%%%%%%%%%%%%%%%%
where \(\Lambda\) is the effective potential scale (adjusted per model, e.g., from \(M^4\) for the linear inflation potentials provided in Eq.~\eqref{eq:U2_Palatini}, and \(M^2\) for both the Brans-Dicke-like potential in Eq.~\eqref{eq:3_potential} and Higgs inflation-like model expressed as in Eq.~\eqref{eq:4_potential}). The parameter \(\varrho\) is a dimensionless factor that suppresses the inflaton mass relative to the fundamental Planck scales, ensuring sub-Planckian effective masses consistent with quantum gravity constraints and observational bounds on inflationary scales. We choose \(\varrho = 10\) and \(100\) to illustrate moderate and stronger suppression, respectively; \(\varrho = 10\) represents typical values in models where the non-minimal coupling flattens the potential without extreme fine-tuning, while \(\varrho = 100\) probes regimes with more pronounced hierarchy, akin to scenarios evading swampland conjectures by larger field-space re-scalings.

The bounds present in the \((T_{\rm rh}, m_\chi)\) planes through Fig.~\ref{fig:dm_1} and Figs.~\ref{fig:dm_2}--~\ref{fig:dm_4} can be provided in more details as the following:
%%%%%%%%%%%%%%%%%%%%%%%%%%%%%%%%%%%%%%%%%%%%%%%%%%%%%%%%%%%%%%%%%%%%%%%%%%%%%%
\begin{itemize}
%%%%%%%%%%%%%%%%%%%%%%%%%%%%%%%%%%%%%%%%%%%%%%%%%%%%%%%%%%%%%%%%%%%%%%%%%%%%%%
    \item \textbf{Max \(T_{\rm rh}\)}: From max \(\lambda_{12}\) (Tables \ref{Table:stability0m}--\ref{Table:stability3})~\cite{Ghoshal:2024ycp}
%%%%%%%%%%%%%%%%%%%%%%%%%%%%%%%%%%%%%%%%%%%%%%%%%%%%%%%%%%%%%%%%%%%%%%%%%%%%%%
\begin{equation}\label{eq:Trh_max}
T_{\rm rh}\Big|_{\rm max} \simeq
\begin{cases}
 \sqrt{\frac{2}{\pi}} \left(\frac{10}{g_*}\right)^{1/4} \sqrt{M_P} \left(\frac{\lambda_{12,{\rm max}}^2}{8\pi\, m_{(m),\xi}} \right)^{1/2} \,, & \text{(metric)},\\
 \sqrt{\frac{\zeta}{2}}  \left(\frac{1}{3 \, \pi^{10}\,\Lambda} \right)^{1/12} \left(\frac{10}{g_*}\right)^{1/4} {M_P}^{1/3} {\lambda_{12,{\rm max}}}^{2/3} \,, & \text{(Palatini)}.
\end{cases}
\end{equation}
%%%%%%%%%%%%%%%%%%%%%%%%%%%%%%%%%%%%%%%%%%%%%%%%%%%%%%%%%%%%%%%%%%%%%%%%%%%%%%
Shown as horizontal green bands. This bound ensures perturbative reheating without destabilizing the inflationary potential.

%%%%%%%%%%%%%%%%%%%%%%%%%%%%%%%%%%%%%%%%%%%%%%%%%%%%%%%%%%%%%%%%%%%%%%%%%%%%%%
    \item \textbf{Max \(y_\chi\)}: From stability~\cite{Ghoshal:2024ycp}
%%%%%%%%%%%%%%%%%%%%%%%%%%%%%%%%%%%%%%%%%%%%%%%%%%%%%%%%%%%%%%%%%%%%%%%%%%%%%%
\begin{equation}
 T_{\rm rh} \simeq
\begin{cases}
 6.49 \times 10^{25}\, y_{\chi, \rm max}^2 \, m_\chi \,, & \text{(metric)},\\
 9.74 \times 10^{25}\, y_{\chi, \rm max}^2 \, m_\chi \,, & \text{(Palatini)}.
\end{cases}
\end{equation}
%%%%%%%%%%%%%%%%%%%%%%%%%%%%%%%%%%%%%%%%%%%%%%%%%%%%%%%%%%%%%%%%%%%%%%%%%%%%%%
cyan wedge at top-left. It prevents excessive quantum corrections from the DM sector.
%%%%%%%%%%%%%%%%%%%%%%%%%%%%%%%%%%%%%%%%%%%%%%%%%%%%%%%%%%%%%%%%%%%%%%%%%%%%%%
    \item \textbf{Max \(m_\chi\)}: \(m_\chi \lesssim m_{(m/P),\xi}/2\). Vertical band (metric); wedge (Palatini, due to \(T_{\rm rh}\) dependence). This kinematic threshold ensures decay is possible.
     
%%%%%%%%%%%%%%%%%%%%%%%%%%%%%%%%%%%%%%%%%%%%%%%%%%%%%%%%%%%%%%%%%%%%%%%%%%%%%%
    \item \textbf{Lyman-\(\alpha\)}: Ensures cold DM~\cite{Bernal:2021qrl, Ghoshal:2022jeo}
%%%%%%%%%%%%%%%%%%%%%%%%%%%%%%%%%%%%%%%%%%%%%%%%%%%%%%%%%%%%%%%%%%%%%%%%%%%%%%
\begin{equation}\label{eq:Lyman-alpha}
    m_{\chi} \gtrsim 2 \times 10^{-6}\frac{m_{(m),\xi}}{T_{\rm rh}}\,{\rm GeV} \quad \text{(metric)}.
\end{equation}
%%%%%%%%%%%%%%%%%%%%%%%%%%%%%%%%%%%%%%%%%%%%%%%%%%%%%%%%%%%%%%%%%%%%%%%%%%%%%%
However, these constraints can be introduced for the Palatini case as~\cite{Ghoshal:2024ycp}
%%%%%%%%%%%%%%%%%%%%%%%%%%%%%%%%%%%%%%%%%%%%%%%%%%%%%%%%%%%%%%%%%%%%%%%%%%%%%%
\begin{equation}
m_\chi \gtrsim \frac{2 \times 10^{-6}}{\varrho} \,  \left(\frac{3 \pi^2}{40 }\right)^{1/4}  \, (g_{reh} \, \Lambda)^{1/4} \,{\rm GeV}\,.
\end{equation}
%%%%%%%%%%%%%%%%%%%%%%%%%%%%%%%%%%%%%%%%%%%%%%%%%%%%%%%%%%%%%%%%%%%%%%%%%%%%%%
Given the negligible magnitude of this constraint in the Palatini framework, it has been excluded from the Palatini case figures in this subsection. In practice, the minimum mass for warm DM particles is established at $\gtrsim 3.5$ keV~\cite{Bernal:2021qrl}, increasing to $\gtrsim 5.3$ keV~\cite{Viel:2013fqw,Palanque-Delabrouille:2019iyz} at 95\% confidence level for thermal warm DM scenarios. Variations in the Universe's thermal evolution could lower this threshold to 1.9 keV~\cite{Garzilli:2019qki}. For other feebly interacting massive particles, the constraints range from 4 to 16 keV~\cite{Murgia:2017lwo, Heeck:2017xbu, Bae:2017dpt, Boulebnane:2017fxw, Baldes:2020nuv, Ballesteros:2020adh, Decant:2021mhj}.
%%%%%%%%%%%%%%%%%%%%%%%%%%%%%%%%%%%%%%%%%%%%%%%%%%%%%%%%%%%%%%%%%%%%%%%%%%%%%%
    \item \textbf{BBN}: \(T_{\rm rh} \gtrsim 4\,{\rm MeV}\), gray band at bottom. Essential for successful BBN.
\end{itemize}
%%%%%%%%%%%%%%%%%%%%%%%%%%%%%%%%%%%%%%%%%%%%%%%%%%%%%%%%%%%%%%%%%%%%%%%%%%%%%%
The results presented in Fig.~\ref{fig:dm_1} and Figs.~\ref{fig:dm_2},--~\ref{fig:dm_4}, informed by the radiative stability bounds in Tables~\ref{Table:stability0m}--\ref{Table:stability3}, delineate the viable parameter space in the $(T_{\rm rh}, m_\chi)$ plane where perturbative inflaton decay can account for the observed cold DM abundance via the non-thermal production of the fermionic candidate $\chi$. These figures, derived for select benchmark values of $(n, \xi)$ to illustrate representative behaviors, reveal unshaded regions satisfying the relic density constraint, which is subject to exclusions from radiative stability, kinematics, BBN, and Lyman-$\alpha$ forest observations as discussed above.

In the large-field linear inflation models (Eqs.~\eqref{eq:U1_Palatini} and \eqref{eq:U2_Palatini}; Figs.~\ref{fig:dm_1} and~\ref{fig:dm_2}), expansive allowed regions emerge, accommodating DM masses $m_\chi$ up to the PeV scale and reheating temperatures $T_{\rm rh}$ spanning from the BBN lower limit of $\sim 4$~MeV to $\sim 10^{13}$~GeV. This broad viability stems from the super-Planckian field excursions during inflation, which, while imposing stringent upper limits on the Yukawa coupling (as presented in Tabs.~\ref{Table:stability0m}--~\ref{Table:stability1}) $y_\chi \lesssim 10^{-3}$ and trilinear $\lambda_{12} \lesssim 10^{-5}$ to preserve potential flatness, permit sufficient branching to $\chi$ pairs without destabilizing the multi-phase inflationary trajectory. The cyan wedge, enforcing the maximum $y_\chi$ from CW corrections which are given by Eqs.~\eqref{Eq:first derivative test1},~\eqref{Eq:second derivative test1}, and~\eqref{eq:constraints}, and the green horizontal stripe, capping $T_{\rm rh}$ via the bounded $\lambda_{12}$, collectively favor higher-mass $\chi$ at elevated $T_{\rm rh}$, where rapid decays minimize backreaction effects. Comparing the two linear potentials, Eq.~\eqref{eq:U2_Palatini} exhibits marginally narrower regions due to its quadratic modulation, which softens the potential's end and reduces the inflationary Hubble scale, thereby tightening kinematic bounds on $m_\chi < m_\phi/2$.

In contrast, the small-field models---Brans-Dicke-like (Eq.~\eqref{eq:3_potential}; Fig.~\ref{fig:dm_3}) and Higgs-like (Eq.~\eqref{eq:4_potential}; Fig.~\ref{fig:dm_4}), yield more constrained parameter spaces, with $m_\chi$ typically below $10^8$~GeV and $T_{\rm rh} \lesssim 10^{10}$~GeV. Sub-Planckian excursions here allow somewhat looser stability bounds ($y_\chi \lesssim 10^{-3}$, $\lambda_{12} \lesssim 10^{-5}$), yet plateau-like structures demand exquisite control over quantum corrections to sustain attractor solutions, resulting in reduced branching ratios and thus lower viable $T_{\rm rh}$. The Higgs-like variant imposes the tightest constraints, aligning with its sensitivity to radiative shifts that could disrupt the CMB-compatible spectral index $n_s \approx 0.96$ and suppressed tensor-to-scalar ratio $r \lesssim 10^{-5}$ as shown in Tab.~\ref{tab:infl-para_4} and Figs.~\ref{fig:r_ns_4_m} and~\ref{fig:r_ns_4_P}.

Distinctions between metric and Palatini formalisms are pronounced in large-field regimes, where Palatini's independent connection yields larger canonical field values, amplifying loop corrections and thus stricter bounds on $y_\chi$ and $\lambda_{12}$ (e.g., $y_\chi \lesssim 4.58 \times 10^{-4}$ vs. $4.91 \times 10^{-4}$ for Eq.~\eqref{eq:U1_Palatini} at $n=0.5$, $\xi=0.1$). This manifests in Palatini plots as steeper wedges and reduced high-$m_\chi$ accessibility, reflecting radiation-like inflaton oscillations that accelerate energy dilution and diminish the production window. In small-field models, however, the formalisms converge, with negligible differences in viable areas owing to subdued non-minimal coupling effects.

These findings underscore that multi-phase, inspired models can unify inflation with DM genesis via feeble inflaton portals, constrained by radiative stability to ensure CMB consistency. Large-field scenarios offer greater flexibility for heavy DM, potentially testable via indirect detection, whereas small-field cases favor lighter candidates amenable to collider probes. Critically, the gravity formalism modulates the interplay between inflationary scales and post-inflationary phenomenology, highlighting Palatini's role in constraining parameter space to evade quantum-gravity tensions. Although illustrated for specific benchmarks, these patterns generalize across parameter choices, informing strategies for integrated cosmology-particle physics frameworks.

%%%%%%%%%%%%%%%%%%%%%%%%%%%%%%%%%%%%%%%%%%%%%%%%%%%%%%%%%%%%%%%%%%%%%%%%%%%%%%
%%%%%%%%%%%%%%%%%%%%%%%%%%%%%%%%%%%%%%%%%%%%%%%%%%%%%%%%%%%%%%%%%%%%%%%%%%%%%%
%%%%%%%%%%%%%%%%%%%%%%%%%%%%%%%%%%%%%%%%%%%%%%%%%%%%%%%%%%%%%%%%%%%%%%%%%%%%%%
\section{Neutrino Masses and Non-Thermal Leptogenesis}
\label{sec:lepto}
As in the previous section, we show here the leptogenesis dynamics for the linear inflation model defined in Eq.~\eqref{eq:U1_Palatini}, while the corresponding figures for the other models, including the second linear potential in Eq.~\eqref{eq:U2_Palatini}, the Brans-Dicke-like model in Eq.~\eqref{eq:3_potential}, and the Higgs-like model in Eq.~\eqref{eq:4_potential}, are provided in App.~\ref{app:other_leptogenesis}.

\subsection{Neutrino Masses via Type-I Seesaw Mechanism}
To explain the tiny neutrino masses observed in oscillation experiments, we invoke the Type-I seesaw mechanism, which introduces RHNs \(N_j\) (gauge singlets under the SM) that couple to the left-handed lepton doublets and the Higgs field. The relevant part of the Lagrangian is~\cite{Trodden:2004mj, Buchmuller:2005eh, Davidson:2008bu, DiBari:2012fz, Fong:2012buy, doi:10.1142/10447}
%%%%%%%%%%%%%%%%%%%%%%%%%%%%%%%%%%%%%%%%%%%%%%%%%%%%%%%%%%%%%%%%%%%%%%%%%%%%%%
\begin{align}\label{eq:nu_Lagrangian}
    \mathcal{L}_{\nu} = i \sum_{j=1,2,3} \bar{N}_j \gamma^\mu \partial_\mu N_j - \frac{M_{N_j}}{2} \bar{N}^c_j N_j - \sum_{\alpha,j} \left( Y_{\alpha j} \bar{L}_\alpha \tilde{H} N_j + \text{H.c.} \right),
\end{align}
%%%%%%%%%%%%%%%%%%%%%%%%%%%%%%%%%%%%%%%%%%%%%%%%%%%%%%%%%%%%%%%%%%%%%%%%%%%%%%
where \(\alpha = e, \mu, \tau\) labels the lepton flavors, \(Y_{\alpha j}\) is the complex Yukawa matrix, \(M_{N_j}\) are the Majorana masses (diagonal basis), \(L_\alpha = (\nu_\alpha, \ell_\alpha^-)\) are the left-handed lepton doublets, and \(\tilde{H} = i \sigma_2 H^\dagger\) with \(H\) the SM Higgs doublet.
After electroweak symmetry breaking, with \(\langle H \rangle = v / \sqrt{2}\) and \(v = 246\) GeV, the Dirac mass matrix is \(m_D = Y v / \sqrt{2}\). Integrating out the heavy \(N_j\) yields the effective Majorana mass matrix for light neutrinos~\cite{Casas:2001sr,Ibarra:2003up}
%%%%%%%%%%%%%%%%%%%%%%%%%%%%%%%%%%%%%%%%%%%%%%%%%%%%%%%%%%%%%%%%%%%%%%%%%%%%%%
\begin{equation} \label{mnu1}
    \widetilde{m}_\nu = - m_D^T M_N^{-1} m_D = - \frac{v^2}{2} Y^T M_N^{-1} Y,
\end{equation}
%%%%%%%%%%%%%%%%%%%%%%%%%%%%%%%%%%%%%%%%%%%%%%%%%%%%%%%%%%%%%%%%%%%%%%%%%%%%%%
where \(M_N = \text{diag}(M_{N_1}, M_{N_2}, M_{N_3})\). The essence of the seesaw mechanism can be captured in its simplest single-generation form: the light neutrino mass is approximately \(m_\nu \approx m_D^2 / M_N\), where a large Majorana mass \(M_N\) naturally suppresses \(m_\nu\) to the sub-eV scale observed, even if the Dirac mass \(m_D\) is comparable to other fermion masses. To visualize this mass-generation process, the tree-level exchange of the heavy RHN is shown in 
Fig.~\ref{fig:seesaw_diagram}. The resulting mass matrix is diagonalized by the Pontecorvo-Maki-Nakagawa-Sakata (PMNS) matrix \(U\)~\cite{Drees:2024hok}
%%%%%%%%%%%%%%%%%%%%%%%%%%%%%%%%%%%%%%%%%%%%%%%%%%%%%%%%%%%%%%%%%%%%%%%%%%%%%%
\begin{equation} \label{eq:PMNS}
    U^T \widetilde{m}_\nu U = m_\nu^{\rm diag} = \text{diag}(m_1, m_2, m_3).
\end{equation}
%%%%%%%%%%%%%%%%%%%%%%%%%%%%%%%%%%%%%%%%%%%%%%%%%%%%%%%%%%%%%%%%%%%%%%%%%%%%%%
Assuming normal hierarchy (\(m_1 < m_2 < m_3\)) and using oscillation data~\cite{Esteban:2018azc}
%%%%%%%%%%%%%%%%%%%%%%%%%%%%%%%%%%%%%%%%%%%%%%%%%%%%%%%%%%%%%%%%%%%%%%%%%%%%%%
\begin{equation} \label{eq:numass2}
\begin{aligned}
    m_1 &\approx 0\,, \\
    m_2 &= \sqrt{\Delta m^2_{\odot}} \approx 8.6 \times 10^{-3} \, \text{eV}\,, \\
    m_3 &= \sqrt{\Delta m^2_{\rm atm}} \approx 5.0 \times 10^{-2} \, \text{eV}\,.
\end{aligned}
\end{equation}
%%%%%%%%%%%%%%%%%%%%%%%%%%%%%%%%%%%%%%%%%%%%%%%%%%%%%%%%%%%%%%%%%%%%%%%%%%%%%%
With \(m_1 = 0\), two RHNs suffice, but we keep three for generality. Using the Casas-Ibarra parametrization~\cite{Casas:2001sr}
%%%%%%%%%%%%%%%%%%%%%%%%%%%%%%%%%%%%%%%%%%%%%%%%%%%%%%%%%%%%%%%%%%%%%%%%%%%%%%
\begin{equation} \label{CIY}
    Y = \frac{i \sqrt{2}}{v} U^* \sqrt{m_\nu^{\rm diag}} R^T \sqrt{M_N},
\end{equation}
%%%%%%%%%%%%%%%%%%%%%%%%%%%%%%%%%%%%%%%%%%%%%%%%%%%%%%%%%%%%%%%%%%%%%%%%%%%%%%
where \(R\) is a complex orthogonal matrix (\(R R^T = \mathbb{1}\)). For \(m_1 = 0\) and \(M_3 \to \infty\)~\cite{Drees:2024hok}
%%%%%%%%%%%%%%%%%%%%%%%%%%%%%%%%%%%%%%%%%%%%%%%%%%%%%%%%%%%%%%%%%%%%%%%%%%%%%%
\begin{equation} \label{Rmatrix}
    R = \begin{pmatrix}
        0 & \cos z & \sin z \\
        0 & -\sin z & \cos z \\
        1 & 0 & 0
    \end{pmatrix},
\end{equation}
%%%%%%%%%%%%%%%%%%%%%%%%%%%%%%%%%%%%%%%%%%%%%%%%%%%%%%%%%%%%%%%%%%%%%%%%%%%%%%
with \(z\) complex. Relevant elements are~\cite{Drees:2024hok}
%%%%%%%%%%%%%%%%%%%%%%%%%%%%%%%%%%%%%%%%%%%%%%%%%%%%%%%%%%%%%%%%%%%%%%%%%%%%%%
\begin{align} \label{eq:Y11}
    (Y^\dagger Y)_{11} &= \frac{2M_{N_1}}{v^2} (m_2 |\cos z|^2 + m_3 |\sin z|^2), \\
    (Y^\dagger Y)_{21} &= \frac{2\sqrt{M_{N_1} M_{N_2}}}{v^2} [-m_2 \sin z^* \cos z + m_3 \sin z \cos z^*]. \label{eq:Y21}
\end{align}
%%%%%%%%%%%%%%%%%%%%%%%%%%%%%%%%%%%%%%%%%%%%%%%%%%%%%%%%%%%%%%%%%%%%%%%%%%%%%%
This setup generates tiny neutrino masses while allowing RHNs to drive leptogenesis.

%%%%%%%%%%%%%%%%%%%%%%%%%%%%%%%%%%%%%%%%%%%%%%%%%%%%%%%%%%%%%%%%%%%%%%%%%%%%%%
%%%%%%%%%%%%%%%%%%%%%%%%%%%%%%%%%%%%%%%%%%%%%%%%%%%%%%%%%%%%%%%%%%%%%%%%%%%%%%
%%%%%%%%%%%%%%%%%%%%%%%%%%%%%%%%%%%%%%%%%%%%%%%%%%%%%%%%%%%%%%%%%%%%%%%%%%%%%%
\subsection{Non-Thermal Leptogenesis}
We consider non-thermal production of RHNs $N_j$ 
from the inflaton decay during reheating, followed by their CP-violating 
decays that generate a lepton asymmetry. This asymmetry is subsequently 
converted into a baryon asymmetry via electroweak sphaleron 
processes~\cite{Fong:2012buy,doi:10.1142/10447,DiBari:2012fz,
Buchmuller:2005eh,Davidson:2008bu,Trodden:2004mj}.

To ensure the genuinely non-thermal nature of leptogenesis, it is 
crucial to impose the condition
%%%%%%%%%%%%%%%%%%%%%%%%%%%%%%%%%%%%%%%%%%%%%%%%%%%%%%%%%%%%%%%%%%%%%%%%%%%%%%
\begin{equation}
M_{N_1} > T_{\rm max},
\end{equation}
%%%%%%%%%%%%%%%%%%%%%%%%%%%%%%%%%%%%%%%%%%%%%%%%%%%%%%%%%%%%%%%%%%%%%%%%%%%%%%
\begin{figure}[t!]
\centering
\begin{tikzpicture}
  \begin{feynman}
    \vertex (v1);
    \vertex [right=3cm of v1] (v2);
    
    \vertex [above left=1.5cm of v1] (L1) {\(L_\alpha\)};
    \vertex [below left=1.5cm of v1] (H1) {\(\langle H \rangle\)};
    
    \vertex [above right=1.5cm of v2] (L2) {\(L_\beta\)};
    \vertex [below right=1.5cm of v2] (H2) {\(\langle H \rangle\)};
    
    \diagram* {

      (L1) -- [fermion] (v1),
      (H1) -- [scalar] (v1),
      
      (v1) -- [plain, edge label=\(N_j\), insertion=0.5] (v2),

      (L2) -- [fermion] (v2),
      (H2) -- [scalar] (v2),
    };
    
    \node[above=0.3cm of v1] {\(Y_{\alpha j}\)};
    \node[above=0.3cm of v2] {\(Y_{\beta j}\)};
    
    \node[below=0.3cm of v1] {}; 
    \node[below=0.2cm of v2] {};
    \node at (1.5, -0.4) {\(M_{N_j}\)};
  \end{feynman}
\end{tikzpicture}
\caption{\it Tree-level Feynman diagram illustrating the Type-I seesaw mechanism. The exchange of a heavy RHN \(N_j\) generates an effective mass for the light active neutrinos after the Higgs field acquires its vacuum expectation value \(\langle H \rangle\). The cross on the internal propagator denotes the Majorana mass insertion \(M_{N_j}\), while the vertices are governed by the Yukawa couplings \(Y_{\alpha j}\) and \(Y_{\beta j}\).}
\label{fig:seesaw_diagram}
\end{figure}
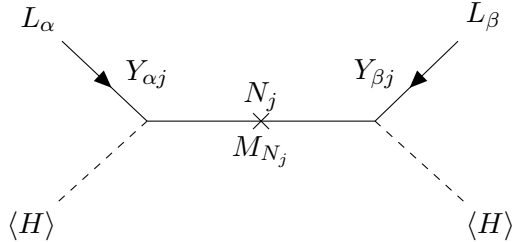
%%%%%%%%%%%%%%%%%%%%%%%%%%%%%%%%%%%%%%%%%%%%%%%%%%%%%%%%%%%%%%%%%%%%%%%%%%%%%%
where $T_{\rm max}$ denotes the maximum temperature reached during 
reheating~\cite{Barman:2021tgt}. If instead \(M_{N_1} < T_{\rm max}\), RHNs could be thermally produced from the SM bath, 
potentially erasing the non-thermal history originating from inflaton decays and effectively 
reducing the scenario to thermal leptogenesis. This requirement is particularly important in 
non-instantaneous reheating, where \(T_{\rm max} \gg T_{\rm rh}\) due to the gradual transfer 
of energy from the inflaton condensate.

\begin{figure}[t!]
    \centering
    \includegraphics[width=0.9\linewidth]{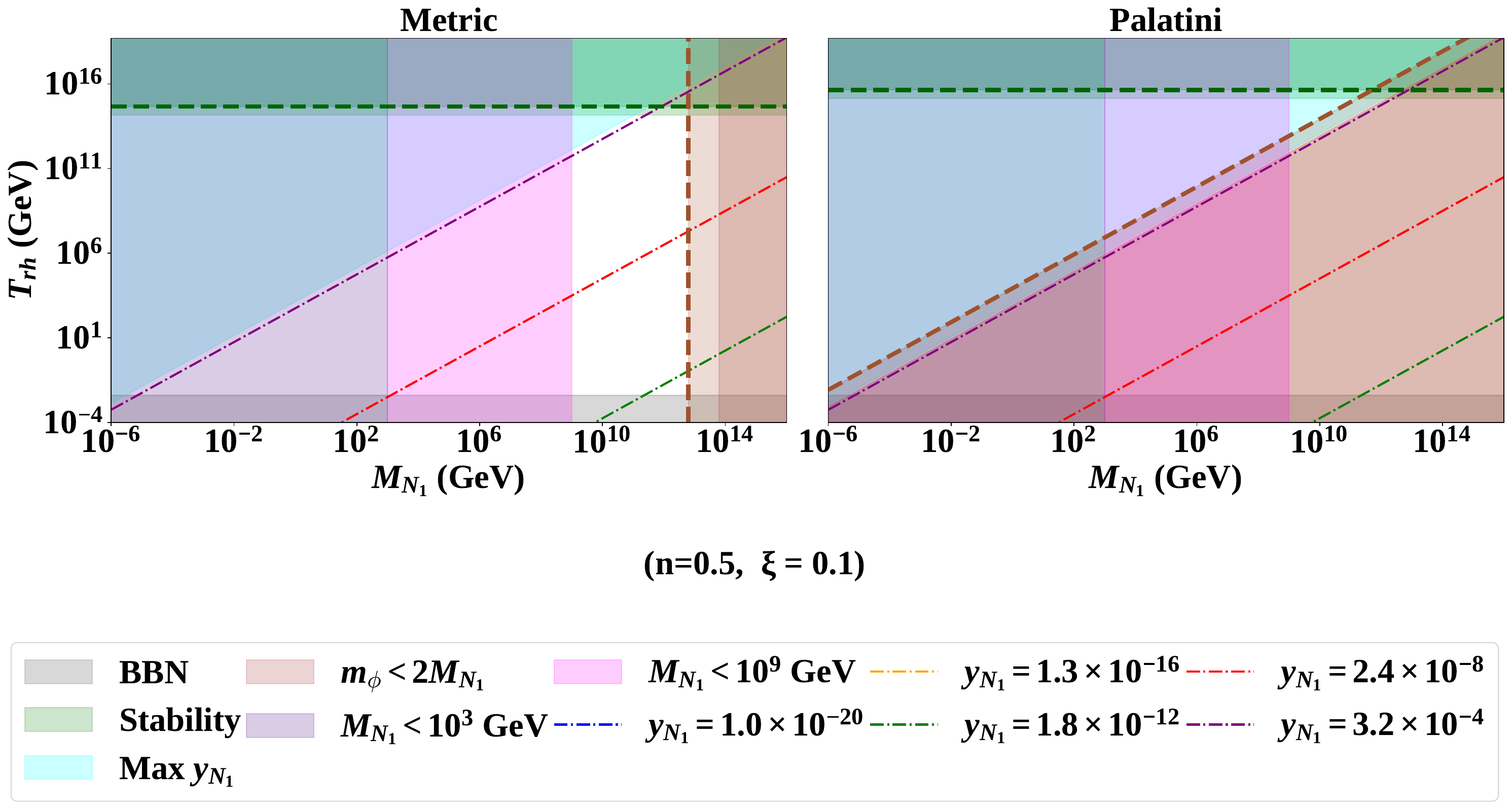}
    \caption{\it Allowed regions (unshaded areas) in the 
    $(T_{\rm rh}, M_{N_1})$ plane for the linear inflation model 
    of Eq.~\eqref{eq:U1_Palatini} within the \textbf{metric} (left) 
    and \textbf{Palatini} (right) formulations. 
    Dotted lines denote the viable ranges of the Yukawa coupling 
    $y_{N_1}$ required for successful non-thermal leptogenesis. 
    Shaded regions correspond to excluded parameter space. 
    Thick dashed curves indicate $\zeta = 100$, while solid curves 
    correspond to $\zeta = 10$, using the stability bounds from 
    Tabs.~\ref{Table:stability0m} and~\ref{Table:stability0}. 
    Inflationary benchmark parameters are given in 
    Tab.~\ref{tab:infl-para_1}.}
    \label{fig:lepto_1}
\end{figure}

In our analysis, we strictly confine the viable parameter space to 
regions satisfying $M_{N_1} > T_{\rm max}$. Although the ratio 
$T_{\rm max}/T_{\rm rh}$ increases at low reheating temperatures 
(cf. Fig.~\ref{fig:tmax_1} and Figs.~\ref{fig:tmax_2}--~\ref{fig:tmax_4}), the allowed (unshaded) 
regions in Fig.~\ref{fig:lepto_1} and Figs.~\ref{fig:lepto_2}--~\ref{fig:lepto_4} correspond to 
sufficiently heavy RHNs such that the inequality 
$M_{N_1} > T_{\rm max}$ is maintained throughout the viable parameter 
space. This guarantees the non-thermal origin of the baryon asymmetry.

It is worth emphasizing that the leptogenesis framework builds upon 
the neutrino-sector Lagrangian $\mathcal{L}_\nu$ introduced in 
Eq.~\eqref{eq:nu_Lagrangian}, supplemented by the inflaton interactions~\cite{Trodden:2004mj, Buchmuller:2005eh, Davidson:2008bu, DiBari:2012fz, Fong:2012buy, doi:10.1142/10447}
%%%%%%%%%%%%%%%%%%%%%%%%%%%%%%%%%%%%%%%%%%%%%%%%%%%%%%%%%%%%%%%%%%%%%%%%%%%%%%
\begin{align}
\mathcal{L}_{\rm lepto} &= \mathcal{L}_\nu
+ \mathcal{L}_{\phi N N}
+ \mathcal{L}_{\phi H H},
\label{eq:Lepto-lagrangian}
\end{align}
%%%%%%%%%%%%%%%%%%%%%%%%%%%%%%%%%%%%%%%%%%%%%%%%%%%%%%%%%%%%%%%%%%%%%%%%%%%%%%
where
%%%%%%%%%%%%%%%%%%%%%%%%%%%%%%%%%%%%%%%%%%%%%%%%%%%%%%%%%%%%%%%%%%%%%%%%%%%%%%
\begin{equation}
\mathcal{L}_{\phi N N} = 
- y_{N_j} \phi \bar{N}_j N_j ,
\qquad
\mathcal{L}_{\phi H H} = 
- \lambda_{12} \phi H^\dagger H 
- \frac{1}{2} \lambda_{22} \phi^2 H^\dagger H 
 \, .
\end{equation}
%%%%%%%%%%%%%%%%%%%%%%%%%%%%%%%%%%%%%%%%%%%%%%%%%%%%%%%%%%%%%%%%%%%%%%%%%%%%%%
The reheating temperature $T_{\rm rh}$ is primarily determined by 
the dominant decay channel $\phi \to H^\dagger H$, as given in 
Eq.~\eqref{tre_metric}.

The partial decay width for $\phi \to N_j N_j$ (analogous to the 
vector-like DM case) is~\cite{Ghoshal:2024ycp}
%%%%%%%%%%%%%%%%%%%%%%%%%%%%%%%%%%%%%%%%%%%%%%%%%%%%%%%%%%%%%%%%%%%%%%%%%%%%%%
\begin{equation}
\label{eq:Gamma:phi-to-NN}
\Gamma_{\phi \to N_j N_j} 
\simeq 
\frac{y_{N_j}^2 \, m_\phi}{4\pi}.
\end{equation}
%%%%%%%%%%%%%%%%%%%%%%%%%%%%%%%%%%%%%%%%%%%%%%%%%%%%%%%%%%%%%%%%%%%%%%%%%%%%%%
The dynamics of non-thermal leptogenesis depend on the competition between the RHN decay rate~\cite{Zhang:2023oyo}
%%%%%%%%%%%%%%%%%%%%%%%%%%%%%%%%%%%%%%%%%%%%%%%%%%%%%%%%%%%%%%%%%%%%%%%%%%%%%%
\begin{equation}
    \Gamma_{N_1}=\frac{1}{8\pi} \left(Y^\dagger Y\right)_{11} M_{N_1}\,,
\end{equation}
%%%%%%%%%%%%%%%%%%%%%%%%%%%%%%%%%%%%%%%%%%%%%%%%%%%%%%%%%%%%%%%%%%%%%%%%%%%%%%
and the Hubble expansion parameter \(H_R\) (evaluated assuming a standard radiation-dominated Universe), typically parameterized by the decay parameter~\cite{Zhang:2023oyo}
%%%%%%%%%%%%%%%%%%%%%%%%%%%%%%%%%%%%%%%%%%%%%%%%%%%%%%%%%%%%%%%%%%%%%%%%%%%%%%
\begin{equation}
    K\equiv \frac{\Gamma_{N_1}}{H_R(M_{N_1})}\,.
\end{equation}
%%%%%%%%%%%%%%%%%%%%%%%%%%%%%%%%%%%%%%%%%%%%%%%%%%%%%%%%%%%%%%%%%%%%%%%%%%%%%%
While neutrino oscillation data naturally point to the strong coupling regime (\(K\gg1\)), where standard thermal leptogenesis would typically suffer from severe efficiency loss due to inverse decays, our non-thermal setup safely avoids this~\cite{Ghoshal:2022fud, Zhang:2023oyo}. Because we strictly enforce the kinematic bound \(M_{N_1}>T_{\rm max}\), the thermal bath does not attain sufficient energy to regenerate RHNs. Consequently, inverse decays and associated washout processes are exponentially suppressed. The \(N_1\) population produced directly from inflaton decays subsequently decays entirely out of equilibrium. This strongly non-thermal environment prevents thermalization and rigorously justifies our use of the direct analytical yield approximation to calculate the final lepton asymmetry.

We focus on the lightest RHN $N_1$, 
assuming a hierarchical mass spectrum 
$M_{N_1} \ll M_{N_2}, M_{N_3}$. 
In this regime, any asymmetry generated by the heavier states 
$N_{2,3}$ is efficiently erased by $N_1$-mediated washout 
processes~\cite{doi:10.1142/10447}. 
The relevant decays of $N_1$ are~\cite{Asaka:1999yd, Fong:2012buy, Co:2022bgh}
%%%%%%%%%%%%%%%%%%%%%%%%%%%%%%%%%%%%%%%%%%%%%%%%%%%%%%%%%%%%%%%%%%%%%%%%%%%%%%
\begin{equation}
N_1 \to L_\alpha + H, 
\qquad 
N_1 \to \bar{L}_\alpha + \tilde{H},
\end{equation}
%%%%%%%%%%%%%%%%%%%%%%%%%%%%%%%%%%%%%%%%%%%%%%%%%%%%%%%%%%%%%%%%%%%%%%%%%%%%%%
which generate a CP asymmetry defined by~\cite{Asaka:1999yd, Hamaguchi:2002vc, Fukuyama:2005us}
%%%%%%%%%%%%%%%%%%%%%%%%%%%%%%%%%%%%%%%%%%%%%%%%%%%%%%%%%%%%%%%%%%%%%%%%%%%%%%
\begin{align}
\label{eq:epsilon-lepto-def}
\epsilon_1^{\rm lep} =
\frac{\Gamma(N_1 \to L_\alpha H)
-
\Gamma(N_1 \to \bar{L}_\alpha \tilde{H})}
{\Gamma(N_1 \to L_\alpha H)
+
\Gamma(N_1 \to \bar{L}_\alpha \tilde{H})}.
\end{align}
%%%%%%%%%%%%%%%%%%%%%%%%%%%%%%%%%%%%%%%%%%%%%%%%%%%%%%%%%%%%%%%%%%%%%%%%%%%%%%
At the tree level, the two decay rates are equal, 
and CP violation arises from the interference between 
tree-level and one-loop amplitudes. 
Including the vertex and self-energy corrections~\cite{Asaka:1999yd,
Fukuyama:2005us,Hamaguchi:2002vc, SravanKumar:2018tgk}, one obtains
%%%%%%%%%%%%%%%%%%%%%%%%%%%%%%%%%%%%%%%%%%%%%%%%%%%%%%%%%%%%%%%%%%%%%%%%%%%%%%
\begin{equation}
\epsilon_1^{\rm lep} =
-\frac{1}{8\pi}
\frac{1}{(Y^\dagger Y)_{11}}
\sum_{j=2,3}
\operatorname{Im}
\!\left[
\big( (Y^\dagger Y)_{1j} \big)^2
\right]
\left[
\mathcal{G}\!\left(\frac{M_{N_j}^2}{M_{N_1}^2}\right)
+
2\, \mathcal{D}\!\left(\frac{M_{N_j}^2}{M_{N_1}^2}\right)
\right],
\label{epsilon_i-exp}
\end{equation}
%%%%%%%%%%%%%%%%%%%%%%%%%%%%%%%%%%%%%%%%%%%%%%%%%%%%%%%%%%%%%%%%%%%%%%%%%%%%%%
where the loop functions $\mathcal{G}(x)$ and $\mathcal{D}(x)$ represent 
the vertex and self-energy contributions, respectively. 
In the hierarchical limit $M_{N_j} \gg M_{N_1}$, 
the expression simplifies and exhibits the expected 
suppression $\epsilon_1^{\rm lep} \propto M_{N_1}/M_{N_j}$. \(\mathcal{G}(x)\) and \(\mathcal{D}(x)\) are given by
%%%%%%%%%%%%%%%%%%%%%%%%%%%%%%%%%%%%%%%%%%%%%%%%%%%%%%%%%%%%%%%%%%%%%%%%%%%%%%

%%%%%%%%%%%%%%%%%%%%%%%%%%%%%%%%%%%%%%%%%%%%%%%%%%%%%%%%%%%%%%%%%%%%%%%%%%%%%%
\begin{align}
    \mathcal{G}(x) &= \sqrt{x} \left[ -1 + (x+1) \ln \left(1 + \frac{1}{x}\right) \right], \quad \mathcal{D}(x) = \frac{\sqrt{x}}{2(x-1)}.
\end{align}
%%%%%%%%%%%%%%%%%%%%%%%%%%%%%%%%%%%%%%%%%%%%%%%%%%%%%%%%%%%%%%%%%%%%%%%%%%%%%%
For \(x = M_{N_j}^2 / M_{N_1}^2 \gg 1\), \(\mathcal{G}(x) \simeq -3/(2 \sqrt{x})\), \(\mathcal{D}(x) \simeq 1/(2 \sqrt{x})\)~\cite{Hamaguchi:2002vc,Fukuyama:2005us}, yielding~\cite{Ghoshal:2024ycp}
%%%%%%%%%%%%%%%%%%%%%%%%%%%%%%%%%%%%%%%%%%%%%%%%%%%%%%%%%%%%%%%%%%%%%%%%%%%%%%
\begin{align} \label{eq:epsilon-step-1}
    \epsilon_1^{\rm lep} = -\frac{3}{16 \pi} \frac{1}{(Y^\dagger Y)_{11}} \sum_{j=2,3} \operatorname{Im} \left[ \left( (Y^\dagger Y)_{1j} \right)^2 \right] \frac{M_{N_1}}{M_{N_j}}.
\end{align}
%%%%%%%%%%%%%%%%%%%%%%%%%%%%%%%%%%%%%%%%%%%%%%%%%%%%%%%%%%%%%%%%%%%%%%%%%%%%%%
Using the seesaw mechanism and the PMNS parametrization, the CP asymmetry can be approximated as~\cite{Asaka:1999yd}
%%%%%%%%%%%%%%%%%%%%%%%%%%%%%%%%%%%%%%%%%%%%%%%%%%%%%%%%%%%%%%%%%%%%%%%%%%%%%%
\begin{align}
    \epsilon_1^{\rm lep} \simeq 
    \frac{3}{16 \pi} 
    \frac{M_{N_1}}{v^2} 
    m_{\nu_3} \, 
    \delta_{\rm eff},
\end{align}
%%%%%%%%%%%%%%%%%%%%%%%%%%%%%%%%%%%%%%%%%%%%%%%%%%%%%%%%%%%%%%%%%%%%%%%%%%%%%%
where $\delta_{\rm eff}$ denotes the effective CP-violating phase~\cite{Hamaguchi:2002vc}
%%%%%%%%%%%%%%%%%%%%%%%%%%%%%%%%%%%%%%%%%%%%%%%%%%%%%%%%%%%%%%%%%%%%%%%%%%%%%%
\begin{equation}
    \delta_{\rm eff} =
    \frac{
    \operatorname{Im} \!\left[
    (\hat{Y}_{13})^2 
    + \frac{m_{\nu_2}}{m_{\nu_3}} (\hat{Y}_{12})^2 
    + \frac{m_{\nu_1}}{m_{\nu_3}} (\hat{Y}_{11})^2
    \right]
    }{
    |\hat{Y}_{13}|^2 
    + |\hat{Y}_{12}|^2 
    + |\hat{Y}_{11}|^2
    },
\end{equation}
%%%%%%%%%%%%%%%%%%%%%%%%%%%%%%%%%%%%%%%%%%%%%%%%%%%%%%%%%%%%%%%%%%%%%%%%%%%%%%
with $0 \leq \delta_{\rm eff} \leq 1$~\cite{Co:2022bgh}, and $\hat{Y} = U^T Y$.

Assuming the instantaneous decay of the produced $N_1$, the resulting lepton yield is~\cite{Hamaguchi:2002vc}
%%%%%%%%%%%%%%%%%%%%%%%%%%%%%%%%%%%%%%%%%%%%%%%%%%%%%%%%%%%%%%%%%%%%%%%%%%%%%%
\begin{align}
\label{eq:Lepton asymmetry}
    Y_L \equiv \frac{n_L}{s}
    = \epsilon_1^{\rm lep} \frac{n_{N_1}}{s}.
\end{align}
%%%%%%%%%%%%%%%%%%%%%%%%%%%%%%%%%%%%%%%%%%%%%%%%%%%%%%%%%%%%%%%%%%%%%%%%%%%%%%
Using the yields from Eqs.~\eqref{eq:Yield of DM} and~\eqref{eq:Gamma:phi-to-NN}
(adapted for $N_1$), we obtain
%%%%%%%%%%%%%%%%%%%%%%%%%%%%%%%%%%%%%%%%%%%%%%%%%%%%%%%%%%%%%%%%%%%%%%%%%%%%%%
\begin{equation}
    \frac{n_{N_1}}{s} \simeq
    \begin{cases}
    \displaystyle 
    2.3258 \times 10^{-2}
    \frac{M_{\rm Pl}}{T_{\rm rh}}
    y_{N_1}^2,
    & \text{(metric formalism),} \\[6pt]
    \displaystyle
    3.4887 \times 10^{-2}
    \frac{M_{\rm Pl}}{T_{\rm rh}}
    y_{N_1}^2,
    & \text{(Palatini formalism).}
    \end{cases}
\end{equation}
%%%%%%%%%%%%%%%%%%%%%%%%%%%%%%%%%%%%%%%%%%%%%%%%%%%%%%%%%%%%%%%%%%%%%%%%%%%%%%
Electroweak sphaleron processes convert the generated lepton asymmetry into a baryon asymmetry~\cite{SravanKumar:2018tgk,Co:2022bgh}
%%%%%%%%%%%%%%%%%%%%%%%%%%%%%%%%%%%%%%%%%%%%%%%%%%%%%%%%%%%%%%%%%%%%%%%%%%%%%%
\begin{align}
    Y_B = \frac{28}{79} \, Y_L.
\end{align}
%%%%%%%%%%%%%%%%%%%%%%%%%%%%%%%%%%%%%%%%%%%%%%%%%%%%%%%%%%%%%%%%%%%%%%%%%%%%%%
From \emph{Planck} 2018~\cite{Planck:2018vyg,doi:10.1142/10447,Co:2022bgh}
%%%%%%%%%%%%%%%%%%%%%%%%%%%%%%%%%%%%%%%%%%%%%%%%%%%%%%%%%%%%%%%%%%%%%%%%%%%%%%
\begin{align}
\label{eq:YB bound}
    Y_B \equiv \frac{n_B - n_{\bar{B}}}{s}
    = \frac{n_\gamma}{s} \, \eta_B
    \simeq 8.7 \times 10^{-11},
\end{align}
%%%%%%%%%%%%%%%%%%%%%%%%%%%%%%%%%%%%%%%%%%%%%%%%%%%%%%%%%%%%%%%%%%%%%%%%%%%%%%
where $n_\gamma \approx 410.73 \,\text{cm}^{-3}$ and 
$s \approx 2891.2 \,\text{cm}^{-3}$.

Combining the above expressions yields~\cite{Ghoshal:2024ycp}
%%%%%%%%%%%%%%%%%%%%%%%%%%%%%%%%%%%%%%%%%%%%%%%%%%%%%%%%%%%%%%%%%%%%%%%%%%%%%%
\begin{equation}
\label{eq:lepto-condition}
    T_{\rm rh}
    \simeq
    5.6073 \times 10^{9}
    \, y_{N_1}^2 \, M_{N_1}
    \qquad
    (\text{valid for both formalisms}),
\end{equation}
%%%%%%%%%%%%%%%%%%%%%%%%%%%%%%%%%%%%%%%%%%%%%%%%%%%%%%%%%%%%%%%%%%%%%%%%%%%%%%
where we have used $\delta_{\rm eff} = 0.5$ and 
$m_{\nu_3} \approx 0.05 \times 10^{-9} \,\text{GeV}$.

Analogous to our DM analysis, specifically the discussion after Eq.~\eqref{eq:Trh_mchi}, the expression in Eq.~\eqref{eq:lepto-condition} serves as the fundamental relation for mapping the viable phenomenological parameter space in the (\(T_{\rm rh}, M_{N_1}\)) plane, explicitly connecting the reheating temperature to the RHN mass required to reproduce the observed baryon asymmetry. Interestingly, unlike DM yield scaling, Eq.~\eqref{eq:lepto-condition} remains analytically identical across both the metric and Palatini formulations due to compensating effects in the underlying yield derivations. However, the geometric boundaries of the allowed regions in the corresponding figures (namely Fig.~\ref{fig:lepto_1} and Figs.~\ref{fig:lepto_2}--~\ref{fig:lepto_4}) are fundamentally dictated by the maximum permissible values of the Yukawa coupling, \(y_{N_1}\), and the effective kinematic limit \(m_\phi < 2M_{N_1}\). Because these stringent \(y_{N_1}\) bounds are derived from the CW radiative stability conditions (Eqs.~\eqref{Eq:first derivative test1},~\eqref{Eq:second derivative test1}, and~\eqref{eq:constraints}), which are highly sensitive to the distinct canonical field excursions, they exhibit strong formalism dependence. Consequently, despite the Universe scaling in Eq.~\eqref{eq:lepto-condition}, it is the application of the tighter Palatini-specific CW bounds and shifted kinematic threshold that ultimately shapes the significantly restricted viable parameter spaces observed in the Palatini framework compared to the metric case.

For $10^{-3.5} \gtrsim y_{N_1} \gtrsim 10^{-12}$, non-thermal leptogenesis successfully reproduces the observed baryon asymmetry of the Universe (BAU), while remaining consistent with the inflationary benchmark parameters and reheating constraints. These parameter ranges incorporate checks ensuring minimal washout and negligible entropy dilution from late RHN decays, as confirmed numerically in the relevant regimes. Consequently, the predicted $Y_B$ agrees with observations without requiring fine-tuning of the effective CP phase $\delta_{\rm eff}$.

The non-thermal leptogenesis framework, based on inflaton decays into RHNs (Eq.~\eqref{eq:Gamma:phi-to-NN}) followed by their CP-violating decays (Eq.~\eqref{epsilon_i-exp}), successfully reproduces the observed BAU,
$Y_B \simeq 8.7 \times 10^{-11}$ (Eq.~\eqref{eq:YB bound}),
while simultaneously accommodating light neutrino masses via the seesaw mechanism (Eq.~\eqref{mnu1}). 
The reheating relation in Eq.~\eqref{eq:lepto-condition}, derived from yields analogous to Eq.~\eqref{eq:Yield of DM}, remains valid across both metric and Palatini formalisms due to compensating effects: matter-like inflaton oscillations in the metric case versus radiation-like scaling in the Palatini case.

Illustrative benchmarks shown in Fig.~\ref{fig:lepto_1} and Figs.~\ref{fig:lepto_2}--~\ref{fig:lepto_4}
(e.g., $n=0.5$, $\xi=0.1$ for linear models; $n=5$, $\xi=0.1$ for small-field models), based on the inflationary parameter sets of Tabs.~\ref{tab:infl-para_1}--\ref{tab:infl-para_4}, exhibit viable regions in the $(T_{\rm rh}, M_{N_1})$ plane.
These regions are constrained by:
radiative stability (cyan regions from $\lambda_{12}^{\max}$, with $y_{N_1}^{\max} \lesssim 10^{-6}$--$10^{-4}$ as given in Tabs.~\ref{Table:stability0m}--\ref{Table:stability3}),
kinematic bounds $m_\phi < 2 M_{N_1}$ (brown regions for $\varrho = 10,100$, see Eqs.~\eqref{eq:baremass-metric}--\eqref{eq:bar mass limit Trh}), and
BBN constraints (gray).
These exclusion regions were discussed in detail in the previous section.

Additional lower bounds $M_{N_1} \gtrsim 10^{3}\,\mathrm{GeV}$ (indigo) ensure sufficiently large $|\epsilon_1^{\rm lep}|$ without requiring fine-tuning of $\delta_{\rm eff}$, while $M_{N_1} \gtrsim 10^{9}\,\mathrm{GeV}$ (magenta) prevents strong washout at high $T_{\rm rh}$ and avoids gravitino overproduction.
Overall, the plots clarify the dynamical differences between formalisms: 
the metric case allows broader parameter coverage ($M_{N_1} \lesssim 10^{14}\,\mathrm{GeV}$, $T_{\rm rh} \lesssim 10^{16}\,\mathrm{GeV}$), with predominantly vertical exclusion boundaries, particularly from the $m_\phi < 2 M_{N_1}$ condition. 
In contrast, the Palatini formulation produces sloped wedge-like regions due to the explicit $T_{\rm rh}$-dependence of $m_b$ (see Eq.~\eqref{eq:bar mass limit Trh}), significantly narrowing the viable space. 
In several cases, strict radiative bounds on $y_{N_1}$, amplified by larger field excursions (cf. Tabs.~\ref{Table:stability0} vs.~\ref{Table:stability0m}), can eliminate otherwise allowed regions.

Across the models, large-field linear potentials (Eqs.~\eqref{eq:U1_Palatini}--\eqref{eq:U2_Palatini}; Figs.~\ref{fig:lepto_1} and~\ref{fig:lepto_2}; Tabs.~\ref{tab:infl-para_1}--\ref{tab:infl-para_2}) favor higher $M_{N_1}$ and $T_{\rm rh}$.
This behavior follows from their larger inflationary scale $H_I$ (Eq.~\eqref{eq:HI}) and higher $T_{\rm max}$ (Figs.~\ref{fig:tmax_1} and~\ref{fig:tmax_2}), consistent with super-Planckian initial field values ($\phi_i \sim 10$--$570\,M_{\rm Pl}$), spectral index $n_s \sim 0.97$--$0.98$, and tensor-to-scalar ratio $r \sim 0.02$--$0.08$.  By contrast, the small-field scenarios, including the Brans–Dicke-like and Higgs-like models (Eqs.~\eqref{eq:3_potential}--\eqref{eq:4_potential}; Figs.~\ref{fig:lepto_3}--\ref{fig:lepto_4}; Tabs.~\ref{tab:infl-para_3}--\ref{tab:infl-para_4}), restrict the parameter space to
$M_{N_1} \lesssim 10^{12}\,\mathrm{GeV}$ and
$T_{\rm rh} \lesssim 10^{13}\,\mathrm{GeV}$.
This reflects their sub-Planckian excursions ($\Delta\phi < 3\,M_{\rm Pl}$), comparatively weaker stability bounds (e.g., $\lambda_{12} \lesssim 10^{-6}$ in Tab.~\ref{Table:stability3m}), and suppressed tensor amplitude ($r < 10^{-4}$) with $n_s \sim 0.95$--$0.96$.

In all cases, the Palatini formulation yields systematically tighter viable regions than the metric case, mirroring the trends observed in the DM analysis (Figs.~\ref{fig:dm_1}-\ref{fig:dm_4}). 
This reduction arises from the combined effects of stronger radiative constraints 
(Tabs.~\ref{Table:stability0}--\ref{Table:stability3}) and radiation-like dilution, 
which accelerates reheating and thereby enhances the overlap of exclusion boundaries.

This interplay highlights the distinct roles of the metric and Palatini formalisms in connecting inflationary cosmology with particle-physics phenomenology. In particular, the Palatini framework exhibits notable advantages in the UV regime, as emphasized in the preceding sections. Its improved high-energy behavior, specifically the flatness of the effective potential, and enhanced radiative control strengthen theoretical consistency, particularly in the plateau-like scenarios. However, this UV robustness is accompanied by reduced flexibility in post-inflationary dynamics. In the Palatini case, radiation-like inflaton oscillations lead to a more rapid energy transfer into the thermal bath, yielding higher $T_{\rm rh}$ for fixed couplings. This behavior amplifies loop effects in stability analyses, resulting in more restrictive bounds on $y_{N_1}^{\max}$ and $\lambda_{12}^{\max}$ (see Tabs.~\ref{Table:stability0}--\ref{Table:stability3}). Furthermore, the explicit $T_{\rm rh}$-dependence of the inflaton mass (Eq.~\eqref{eq:bar mass limit Trh}) induces sloped exclusion boundaries in parameter space, frequently reducing the viable region for non-thermal leptogenesis relative to the metric formulation. 

By contrast, the metric case permits broader parameter coverage, extending up to $M_{N_1} \sim 10^{14}\,\mathrm{GeV}$ and $T_{\rm rh} \sim 10^{16}\,\mathrm{GeV}$ in representative benchmarks. This enlarged viable domain facilitates richer low-energy connections, linking CMB observables (Sec.~\ref{Sec:Slow_roll}), reheating dynamics (Sec.~\ref{Sec:reheating}), DM production (Subsec.~\ref{subsec:dm_decay}), and baryogenesis, without requiring a preheating stage (Subsec.~\ref{preheating}).

Overall, the comparison reveals a clear dichotomy: the Palatini formulation emphasizes 
high-energy theoretical consistency, whereas the metric formulation allows greater 
phenomenological flexibility in the post-inflationary epoch. Discriminating between these 
gravitational frameworks, therefore, require future observational probes, including searches 
for primordial gravitational waves and precision neutrino measurements, which could test 
their distinct predictions across both cosmological and particle-physics scales.

%%%%%%%%%%%%%%%%%%%%%%%%%%%%%%%%%%%%%%%%%%%%%%%%%%%%%%%%%%%%%%%%%%%%%%%%%%%%%%
%%%%%%%%%%%%%%%%%%%%%%%%%%%%%%%%%%%%%%%%%%%%%%%%%%%%%%%%%%%%%%%%%%%%%%%%%%%%%%
%%%%%%%%%%%%%%%%%%%%%%%%%%%%%%%%%%%%%%%%%%%%%%%%%%%%%%%%%%%%%%%%%%%%%%%%%%%%%%
\section{Discussion and Conclusions}
\label{sec:conclusion}
%%%%%%%%%%%%%%%%%%%%%%%%%%%%%%%%%%%%%%%%%%%%%%%%%%%%%%%%%%%%%%%%%%%%%%%%%%%%%%
%%%%%%%%%%%%%%%%%%%%%%%%%%%%%%%%%%%%%%%%%%%%%%%%%%%%%%%%%%%%%%%%%%%%%%%%%%%%%%
%%%%%%%%%%%%%%%%%%%%%%%%%%%%%%%%%%%%%%%%%%%%%%%%%%%%%%%%%%%%%%%%%%%%%%%%%%%%%%
\subsection{Discussion}
In this work, we have established a unified cosmological framework 
based on induced multi-phase inflation, exploring linear, 
Brans--Dicke-like, and Higgs-like potentials within both metric 
and Palatini formulations of gravity. A pivotal finding of our 
analysis is the role of the gravitational formalism as a 
``phenomenological filter''.

In particular, the Palatini formulation, where the affine 
connection is treated as an independent variable, effectively 
flattens the Einstein-frame potential. This mechanism preserves 
UV consistency by keeping field excursions sub-Planckian, even 
for relatively large non-minimal couplings, thereby alleviating the tension with the most stringent Swampland distance conjectures 
that typically challenge metric realizations.

We further demonstrated that the post-inflationary evolution of the homogeneous inflaton condensate 
is governed by perturbative inflaton decays. Within the radiatively stable parameter space 
required to preserve the flatness of the inflationary plateau, reheating proceeds through a 
controlled and calculable approach. This establishes a direct connection between the 
inflationary scale, the reheating temperature, and low-energy observables.

Finally, we identify a compelling direction for future research, which we term “Affine Reheating.” 
Throughout this work, we have assumed that matter fields couple exclusively to the metric. 
However, within a fully general metric-affine gravity framework, fermions, such as DM or RHNs, may 
couple directly to the independent connection. Such interactions could open novel reheating channels, 
potentially modifying relic abundance predictions and providing new observational handles to 
distinguish between competing gravitational descriptions of the early Universe. This possibility 
warrants dedicated future study.

%%%%%%%%%%%%%%%%%%%%%%%%%%%%%%%%%%%%%%%%%%%%%%%%%%%%%%%%%%%%%%%%%%%%%%%%%%%%%%
%%%%%%%%%%%%%%%%%%%%%%%%%%%%%%%%%%%%%%%%%%%%%%%%%%%%%%%%%%%%%%%%%%%%%%%%%%%%%%
%%%%%%%%%%%%%%%%%%%%%%%%%%%%%%%%%%%%%%%%%%%%%%%%%%%%%%%%%%%%%%%%%%%%%%%%%%%%%%
\subsection{Conclusions}
We have constructed a consistent cosmological history from inflation 
to baryogenesis. The core results of our analysis are summarized below:
%%%%%%%%%%%%%%%%%%%%%%%%%%%%%%%%%%%%%%%%%%%%%%%%%%%%%%%%%%%%%%%%%%%%%%%%%%%%%%
\begin{itemize}
%%%%%%%%%%%%%%%%%%%%%%%%%%%%%%%%%%%%%%%%%%%%%%%%%%%%%%%%%%%%%%%%%%%%%%%%%%%%%%
\item \textbf{Inflationary Observables:} 
The models predict a scalar spectral index in the range 
$n_s \simeq 0.93$--$0.98$, consistent with 
\emph{Planck}~\cite{Planck:2018jri} and 
\emph{Planck}+ACT~\cite{ACT:2025tim} data. 
The gravitational formulation plays a decisive role in determining 
the tensor-to-scalar ratio. While metric scenarios allow 
$r \lesssim 0.08$, the Palatini formalism generically suppresses them to \(r< 10^{-5}\). The analytical behavior of \(r\) is derived across all models in Eqs.~\eqref{eq:r_1}, \eqref{eq:r_2}, \eqref{eq:r_3}, and~\eqref{eq:r_4}, and the resulting (\(n_s,\,r\)) parameter spaces are illustrated extensively for the linear (Figs.~\ref{fig:r_ns_1_m}--~\ref{fig:r_ns_2_P}), Brans-Dicke-like (Figs.~\ref{fig:r_ns_3_m}--~\ref{fig:r_ns_3_P}), and Higgs-like (Figs.~\ref{fig:r_ns_4_m}--~\ref{fig:r_ns_4_P}) potentials, with specific numerical benchmarks summarized in Tabs.~\ref{tab:infl-para_1}--~\ref{tab:infl-para_4}. Consequently, while the metric predictions fall within the reach of future experiments like SO~\cite{SimonsObservatory:2018koc} and LiteBIRD~\cite{Hazumi_2020} (which target \(r\sim 10^{-3}\)), the Palatini predictions sit securely below these projected detection thresholds.
%%%%%%%%%%%%%%%%%%%%%%%%%%%%%%%%%%%%%%%%%%%%%%%%%%%%%%%%%%%%%%%%%%%%%%%%%%%%%%
\item \textbf{Radiative Stability and Reheating:} 
Maintaining the inflationary plateau imposes stringent upper bounds 
on the inflaton couplings to both the dark sector ($y_\chi$) and 
the Higgs portal ($\lambda_{12}$), typically in the range 
$\mathcal{O}(10^{-7})$--$\mathcal{O}(10^{-3})$ 
(see Tabs.~\ref{Table:stability0m}--\ref{Table:stability3}). 
These limits translate into a reheating temperature window 
$4\,\text{MeV} \lesssim T_{\rm rh} \lesssim 10^{15}\,\text{GeV}$, 
illustrated in Figs.~\ref{fig:tmax_1}--\ref{fig:tmax_4}. 
Palatini models, characterized by modified canonical dynamics, 
require tighter coupling bounds and therefore exhibit a narrower 
viable reheating window compared to the metric case.
%%%%%%%%%%%%%%%%%%%%%%%%%%%%%%%%%%%%%%%%%%%%%%%%%%%%%%%%%%%%%%%%%%%%%%%%%%%%%%
\item \textbf{Dark Matter:} 
We identify a viable parameter space for non-thermal DM 
production, accommodating fermion masses from the warm DM limit, constrained by Lyman-\(\alpha\) forest observations via Eq.~\eqref{eq:Lyman-alpha} up to the PeV scale \(m_\chi \sim 10^6 \, \text{GeV}\). The relic abundance is generated exclusively through perturbative inflaton decays (Eq.~\eqref{eq:decay_chi}), requiring highly suppressed Yukawa couplings \(y_\chi \lesssim\mathcal{O}(10^{-4}-10^{-3})\) to preserve the radiative stability of the inflationary potential, which is analyzed in detail in Subsec.~\ref{Subsec:pert_decay}. The resulting DM yield \(Y_\chi\) (Eq.~\eqref{eq:Yield of DM}), matched against the observed present-day cold DM abundance (Eq.~\eqref{Eq:present day CDM yield}), is highly sensitive to the underlying gravitational formulation. Because the Palatini formalism induces radiation-like inflaton oscillations, it alters the effective equation of state during reheating and dynamically modifies the yield coefficient as shown in Eq.~\eqref{eq:Trh_mchi}.  Crucially, the relation shown in Eq.~\ref{eq:Trh_mchi} is explicitly formulation-dependent; when this dynamic scaling is combined with the distinct CW radiative constraints (Eqs.~\eqref{Eq:first derivative test1},~\eqref{Eq:second derivative test1}, and~\eqref{eq:constraints}) on \(y_\chi\), it directly dictates the significantly narrower viable (\(T_{\rm rh}, m_\chi\)) parameters spaces observed in the Palatini framework compared to its metric counterpart.  Furthermore, these constraints vary markedly across the specific inflationary potentials; while large-field linear models readily accommodate PeV-scale DM at high reheating temperature, the small-field plateau scenarios (Brans-Dicke-like and Higgs-like) tightly restrict the viable space to lighter masses (\(m_\chi \lesssim 10^8\,\text{GeV}\))  due to their heightened sensitivity to quantum corrections, as extensively mapped across all models in Fig.~\ref{fig:dm_1} and Figs.~\ref{fig:dm_2}--~\ref{fig:dm_4}.
%%%%%%%%%%%%%%%%%%%%%%%%%%%%%%%%%%%%%%%%%%%%%%%%%%%%%%%%%%%%%%%%%%%%%%%%%%%%%%
\item \textbf{Leptogenesis:} 
The framework successfully accommodates non-thermal leptogenesis 
via the Type-I seesaw mechanism (Eq.~\eqref{mnu1}), driven by the CP-violating decays of the lightest RHN (Eq.~\eqref{epsilon_i-exp}). By strictly enforcing the kinematic blocking condition \(M_{N_1}>T_{\rm max}\), the thermal bath is prevented from regenerating \(N_1\) states, which exponentially suppresses washout processes and rigorously justifies the direct analytical yield approximation (Eq.~\eqref{eq:Lepton asymmetry}). We reproduce the observed baryon asymmetry \(Y_B \simeq 8.7\times10^{-11}\) (Eq.~\eqref{eq:YB bound}) for inflaton-neutrino Yukawa couplings in the range \(10^{-12}\lesssim y_{N_1}\lesssim 10^{-3.5}\). This establishes a robust analytical constraint connecting the reheating temperature to the RHN mass as shown in Eq.~\eqref{eq:lepto-condition}. Crucially, although this relation connecting the reheating temperature to the RHN mass is universally valid across both formalisms, its combination with the strictly formulation-dependent CW radiative constraints on \(y_{N_1}\) and the modified kinematic boundaries directly dictates the significantly narrower (\(T_{\rm rh}, M_{N_1}\)) parameter spaces observed in the Palatini framework compared to the metric case. Successful baryogenesis is achieved within a broad mass window of \(10^9 \, \text{GeV} \lesssim M_{N_1}\lesssim 10^{14}\,\text{GeV}\), comfortably avoiding CP-phase fine-tuning at the lower end, while bounded from above by the kinematic limit \(m_\phi > 2M_{N_1}\). As observed in the DM analysis, the Palatini formulation significantly narrows the viable (\(T_{\rm rh},\, M_{N_1}\)) parameter space compared to the metric counterparts. This available parameter space is also highly sensitive to the choice of the inflationary potential, exhibiting a behavioral trend similar to that discussed in the DM analysis as comprehensively illustrated in Fig.~\ref{fig:lepto_1} and Figs.~\ref{fig:lepto_2}--~\ref{fig:lepto_4}.
\end{itemize}
%%%%%%%%%%%%%%%%%%%%%%%%%%%%%%%%%%%%%%%%%%%%%%%%%%%%%%%%%%%%%%%%%%%%%%%%%%%%%%
In summary, induced multi-phase  inflation provides a robust and 
self-consistent framework that unifies inflation, DM 
production, and baryogenesis within a single cosmological setup. 
The distinct phenomenological signatures of the metric and Palatini 
formulations, particularly in the tensor-to-scalar ratio and in the allowed dark sector parameter space, offer concrete targets 
for next-generation cosmological observations aimed at probing the 
fundamental nature of gravity.
%%%%%%%%%%%%%%%%%%%%%%%%%%%%%%%%%%%%%%%%%%%%%%%%%%%%%%%%%%%%%%%%%%%%%%%%%%%%%%
%%%%%%%%%%%%%%%%%%%%%%%%%%%%%%%%%%%%%%%%%%%%%%%%%%%%%%%%%%%%%%%%%%%%%%%%%%%%%%
%%%%%%%%%%%%%%%%%%%%%%%%%%%%%%%%%%%%%%%%%%%%%%%%%%%%%%%%%%%%%%%%%%%%%%%%%%%%%%
\acknowledgments
A.G. thanks University of Warsaw where a major part of the project was carried out.
\clearpage\newpage
\appendix
%%%%%%%%%%%%%%%%%%%%%%%%%%%%%%%%%%%%%%%%%%%%%%%%%%%%%%%%%%%%%%%%%%%%%%%%%%%%%%
%%%%%%%%%%%%%%%%%%%%%%%%%%%%%%%%%%%%%%%%%%%%%%%%%%%%%%%%%%%%%%%%%%%%%%%%%%%%%%
%%%%%%%%%%%%%%%%%%%%%%%%%%%%%%%%%%%%%%%%%%%%%%%%%%%%%%%%%%%%%%%%%%%%%%%%%%%%%%
\section{Supplementary Inflationary Observables}
\label{app:Inflationaries}
In this appendix, we present additional plots for the inflationary observables, including the spectral running \(\alpha_s\) and the initial field value \(\phi_i\) as functions of the non-minimal coupling \(\xi\).  Specifically, these figures correspond to the linear potentials (Eqs.~\eqref{eq:U1_Palatini} and~\eqref{eq:U2_Palatini}), the Brans-Dicke-like model (Eq.~\eqref{eq:3_potential}), and the Higgs-like model (Eq.~\eqref{eq:4_potential}), respectively, evaluated across both the metric and Palatini formalisms. These complement the primary figures in the main text, which focus on \(r\) versus \(n_s\) and \(r\) versus \(\xi\).
%%%%%%%%%%%%%%%%%%%%%%%%%%%%%%%%%%%%%%%%%%%%%%%%%%%%%%%%%%%%%%%%%%%%%%%%%%%%%%
\begin{figure}[H]
\centering
\includegraphics[width=0.46\linewidth]{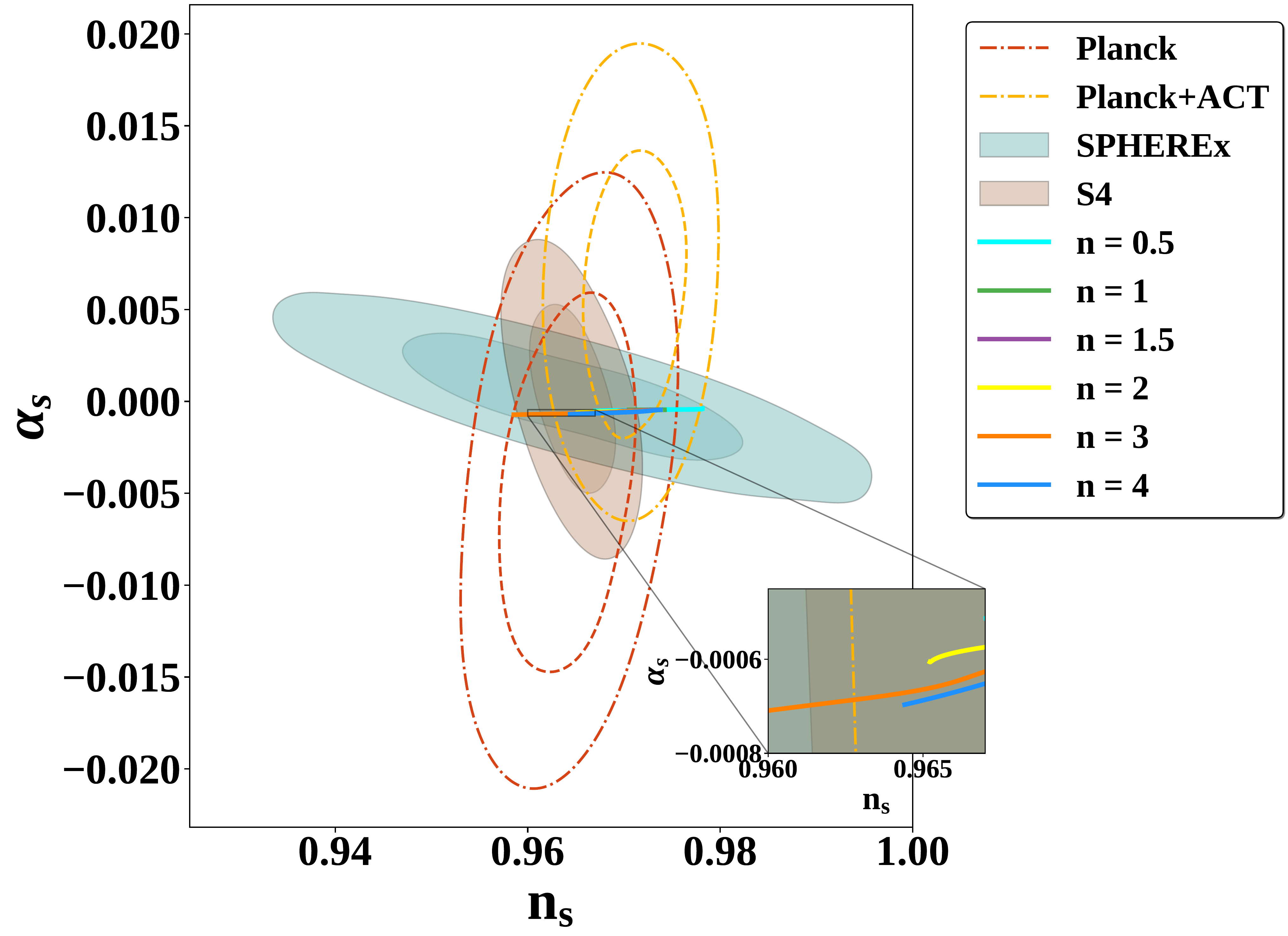}\hfill
\includegraphics[width=0.5\linewidth]{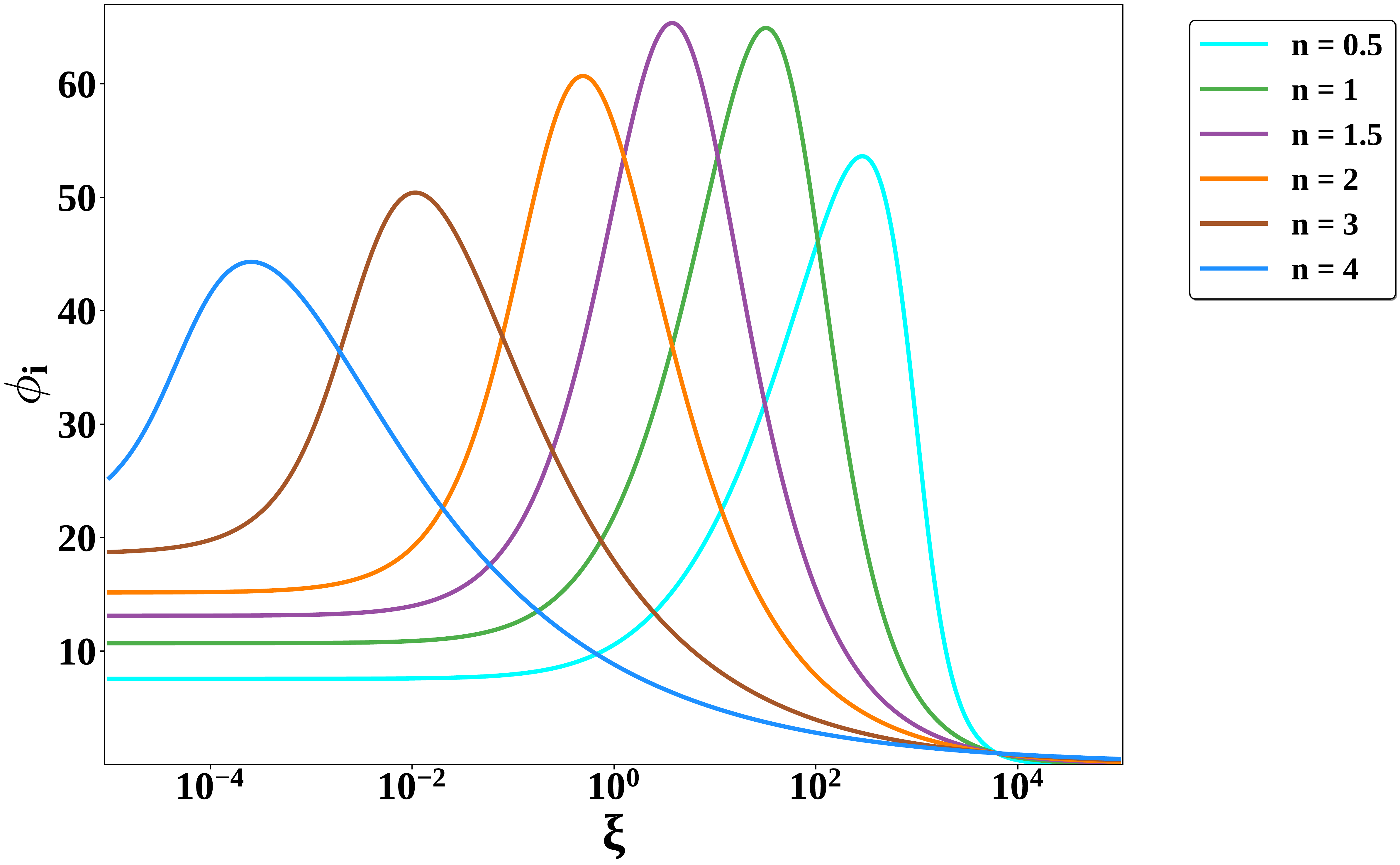}
\caption{\it Supplementary inflationary observables for the linear inflation model defined by \eqref{eq:U1_Palatini} in the \textbf{metric} formalism: spectral running \(\alpha_s\) versus \(\xi\) (left) and initial field value \(\phi_i\) versus \(\xi\) (right).}
\label{fig:supp_1_m}
\end{figure}
%%%%%%%%%%%%%%%%%%%%%%%%%%%%%%%%%%%%%%%%%%%%%%%%%%%%%%%%%%%%%%%%%%%%%%%%%%%%%%
\begin{figure}[H]
\centering
\includegraphics[width=0.46\linewidth]{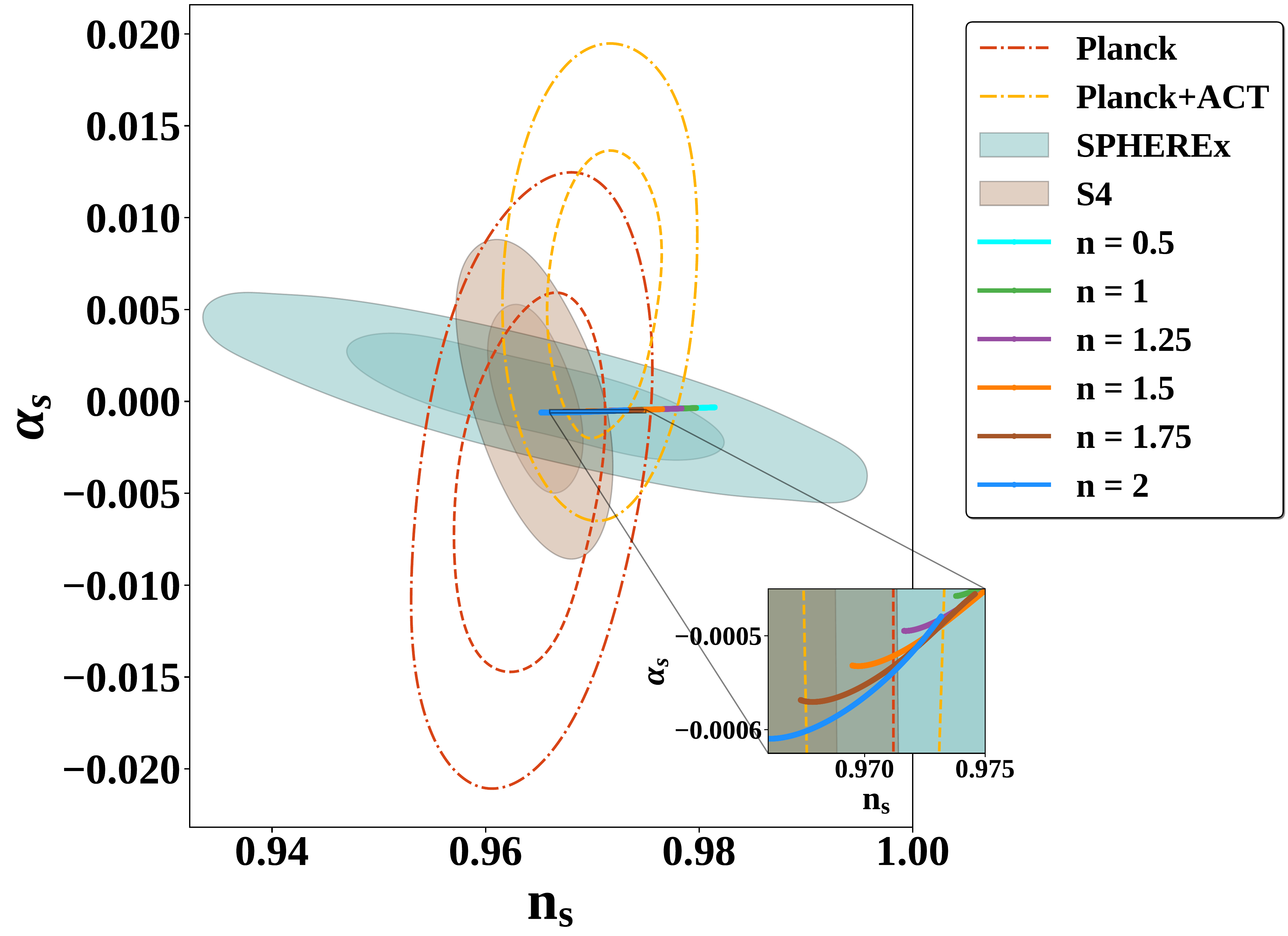}\hfill
\includegraphics[width=0.5\linewidth]{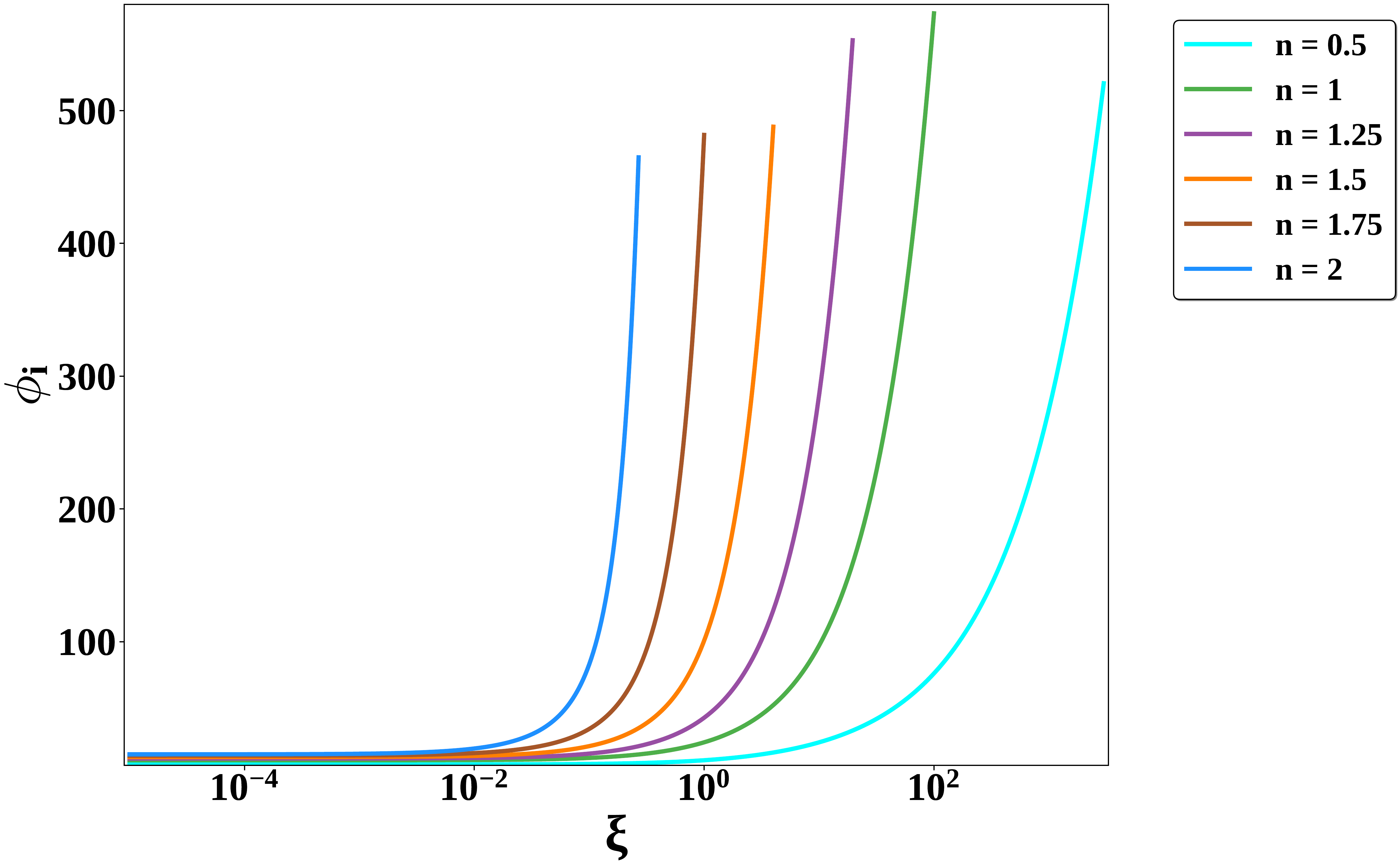}
\caption{\it Same as Fig.~\ref{fig:supp_1_m} but in the \textbf{Palatini} formalism.}
\label{fig:supp_1_P}
\end{figure}
%%%%%%%%%%%%%%%%%%%%%%%%%%%%%%%%%%%%%%%%%%%%%%%%%%%%%%%%%%%%%%%%%%%%%%%%%%%%%%
\begin{figure}[t!]
\centering
\includegraphics[width=0.46\linewidth]{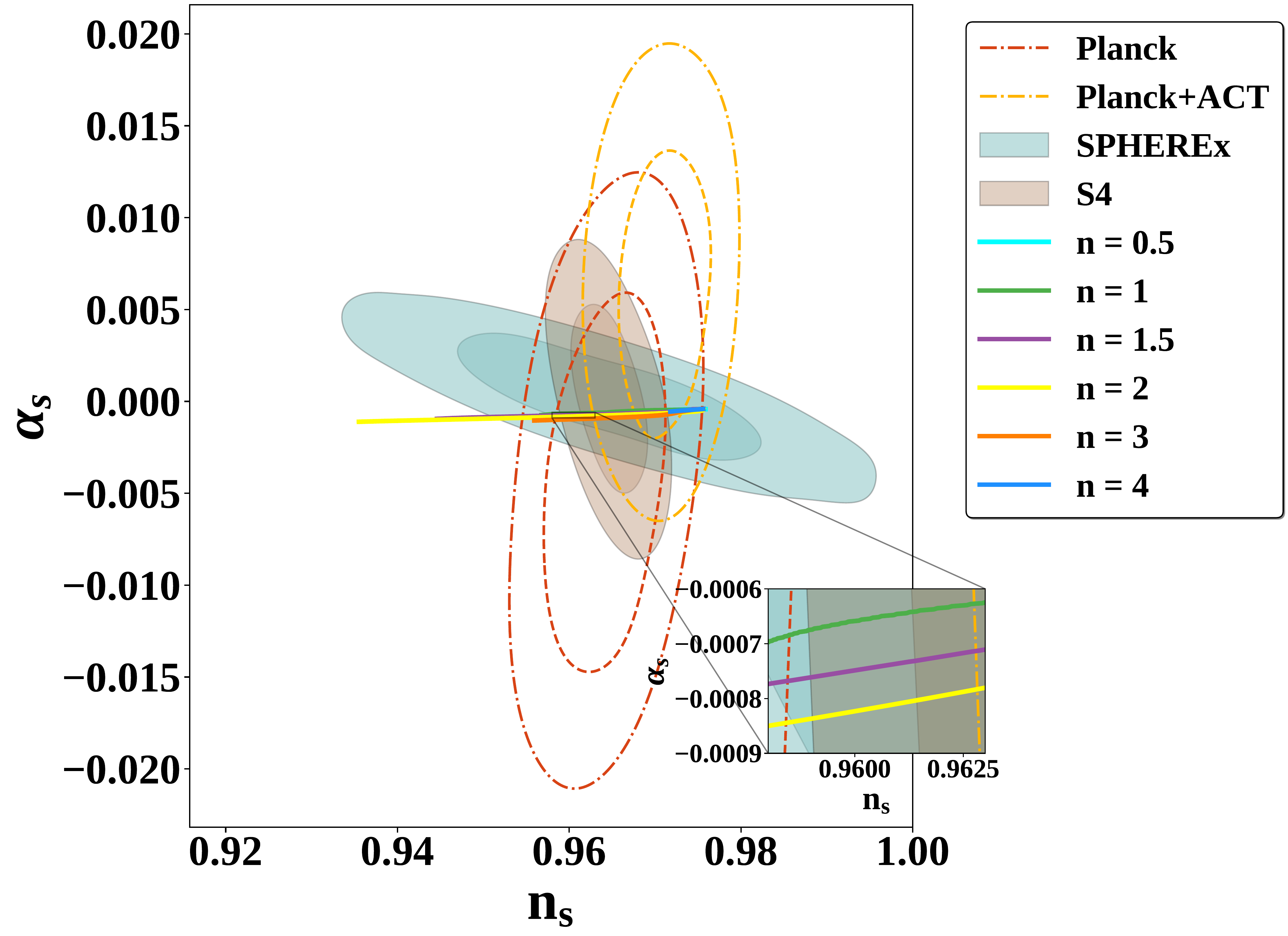}\hfill
\includegraphics[width=0.5\linewidth]{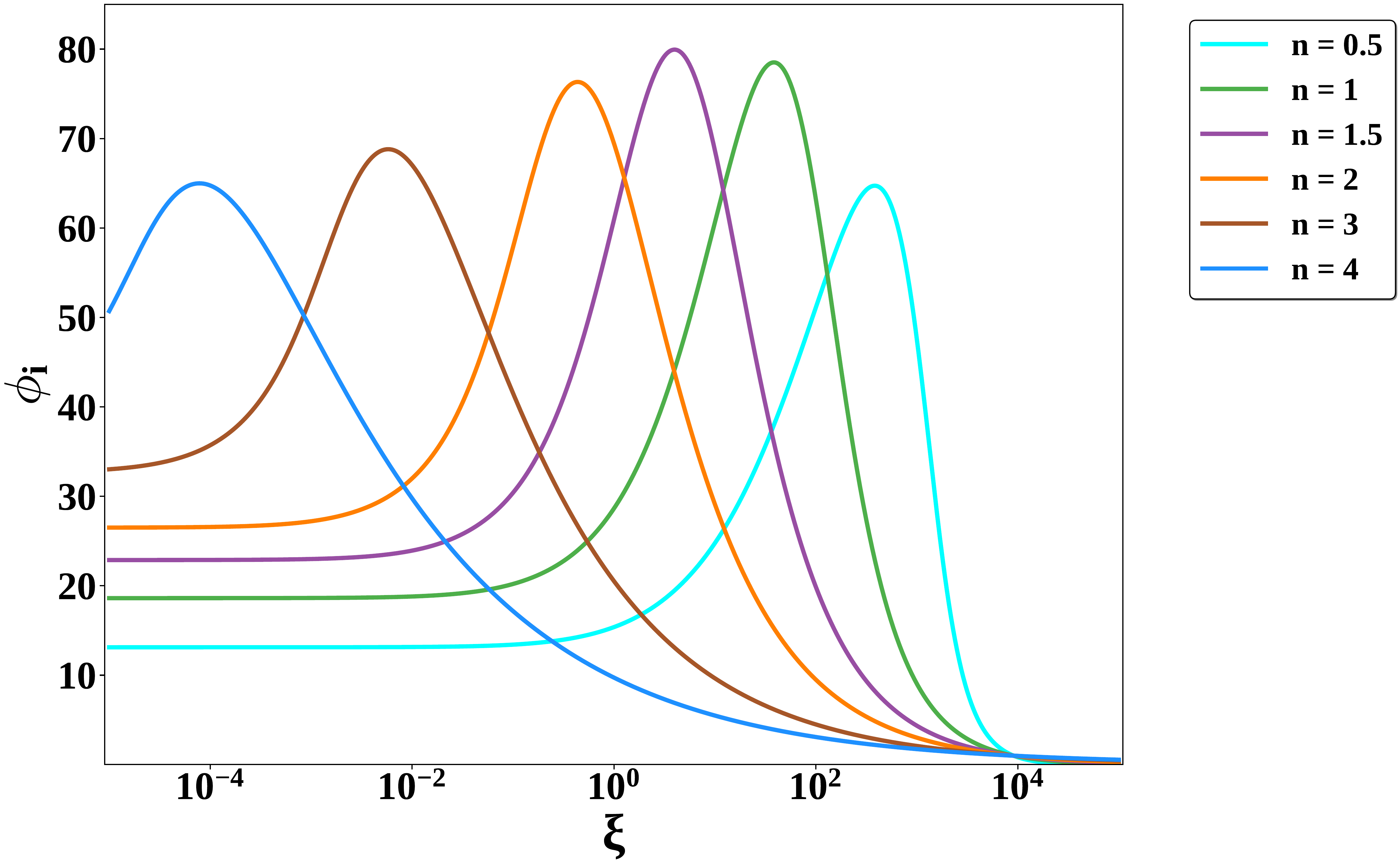}
\caption{\it Supplementary inflationary observables for the linear inflation model defined by \eqref{eq:U2_Palatini} in the \textbf{metric} formalism: spectral running \(\alpha_s\) versus \(\xi\) (left) and initial field value \(\phi_i\) versus \(\xi\) (right).}
\label{fig:supp_2_m}
\end{figure}
%%%%%%%%%%%%%%%%%%%%%%%%%%%%%%%%%%%%%%%%%%%%%%%%%%%%%%%%%%%%%%%%%%%%%%%%%%%%%%
\begin{figure}[t!]
\centering
\includegraphics[width=0.46\linewidth]{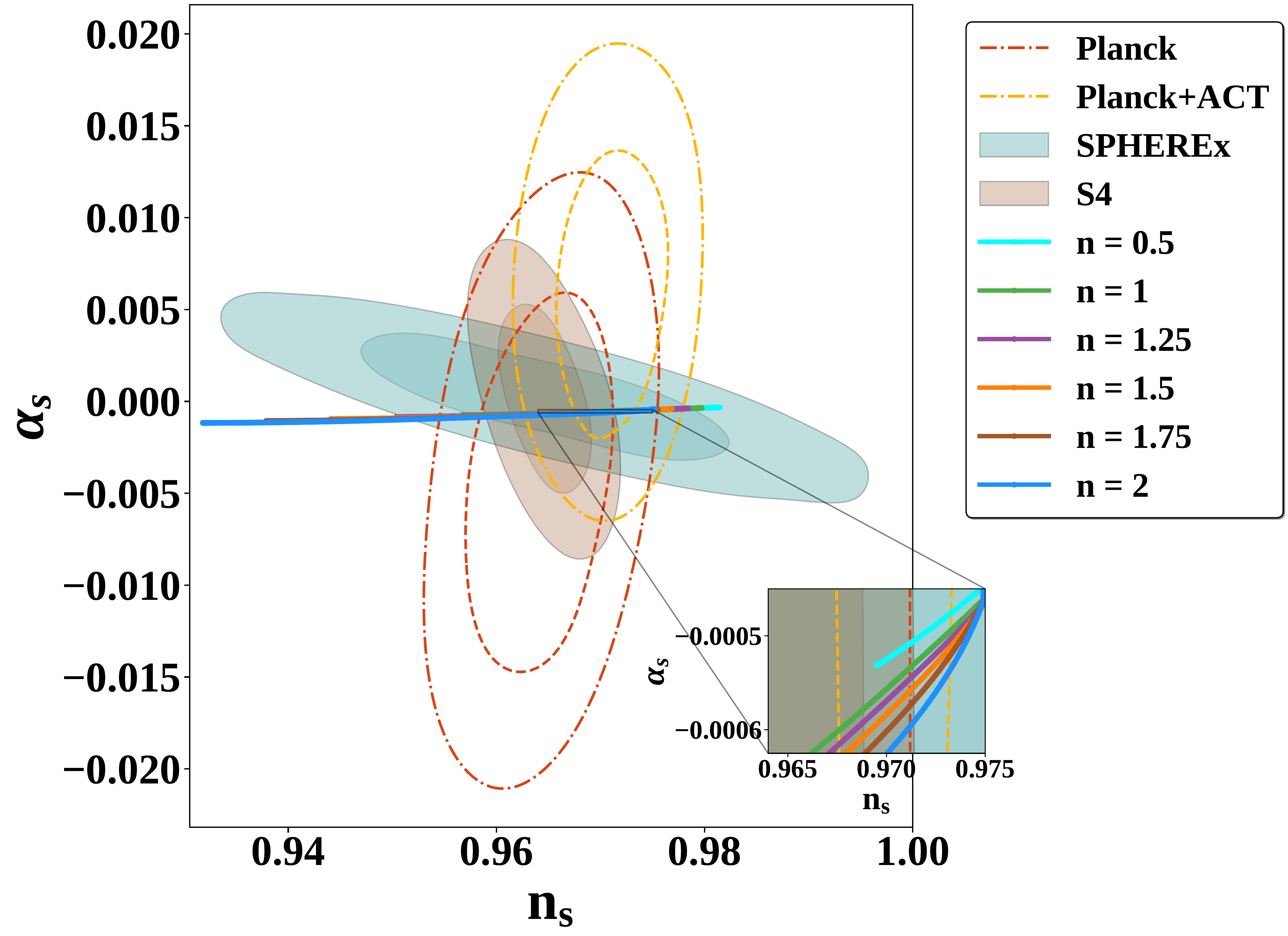}\hfill
\includegraphics[width=0.5\linewidth]{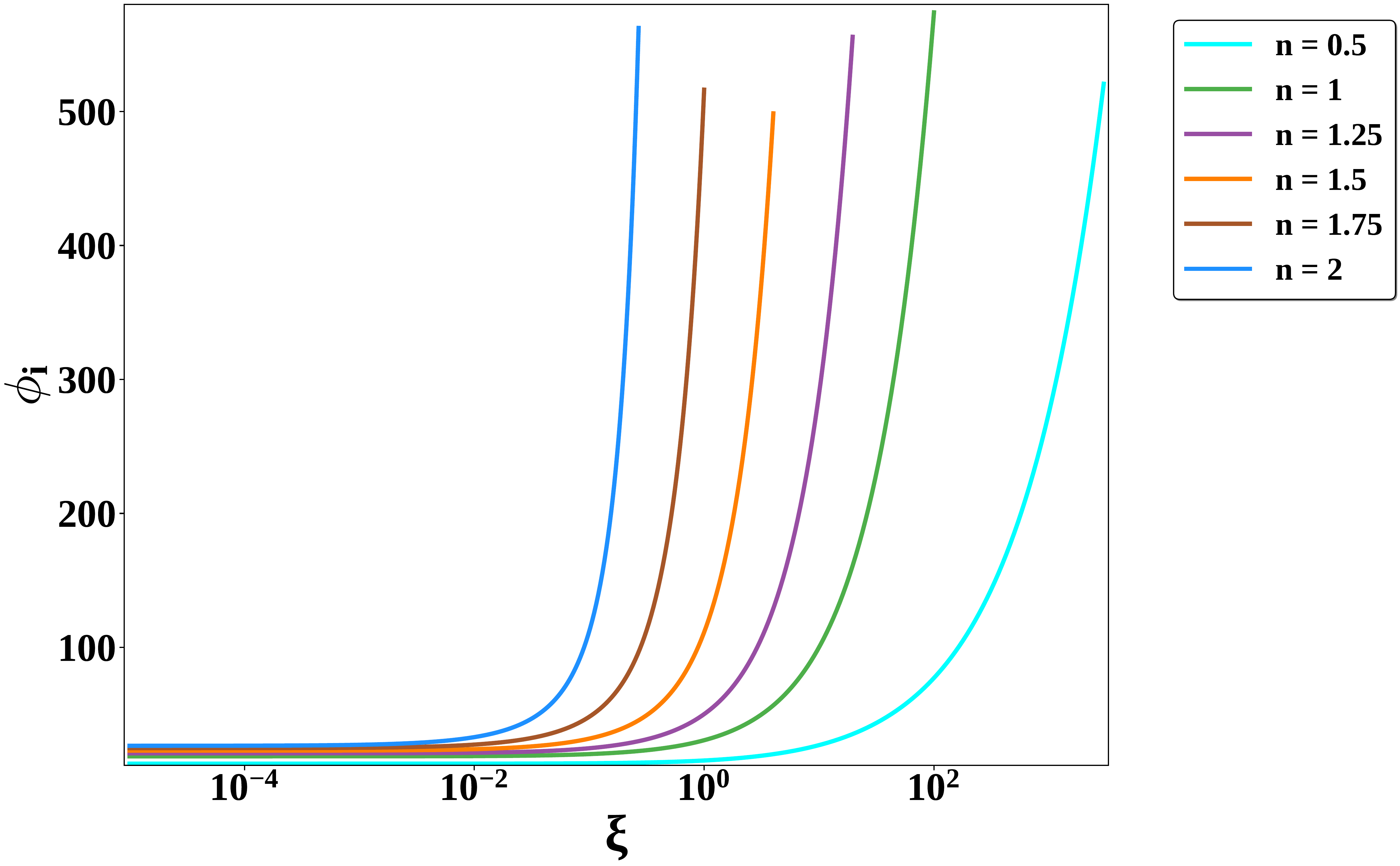}
\caption{\it Same as Fig.~\ref{fig:supp_2_m} but in the \textbf{Palatini} formalism.}
\label{fig:supp_2_P}
\end{figure}
%%%%%%%%%%%%%%%%%%%%%%%%%%%%%%%%%%%%%%%%%%%%%%%%%%%%%%%%%%%%%%%%%%%%%%%%%%%%%%
\clearpage
\newpage
%%%%%%%%%%%%%%%%%%%%%%%%%%%%%%%%%%%%%%%%%%%%%%%%%%%%%%%%%%%%%%%%%%%%%%%%%%%%%%
\begin{figure}[H]
\centering
\includegraphics[width=0.46\linewidth]{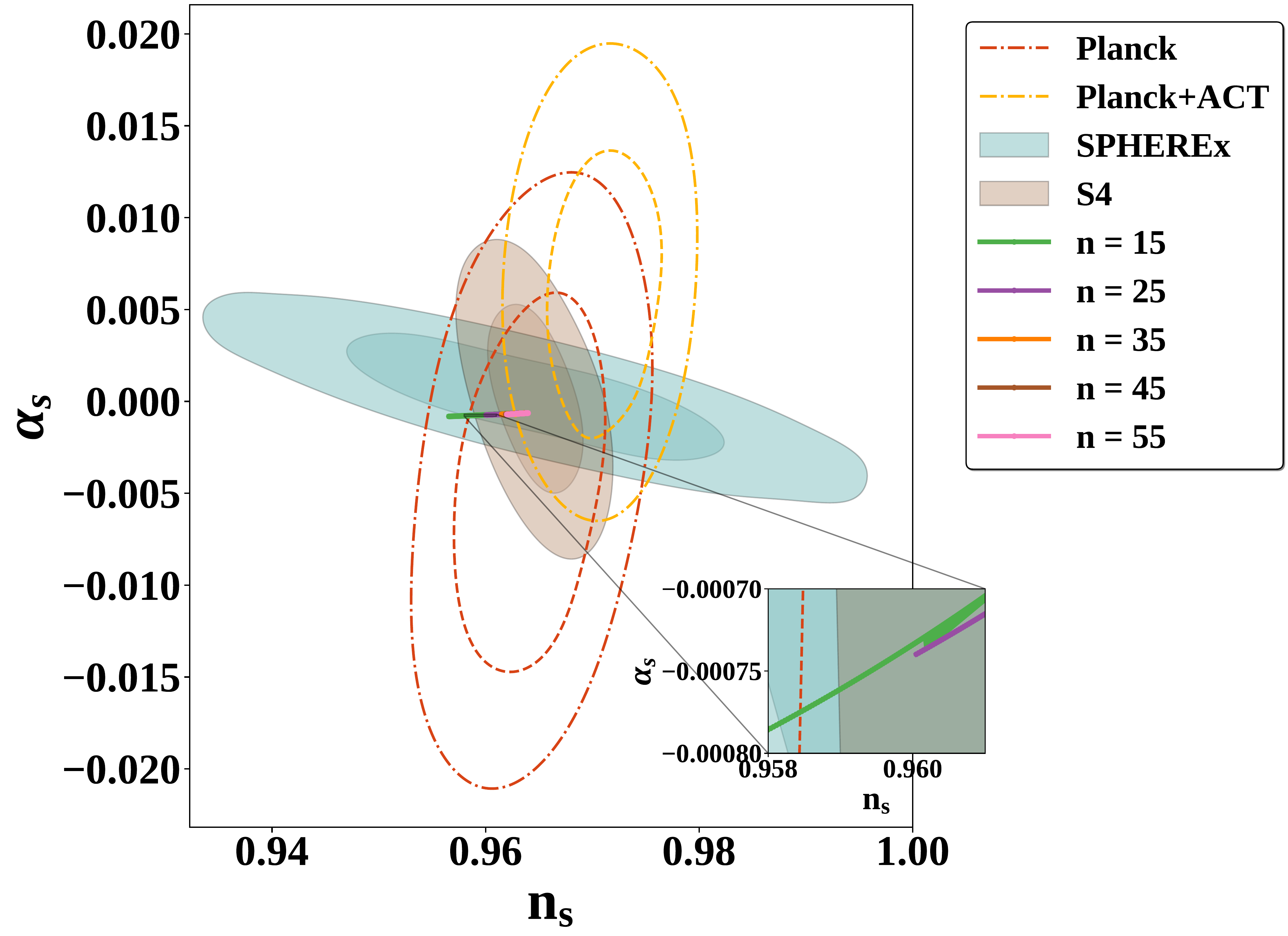}\hfill
\includegraphics[width=0.5\linewidth]{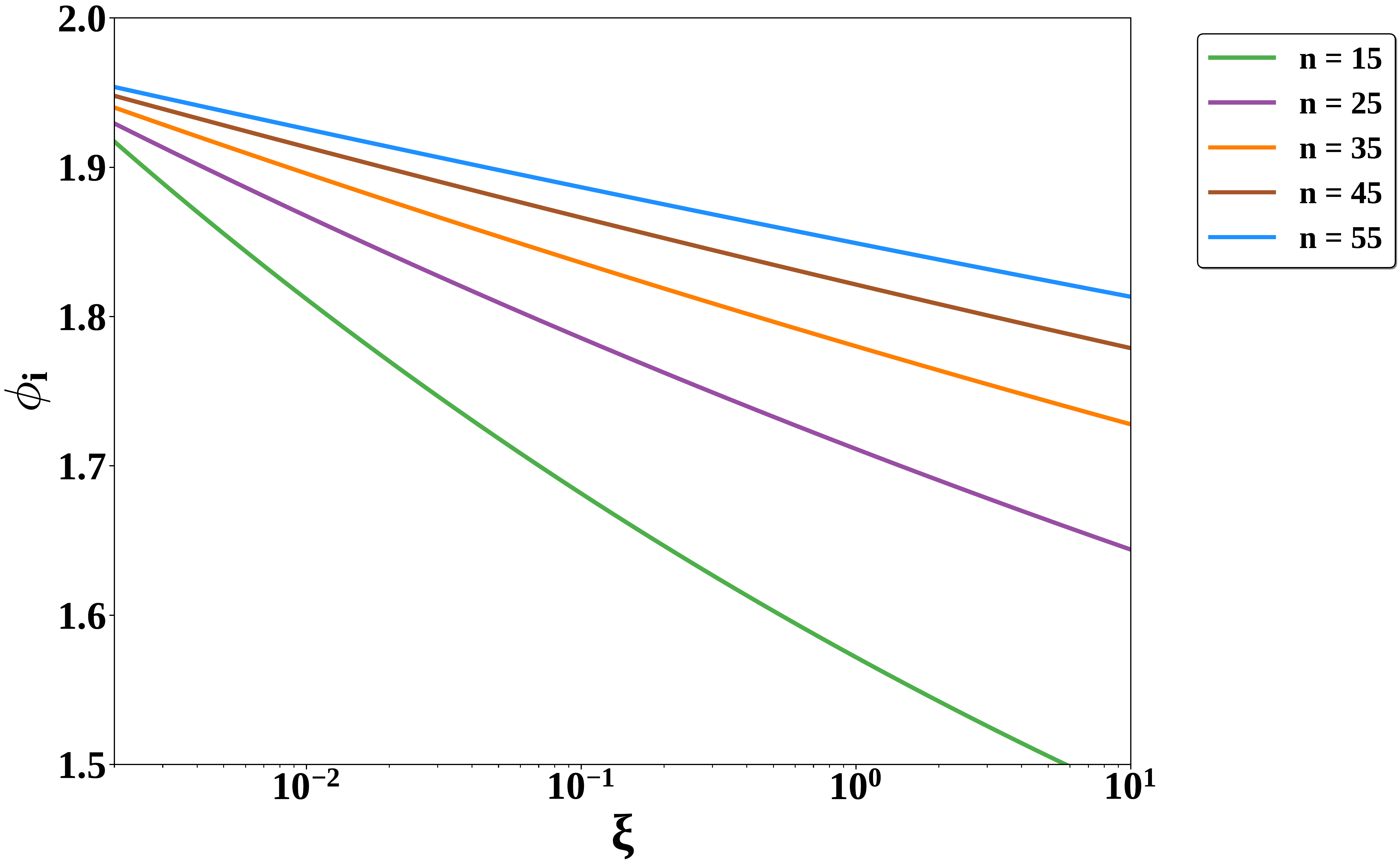}
\caption{\it Supplementary inflationary observables for the Brans-Dicke-like inflation model defined by \eqref{eq:3_potential} in the \textbf{metric} formalism: spectral running \(\alpha_s\) versus \(\xi\) (left) and initial field value \(\phi_i\) versus \(\xi\) (right).}
\label{fig:supp_3_m}
\end{figure}
%%%%%%%%%%%%%%%%%%%%%%%%%%%%%%%%%%%%%%%%%%%%%%%%%%%%%%%%%%%%%%%%%%%%%%%%%%%%%%
\begin{figure}[H]
\centering
\includegraphics[width=0.46\linewidth]{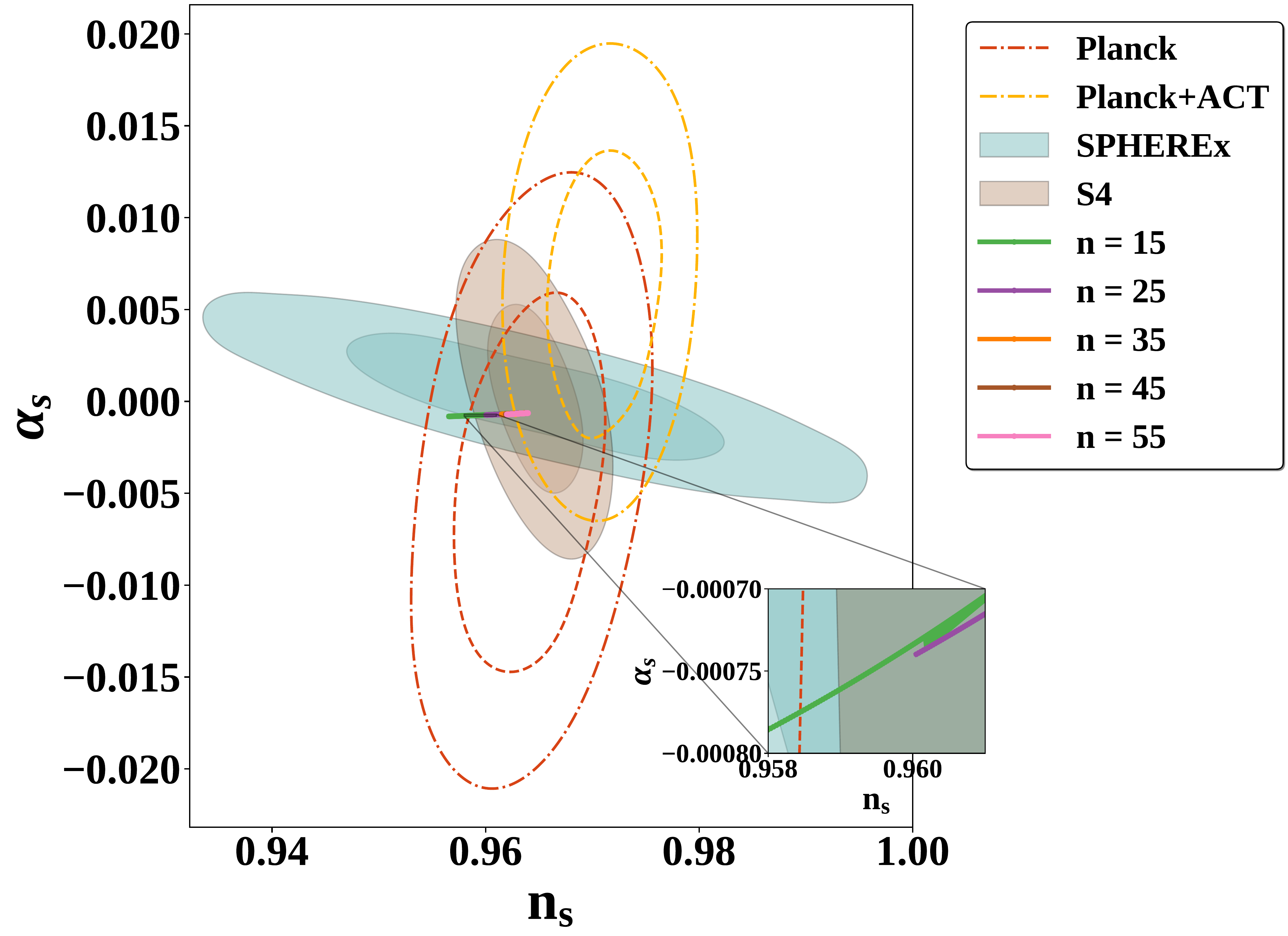}\hfill
\includegraphics[width=0.5\linewidth]{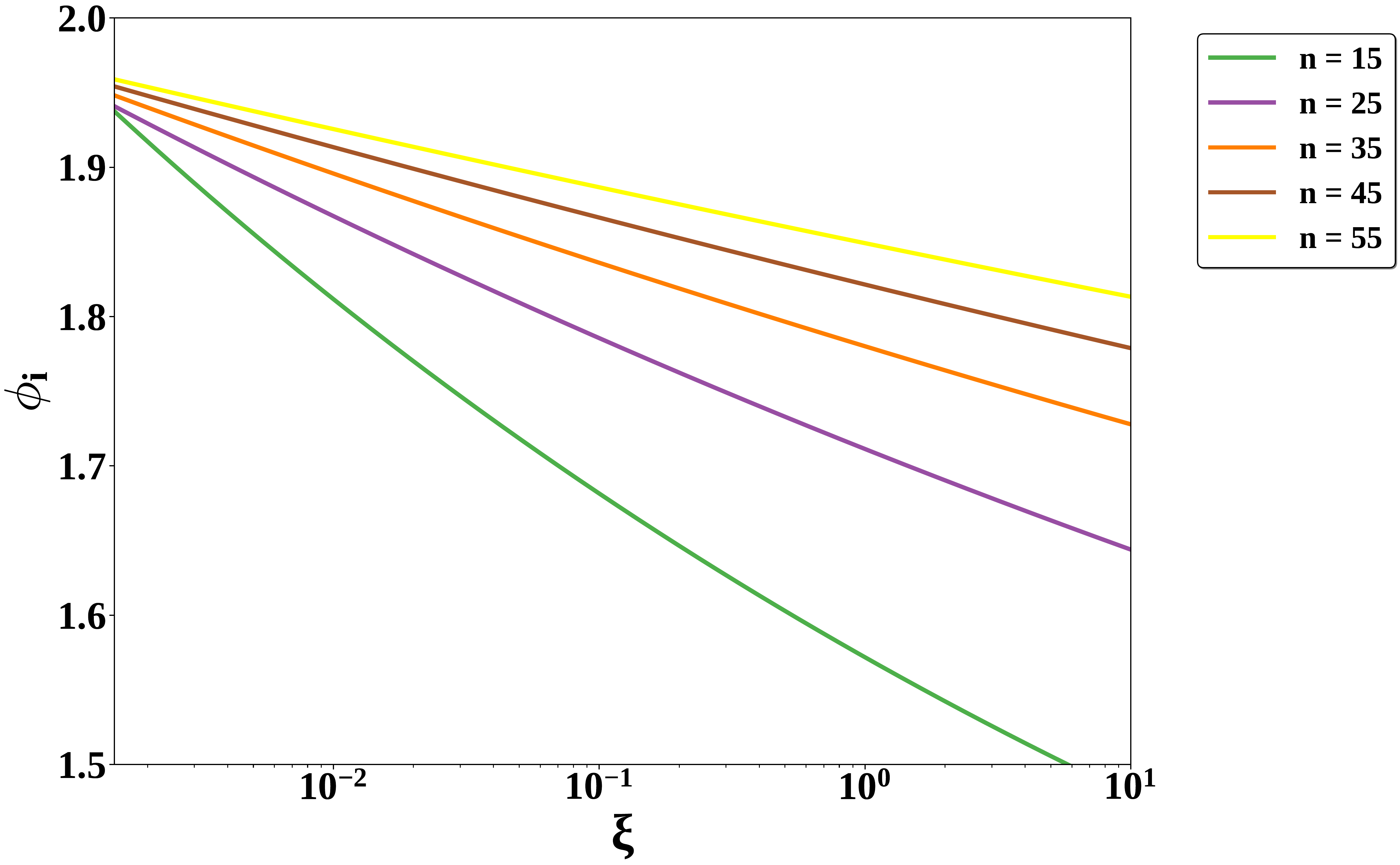}
\caption{\it Same as Fig.~\ref{fig:supp_3_m} but in the \textbf{Palatini} formalism.}
\label{fig:supp_3_P}
\end{figure}
\clearpage
\newpage
%%%%%%%%%%%%%%%%%%%%%%%%%%%%%%%%%%%%%%%%%%%%%%%%%%%%%%%%%%%%%%%%%%%%%%%%%%%%%%
\begin{figure}[H]
\centering
\includegraphics[width=0.46\linewidth]{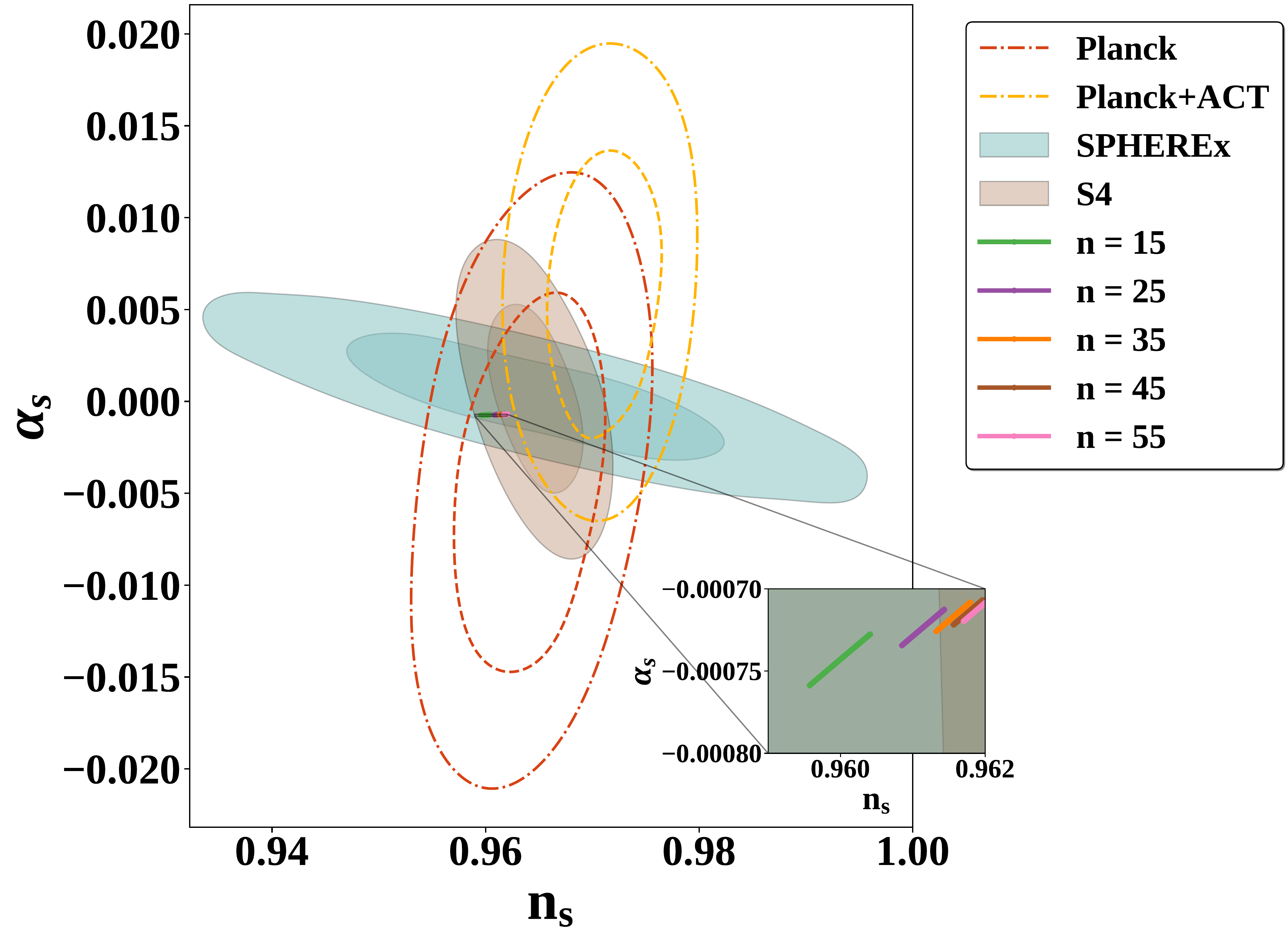}\hfill
\includegraphics[width=0.5\linewidth]{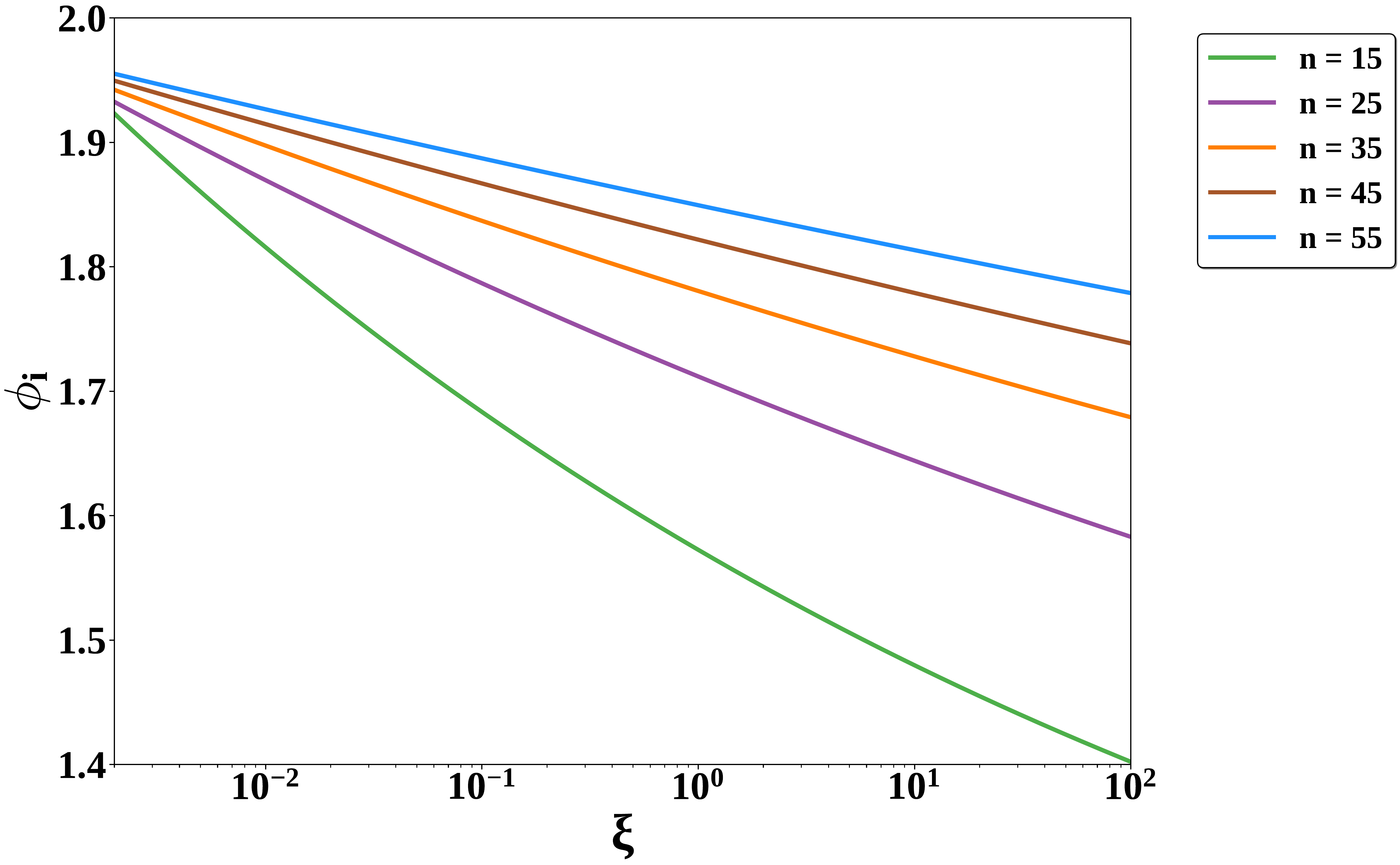}
\caption{\it Supplementary inflationary observables for the Higgs-like inflation defined by \eqref{eq:3_potential} in the \textbf{metric} formalism: spectral running \(\alpha_s\) versus \(\xi\) (left) and initial field value \(\phi_i\) versus \(\xi\) (right).}
\label{fig:supp_4_m}
\end{figure}
%%%%%%%%%%%%%%%%%%%%%%%%%%%%%%%%%%%%%%%%%%%%%%%%%%%%%%%%%%%%%%%%%%%%%%%%%%%%%%
\begin{figure}[H]
\centering
\includegraphics[width=0.46\linewidth]{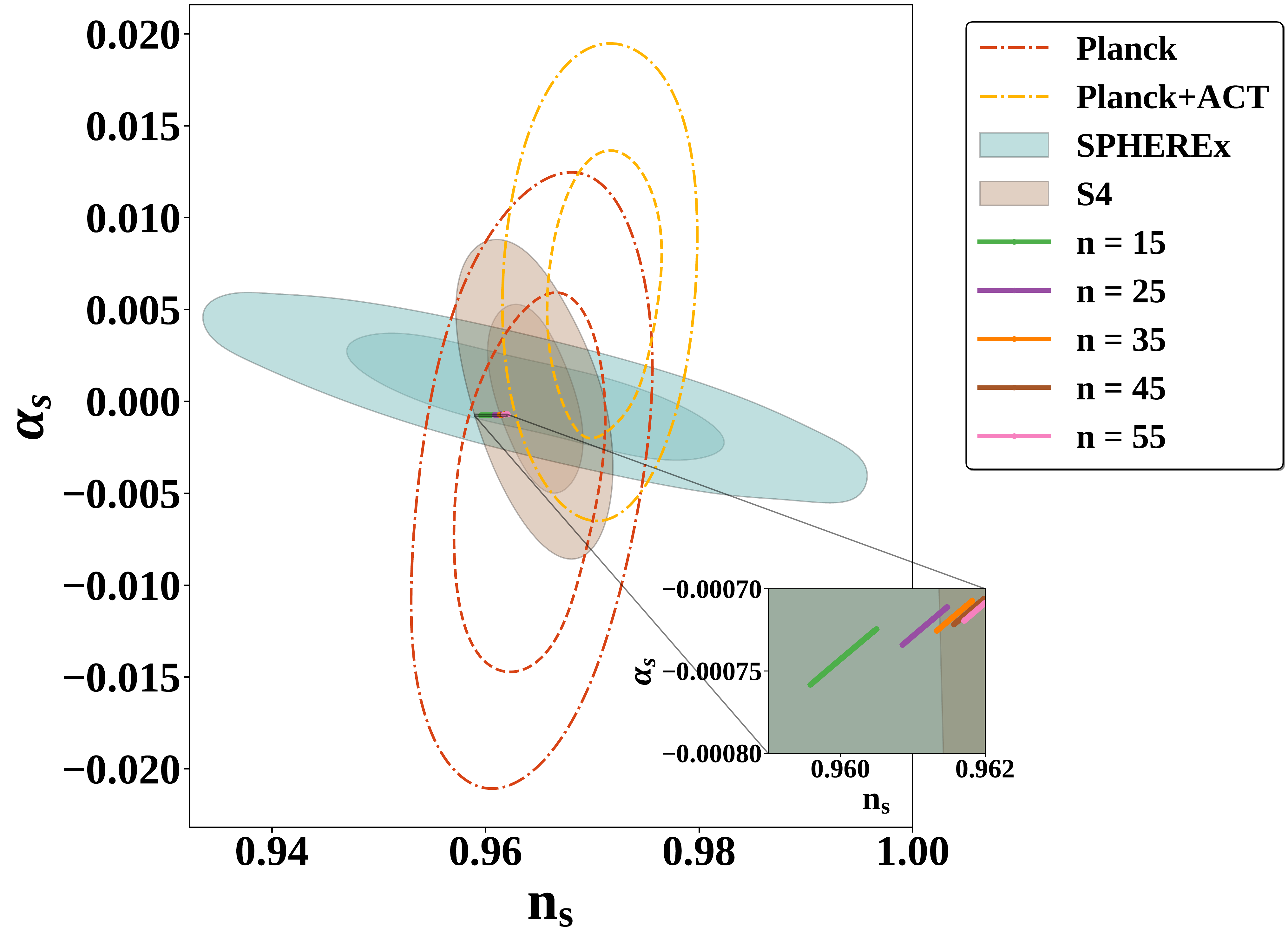}\hfill
\includegraphics[width=0.5\linewidth]{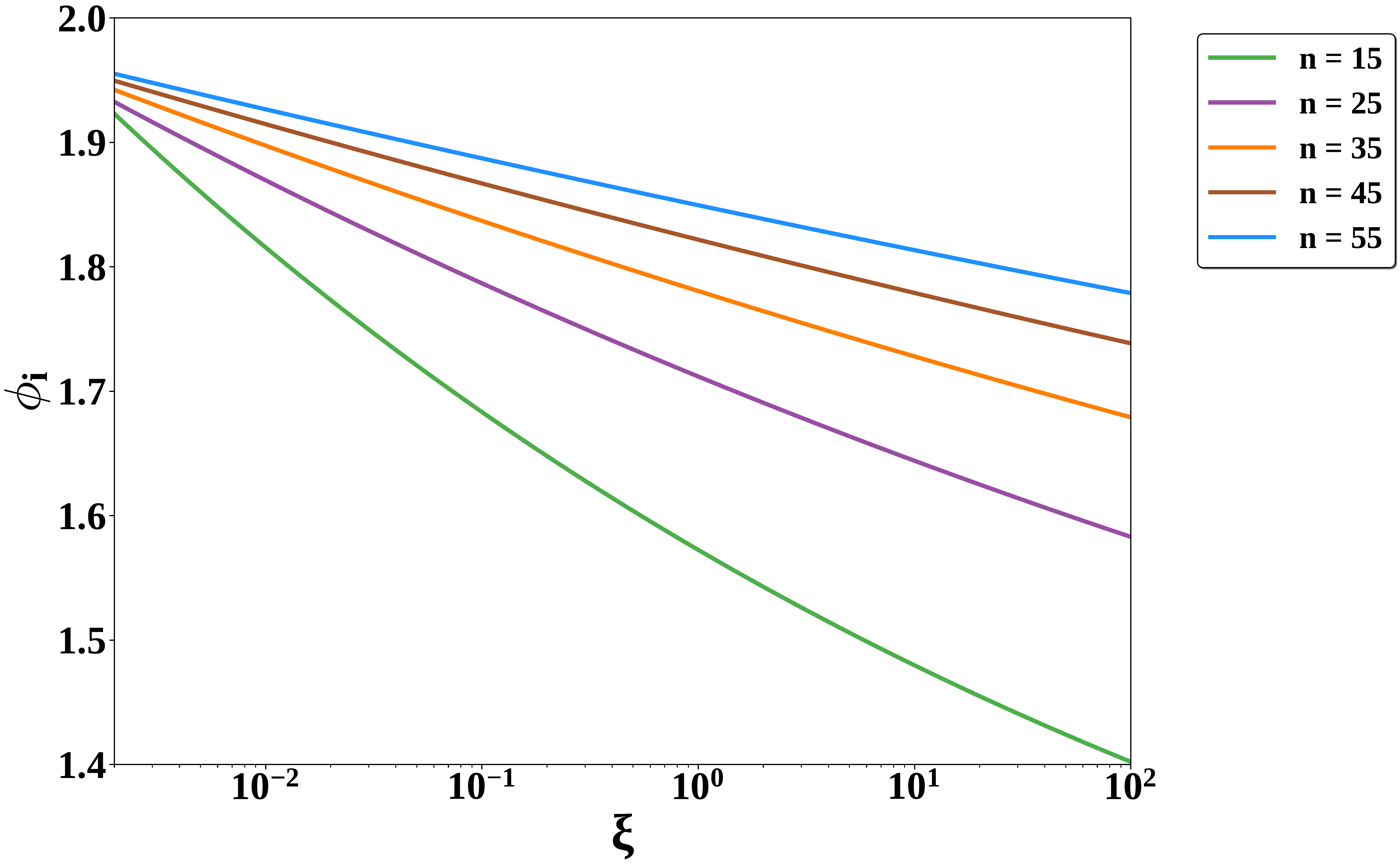}
\caption{\it Same as Fig.~\ref{fig:supp_4_m} but in the \textbf{Palatini} formalism.}
\label{fig:supp_4_P}
\end{figure}
%%%%%%%%%%%%%%%%%%%%%%%%%%%%%%%%%%%%%%%%%%%%%%%%%%%%%%%%%%%%%%%%%%%%%%%%%%%%%%
%%%%%%%%%%%%%%%%%%%%%%%%%%%%%%%%%%%%%%%%%%%%%%%%%%%%%%%%%%%%%%%%%%%%%%%%%%%%%%
%%%%%%%%%%%%%%%%%%%%%%%%%%%%%%%%%%%%%%%%%%%%%%%%%%%%%%%%%%%%%%%%%%%%%%%%%%%%%%
\section{Supplementary Reheating Analyses}
\label{app:other_reheating}
In this appendix, we present the reheating dynamics in both the metric and Palatini formalism for the remaining models: the second linear inflation model defined in Eq.~\eqref{eq:U2_Palatini}, the Brans-Dicke-like model in Eq.~\eqref{eq:3_potential}, and the Higgs-like model in Eq.~\eqref{eq:4_potential}, respectively.
%%%%%%%%%%%%%%%%%%%%%%%%%%%%%%%%%%%%%%%%%%%%%%%%%%%%%%%%%%%%%%%%%%%%%%%%%%%%%%
 \begin{figure}[H]
     \centering
     \includegraphics[width=0.7\linewidth]{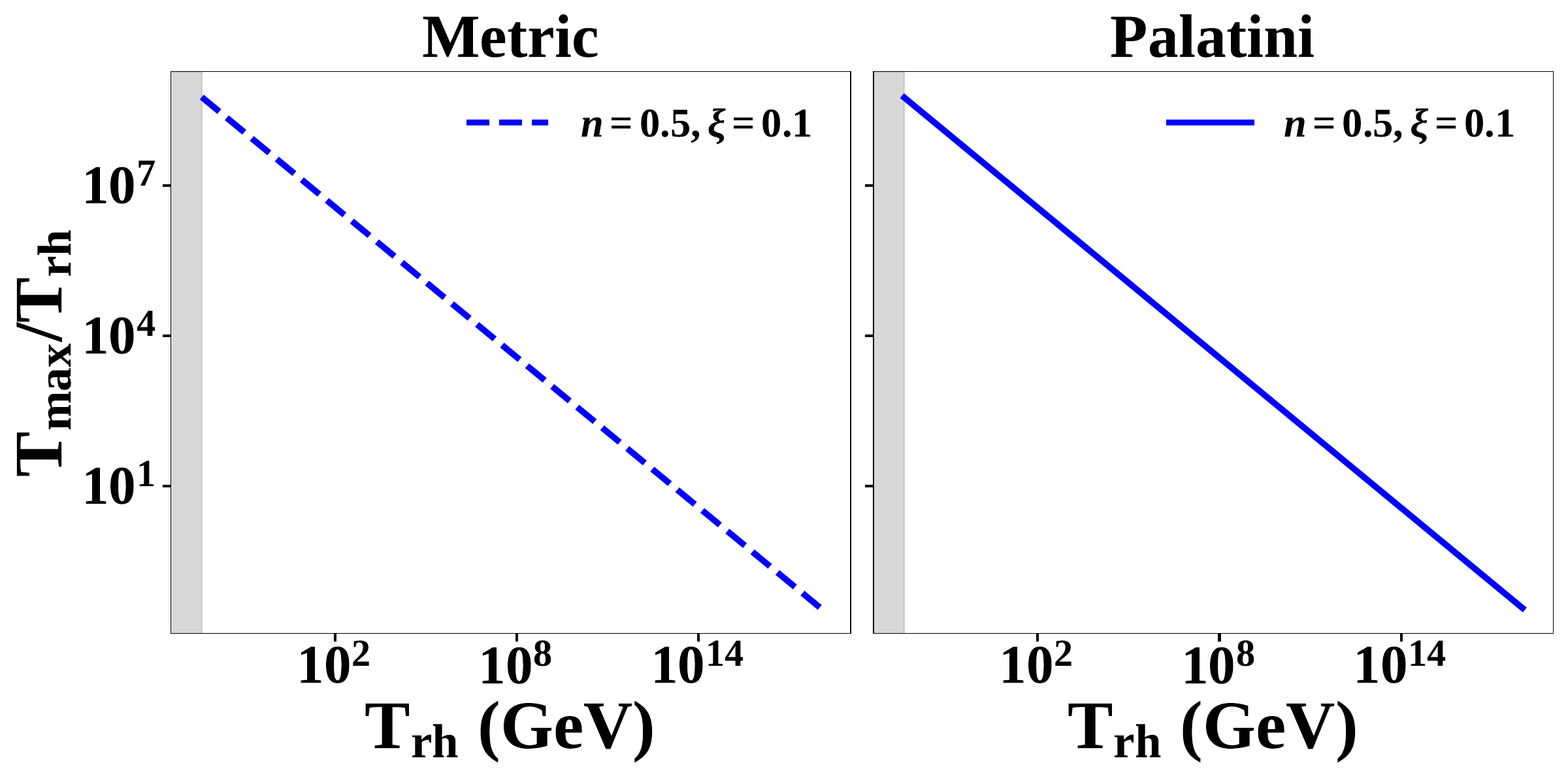}
     \caption{\it Same as Fig.~\ref{fig:tmax_1} but for the model in Eq.~\eqref{eq:U2_Palatini} and corresponds to Tabs.~\ref{Table:stability1m} and~\ref{Table:stability1}.}
     \label{fig:tmax_2}
 \end{figure}
%%%%%%%%%%%%%%%%%%%%%%%%%%%%%%%%%%%%%%%%%%%%%%%%%%%%%%%%%%%%%%%%%%%%%%%%%%%%%%
 \begin{figure}[H]
     \centering
     \includegraphics[width=0.7\linewidth]{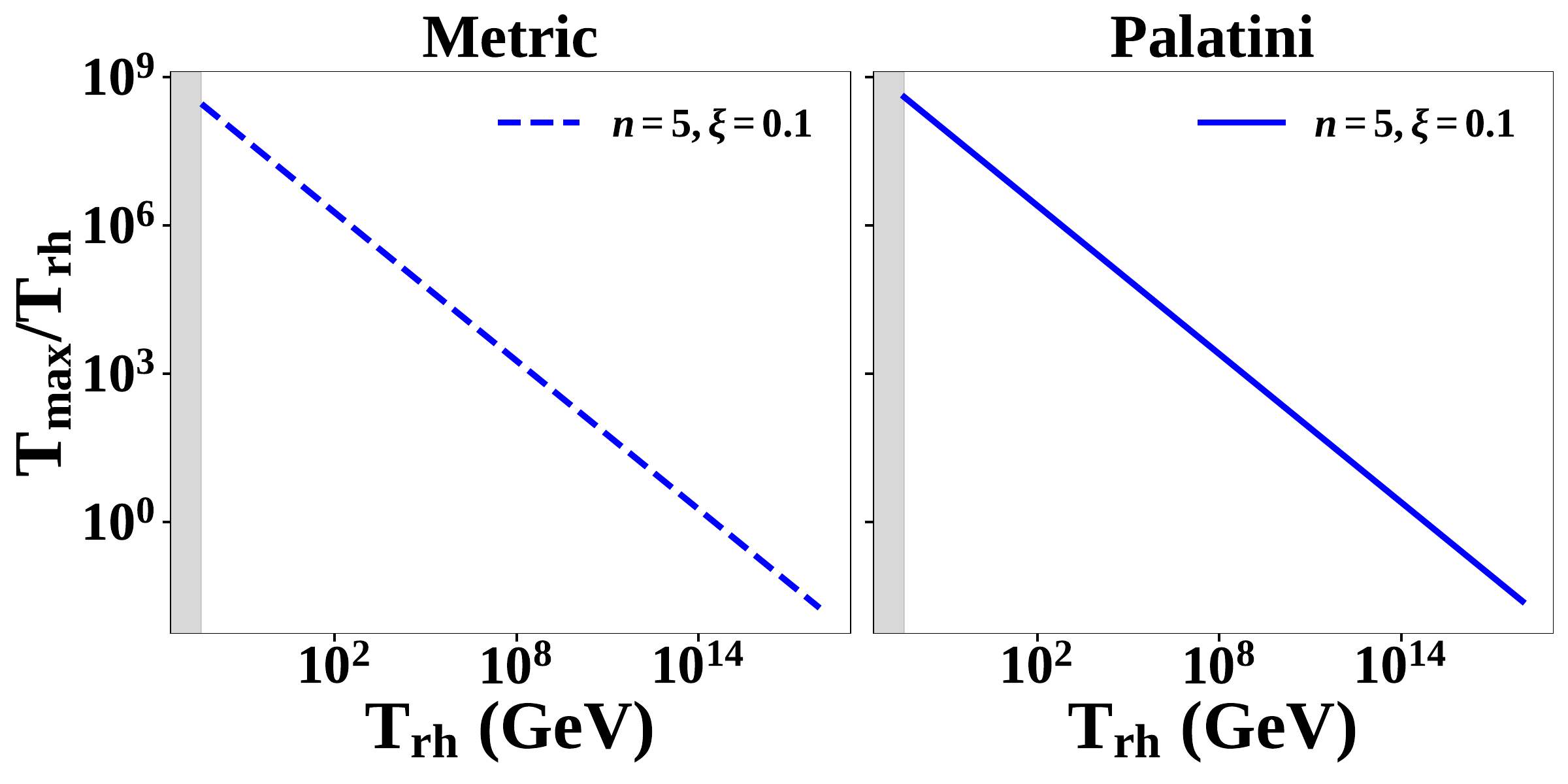}
     \caption{\it Same as Fig.~\ref{fig:tmax_1} but for the model in Eq.~\eqref{eq:3_potential} and corresponds to Tabs.~\ref{Table:stability2m} and~\ref{Table:stability2}.}
     \label{fig:tmax_3}
 \end{figure}
 %%%%%%%%%%%%%%%%%%%%%%%%%%%%%%%%%%%%%%%%%%%%%%%%%%%%%%%%%%%%%%%%%%%%%%%%%%%%%%
\begin{figure}[H]
     \centering
     \includegraphics[width=0.7\linewidth]{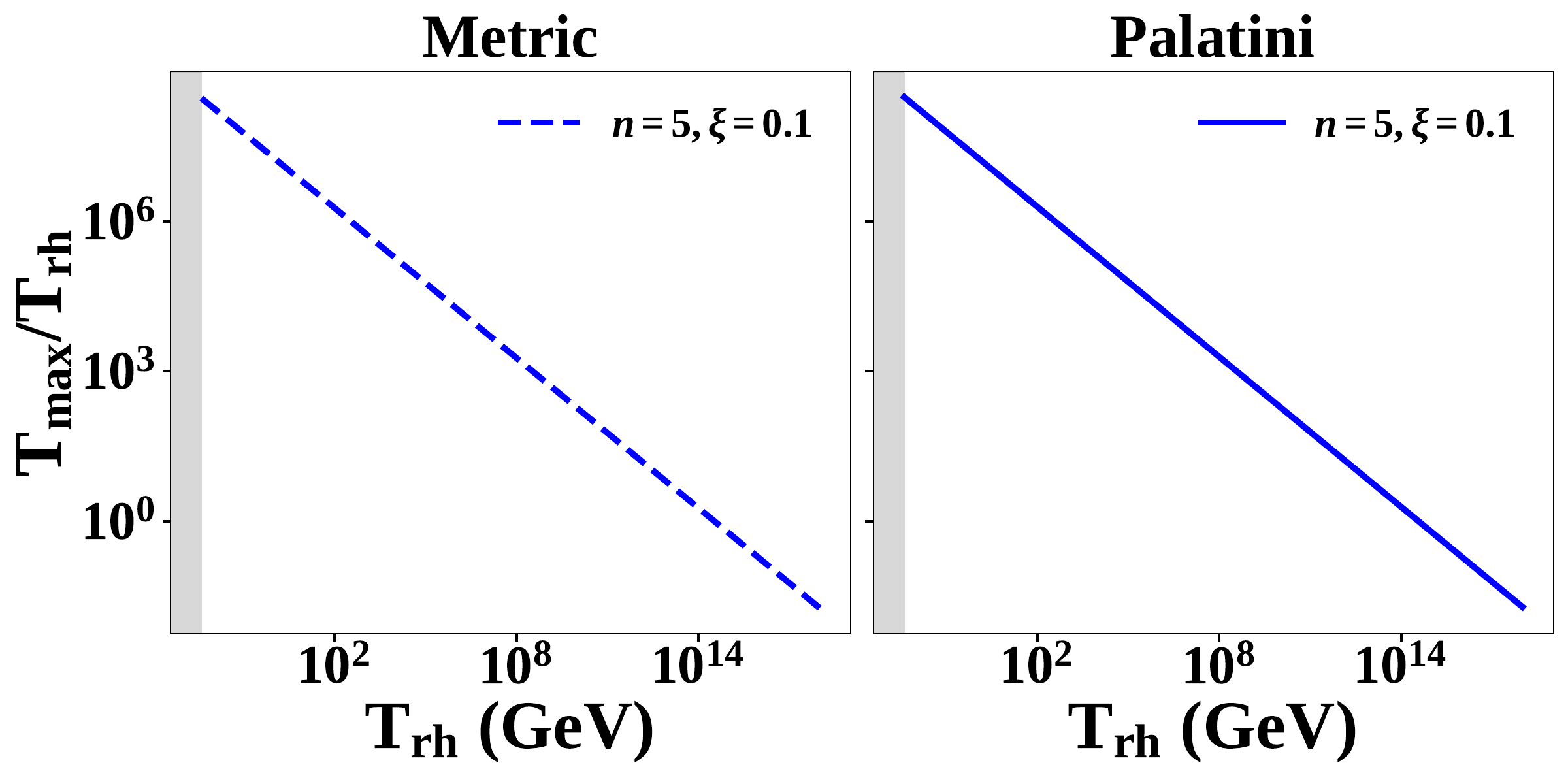}
     \caption{\it Same as Fig.~\ref{fig:tmax_1} but for the model in Eq.~\eqref{eq:4_potential} and corresponds to Tabs.~\ref{Table:stability3m} and~\ref{Table:stability3}.}
     \label{fig:tmax_4}
 \end{figure}
 %%%%%%%%%%%%%%%%%%%%%%%%%%%%%%%%%%%%%%%%%%%%%%%%%%%%%%%%%%%%%%%%%%%%%%%%%%%%%%
\section{Supplementary Dark Matter Production Analyses}
\label{app:other_DM}
In this appendix, we present the rest of the figures that depict the results related to the DM production via inflaton decay which is discussed in details in Subsec.~\ref{subsec:dm_decay}, we particularly show the figures of the remaining models for both metric and Palatini formulations: the second linear inflation model defined in Eq.~\eqref{eq:U2_Palatini}, the Brans-Dicke-like model in Eq.~\eqref{eq:3_potential}, and the Higgs-like model in Eq.~\eqref{eq:4_potential}, respectively.
%%%%%%%%%%%%%%%%%%%%%%%%%%%%%%%%%%%%%%%%%%%%%%%%%%%%%%%%%%%%%%%%%%%%%%%%%%%%%%
\begin{figure}[H]
     \centering
     \includegraphics[width=0.7\linewidth]{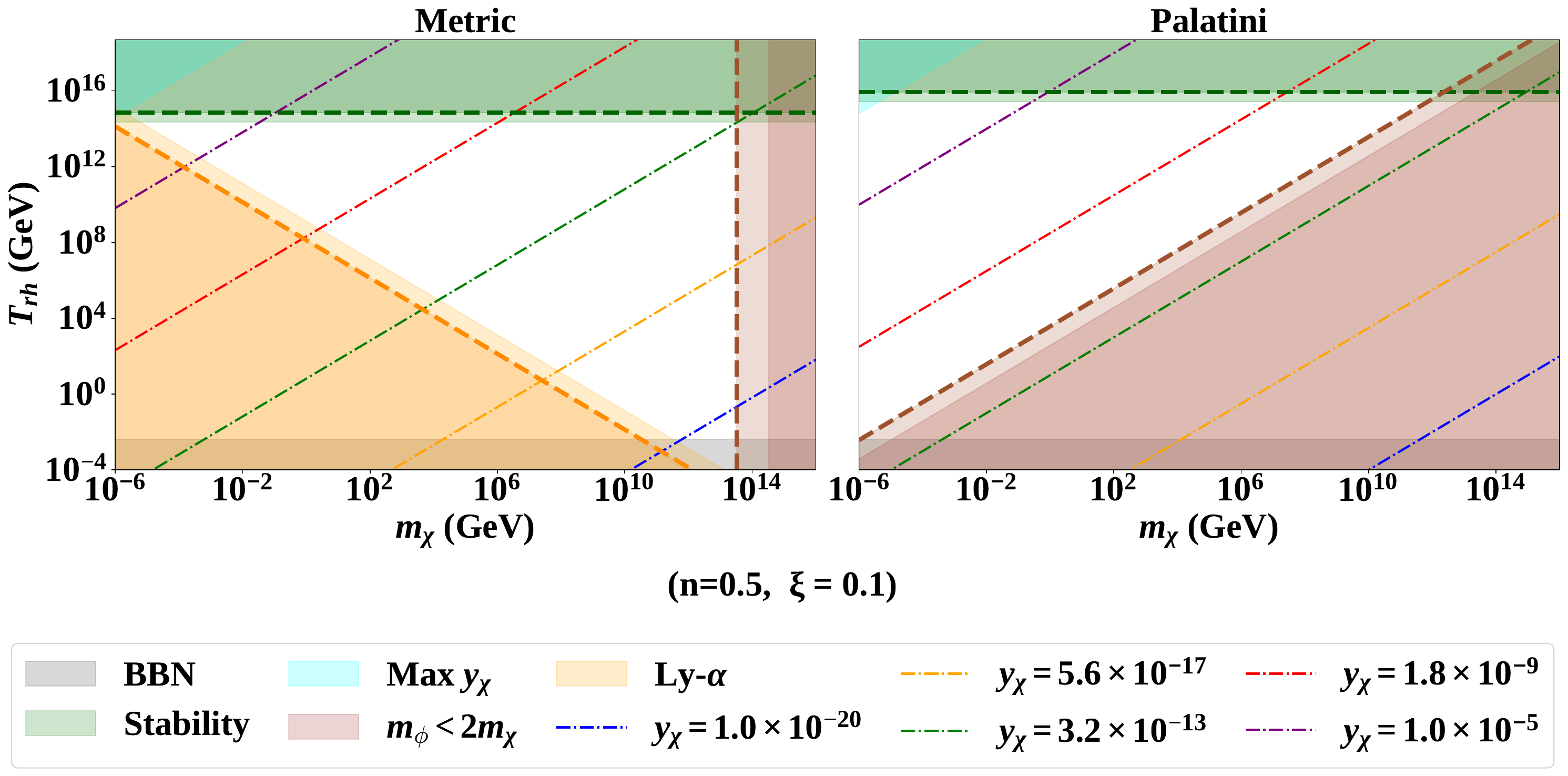}
     \caption{\it Same as Fig.~\ref{fig:dm_1} for the linear inflation model expressed in Eq.~\eqref{eq:U2_Palatini}. The values are as given in Tabs.~\ref{Table:stability1m} and~\ref{Table:stability1}, achieving viable inflationary parameters as detailed in Tabs.~\ref{tab:infl-para_2}.}
     \label{fig:dm_2}
 \end{figure}
%%%%%%%%%%%%%%%%%%%%%%%%%%%%%%%%%%%%%%%%%%%%%%%%%%%%%%%%%%%%%%%%%%%%%%%%%%%%%%
 \begin{figure}[H]
     \centering
     \includegraphics[width=0.7\linewidth]{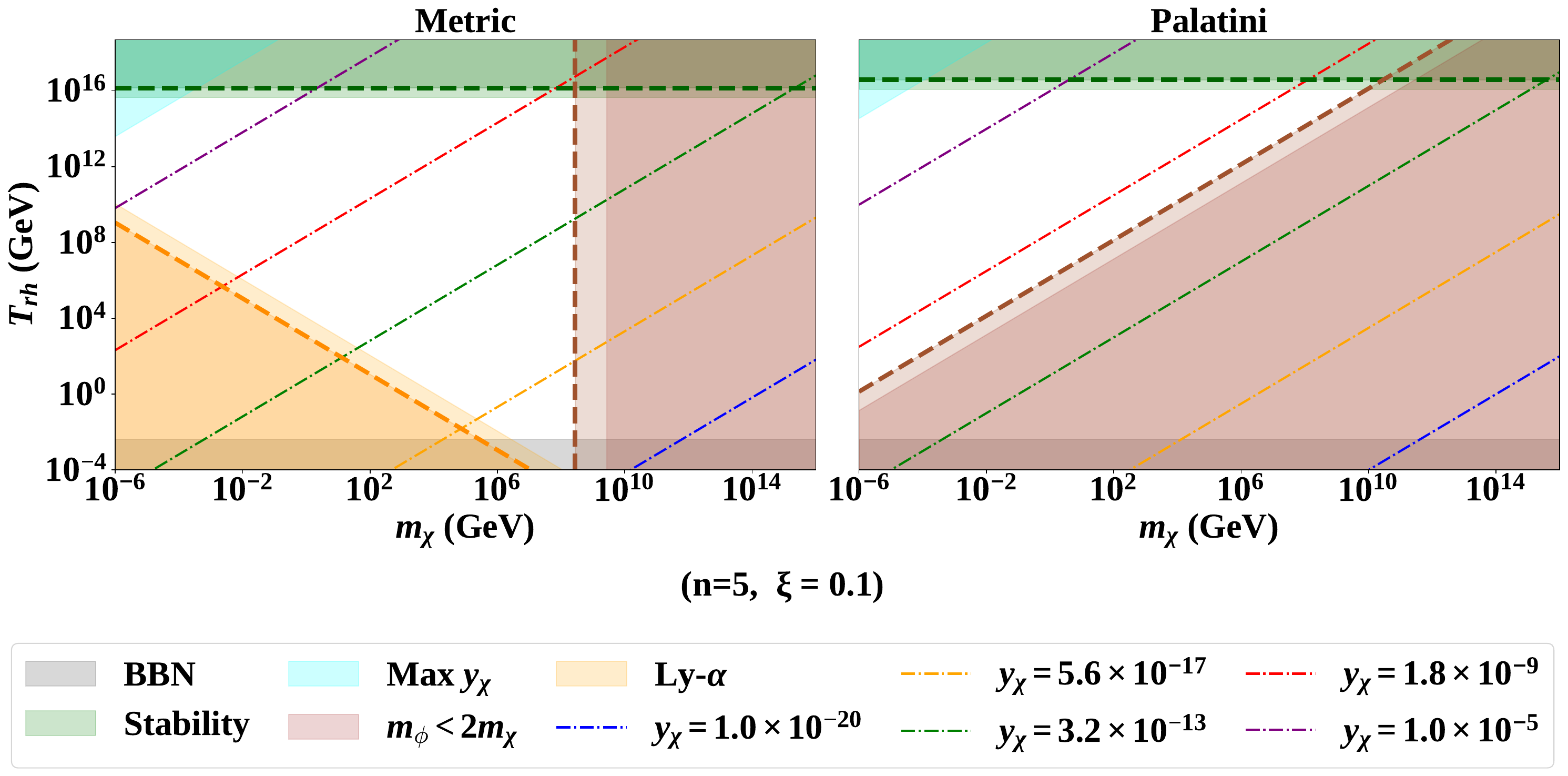}
     \caption{\it Same as Fig.~\ref{fig:dm_1} for the Brans-Dicke-like inflation model expressed in Eq.~\eqref{eq:3_potential}. The values are as given in Tabs.~\ref{Table:stability2m} and~\ref{Table:stability2}, achieving viable inflationary parameters as detailed in Tabs.~\ref{tab:infl-para_3}.}
     \label{fig:dm_3}
 \end{figure}
%%%%%%%%%%%%%%%%%%%%%%%%%%%%%%%%%%%%%%%%%%%%%%%%%%%%%%%%%%%%%%%%%%%%%%%%%%%%%%
   \begin{figure}[H]
     \centering
     \includegraphics[width=0.7\linewidth]{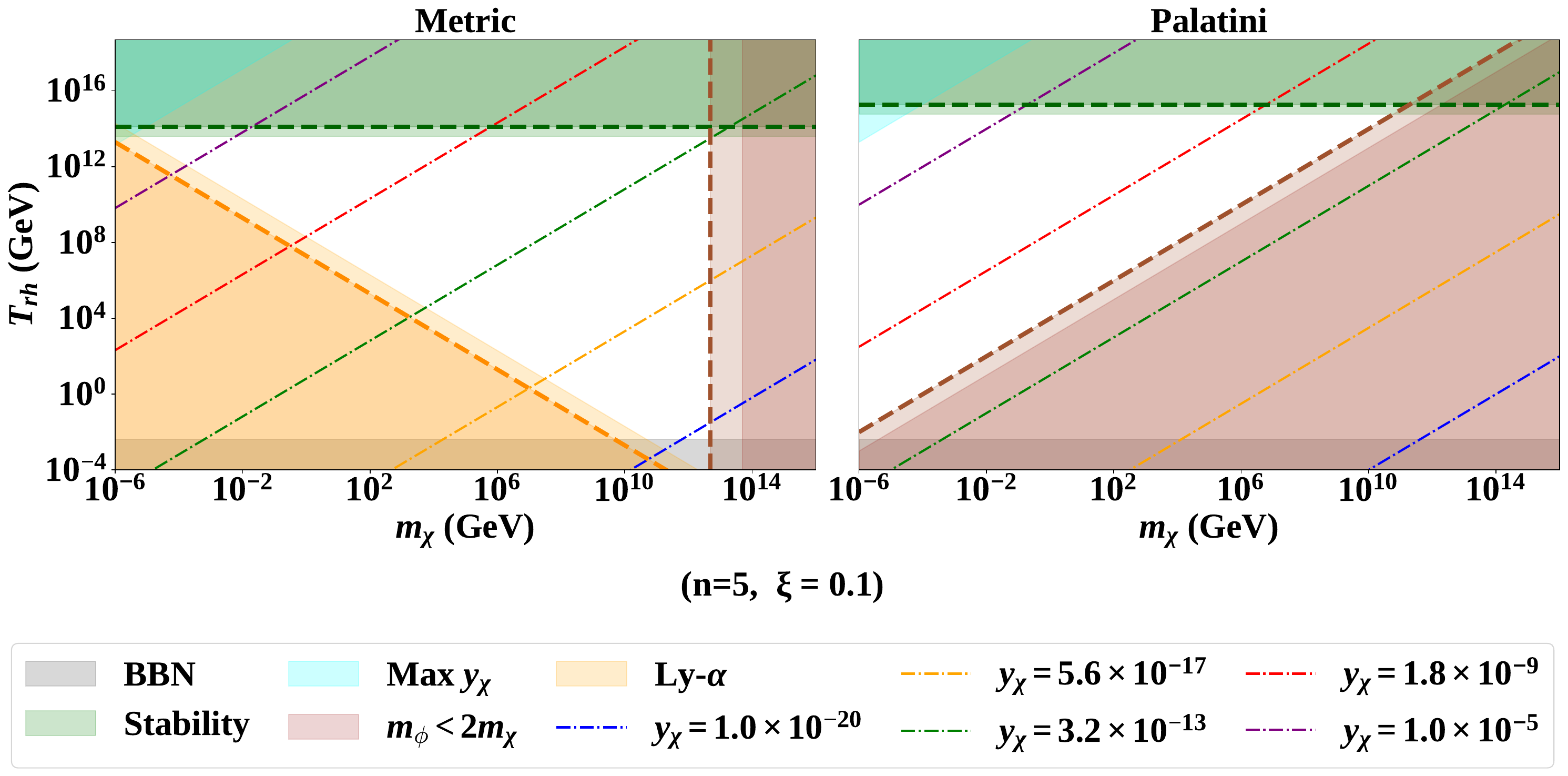}
     \caption{\it Same as Fig.~\ref{fig:dm_1} for the Higgs-like inflation model expressed in Eq.~\eqref{eq:4_potential}. The values are as given in Tabs.~\ref{Table:stability3m} and~\ref{Table:stability3}, achieving viable inflationary parameters as detailed in Tabs.~\ref{tab:infl-para_4}.}
     \label{fig:dm_4}
 \end{figure}
%%%%%%%%%%%%%%%%%%%%%%%%%%%%%%%%%%%%%%%%%%%%%%%%%%%%%%%%%%%%%%%%%%%%%%%%%%%%%%
%%%%%%%%%%%%%%%%%%%%%%%%%%%%%%%%%%%%%%%%%%%%%%%%%%%%%%%%%%%%%%%%%%%%%%%%%%%%%%
%%%%%%%%%%%%%%%%%%%%%%%%%%%%%%%%%%%%%%%%%%%%%%%%%%%%%%%%%%%%%%%%%%%%%%%%%%%%%%
\section{Supplementary Leptogenesis Analyses}
\label{app:other_leptogenesis}
Following the same logic in App.~\ref{app:other_DM} 
In this appendix, we present the non-thermal leptogenesis analyses for the remaining inflation models.
%%%%%%%%%%%%%%%%%%%%%%%%%%%%%%%%%%%%%%%%%%%%%%%%%%%%%%%%%%%%%%%%%%%%%%%%%%%%%%
\begin{figure}[H]
     \centering
     \includegraphics[width=0.7\linewidth]{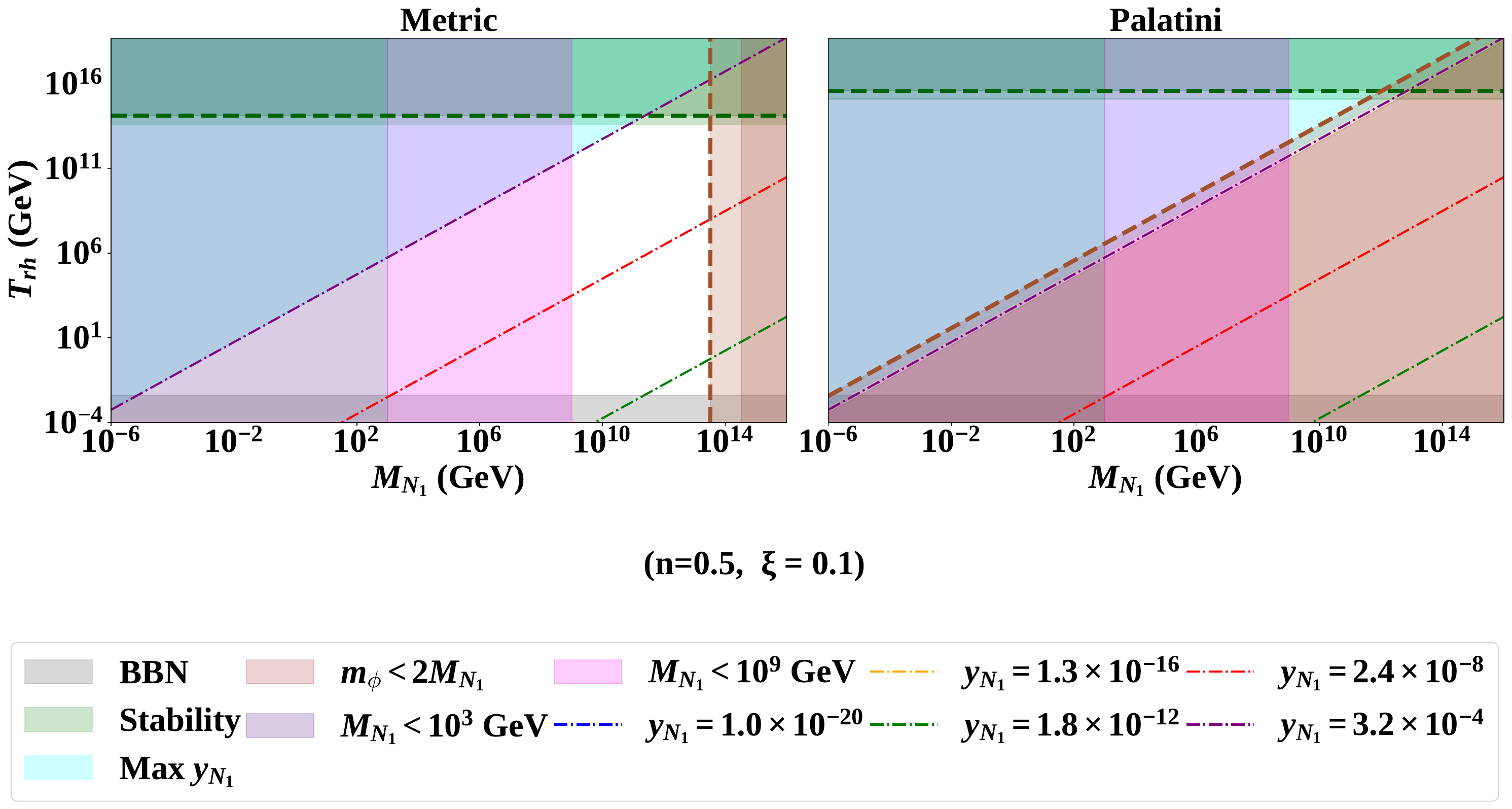}
     \caption{\it Same as Fig.~\ref{fig:lepto_1} for the linear inflation model expressed in Eq.~\eqref{eq:U2_Palatini}. The values are as given in Tabs.~\ref{Table:stability1m} and~\ref{Table:stability1}, achieving viable inflationary parameters as detailed in Tabs.~\ref{tab:infl-para_2}.}
     \label{fig:lepto_2}
 \end{figure}
%%%%%%%%%%%%%%%%%%%%%%%%%%%%%%%%%%%%%%%%%%%%%%%%%%%%%%%%%%%%%%%%%%%%%%%%%%%%%%
\begin{figure}[H]
     \centering
     \includegraphics[width=0.7\linewidth]{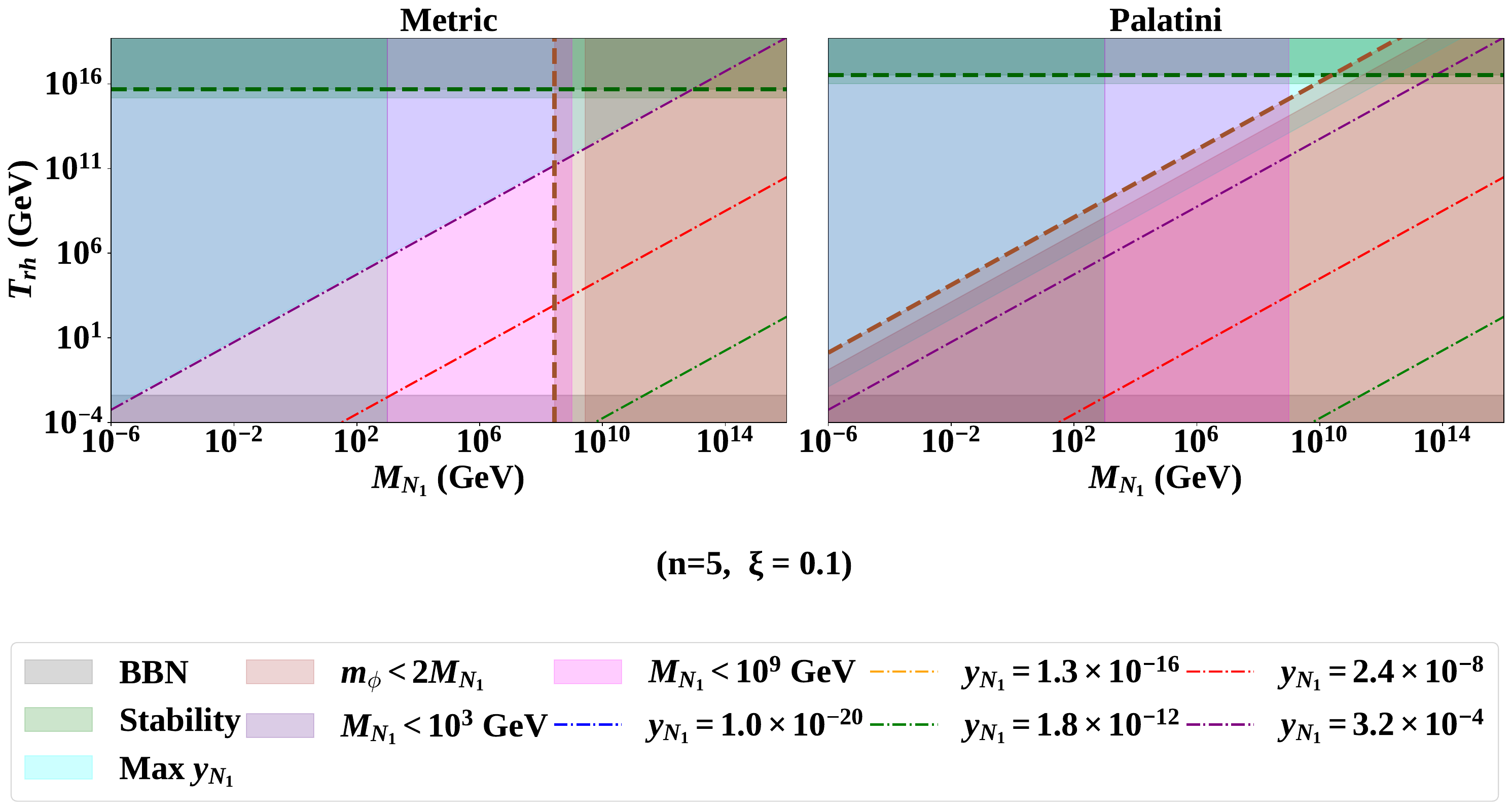}
     \caption{\it Same as Fig.~\ref{fig:lepto_1} for the Brans-Dicke-like inflation model expressed in Eq.~\eqref{eq:3_potential}. The values are as given in Tabs.~\ref{Table:stability2m} and~\ref{Table:stability2}, achieving viable inflationary parameters as detailed in Tabs.~\ref{tab:infl-para_3}.}
     \label{fig:lepto_3}
 \end{figure}
%%%%%%%%%%%%%%%%%%%%%%%%%%%%%%%%%%%%%%%%%%%%%%%%%%%%%%%%%%%%%%%%%%%%%%%%%%%%%%
\begin{figure}[H]
     \centering
     \includegraphics[width=0.7\linewidth]{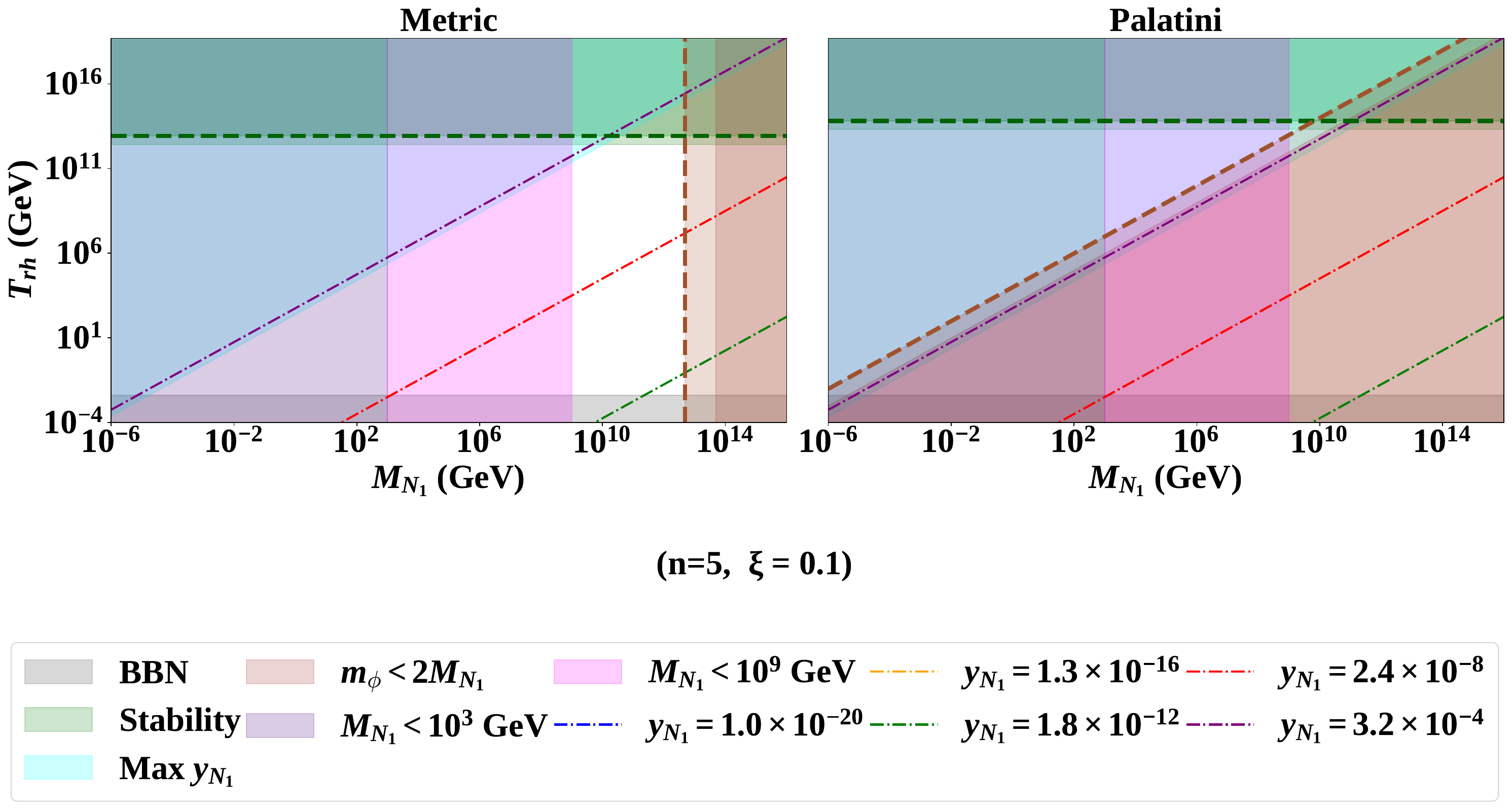}
     \caption{\it Same as Fig.~\ref{fig:lepto_1} for the Higgs-like inflation model expressed in Eq.~\eqref{eq:4_potential}. The values are as given in Tabs.~\ref{Table:stability3m} and~\ref{Table:stability3}, achieving viable inflationary parameters as detailed in Tabs.~\ref{tab:infl-para_4}.}
     \label{fig:lepto_4}
 \end{figure}
 %%%%%%%%%%%%%%%%%%%%%%%%%%%%%%%%%%%%%%%%%%%%%%%%%%%%%%%%%%%%%%%%%%%%%%%%%%%%%%
 %%%%%%%%%%%%%%%%%%%%%%%%%%%%%%%%%%%%%%%%%%%%%%%%%%%%%%%%%%%%%%%%%%%%%%%%%%%%%%
 %%%%%%%%%%%%%%%%%%%%%%%%%%%%%%%%%%%%%%%%%%%%%%%%%%%%%%%%%%%%%%%%%%%%%%%%%%%%%%
 %%%%%%%%%%%%%%%%%%%%%%%%%%%%%%%%%%%%%%%%%%%%%%%%%%%%%%%%%%%%%%%%%%%%%%%%%%%%%%
 %%%%%%%%%%%%%%%%%%%%%%%%%%%%%%%%%%%%%%%%%%%%%%%%%%%%%%%%%%%%%%%%%%%%%%%%%%%%%%
 %%%%%%%%%%%%%%%%%%%%%%%%%%%%%%%%%%%%%%%%%%%%%%%%%%%%%%%%%%%%%%%%%%%%%%%%%%%%%%
 %%%%%%%%%%%%%%%%%%%%%%%%%%%%%%%%%%%%%%%%%%%%%%%%%%%%%%%%%%%%%%%%%%%%%%%%%%%%%%

\bibliography{main}
\bibliographystyle{JHEP}
%\bibliography{reference}
%\bibliographystyle{unsrt}
% \input{ref.bib}
\end{document}